\documentclass[a4paper,11pt]{article}
\usepackage{amsmath}
\usepackage{amssymb}
\usepackage{graphicx}
\usepackage{parskip}
\usepackage{amsthm}
\usepackage{fancyhdr}
\usepackage{dsfont}
\usepackage{algorithm}
\usepackage{algorithmic}
\usepackage{hyperref}
\usepackage{bm}
\usepackage[authoryear]{natbib}

\theoremstyle{definition}

\newtheorem{rmk}{Remark}

\newcommand{\N}{ \mathbb{ N } }
\newcommand{\E}{ \mathbb{ E } }

\newcommand{\Prb}{ \mathbb{ P } }

\newcommand{\bd}{\begin{displaymath}}
\newcommand{\ed}{\end{displaymath}}
\newcommand{\be}{\begin{equation}}
\newcommand{\ee}{\end{equation}}

\title{Multi-locus data distinguishes between population growth and multiple merger coalescents}

\author{Jere Koskela \\
	\texttt{j.koskela@warwick.ac.uk}\\
	\small Department of Statistics\\
	\small University of Warwick \\
	\small Coventry, CV4 7AL \\
	\small United Kingdom
}

\date{\today}

\begin{document}

\maketitle

\begin{abstract}
We introduce a low dimensional function of the site frequency spectrum that is tailor-made for distinguishing coalescent models with multiple mergers from Kingman coalescent models with population growth, and use this function to construct a hypothesis test between these model classes.
The null and alternative sampling distributions of the statistic are intractable, but its low dimensionality renders them amenable to Monte Carlo estimation.
We construct kernel density estimates of the sampling distributions based on simulated data, and show that the resulting hypothesis test dramatically improves on the statistical power of a current state-of-the-art method.
A key reason for this improvement is the use of multi-locus data, in particular averaging observed site frequency spectra across unlinked loci to reduce sampling variance.
We also demonstrate the robustness of our method to nuisance and tuning parameters.
Finally we show that the same kernel density estimates can be used to conduct parameter estimation, and argue that our method is readily generalisable for applications in model selection, parameter inference and experimental design.
\end{abstract}

\section{Introduction} \label{intro}

The Kingman coalescent \citep{K82,K82b,Ki82c, H1983a,H1983b, T1983} describes the random ancestral relations among DNA sequences sampled from large populations, and is a prominent model with which to make predictions about genetic diversity.
This popularity derives from its robustness: a wide class of genealogical models all have the Kingman coalescent, or a variant of it, as their limiting process when the population size is large \citep{M1998}.
Together, the Kingman coalescent and the infinitely-many-sites (IMS) model also form a tractable model of genetic evolution \citep{W75}.
Hence, many inference methods based on the Kingman coalescent have been developed; see e.g.~\cite{DT1995}, \cite{H1990}, \cite{N2001}, \cite{HSW2005} or \cite{W2007} for reviews.

The site frequency spectrum (SFS) at a given locus is an important and popular statistic by which to summarise genetic data under the IMS model and a coalescent process. 
Quantities of interest, such as the expectations and covariances of the SFS, are easily computed under the Kingman coalescent \citep{Fu1995}.

Despite its robustness, many evolutionary histories can also lead to significant deviations from the Kingman coalescent.
A variety of statistical tools is available for detecting such deviations, e.g.~Tajima's $D$ \citep{T1989}, Fu and Li's $D$ \citep{FL1993} or Fay and Wu's $H$ \citep{fay2000hitchhiking}, which are all functions of the SFS. 
However, they do not always allow the identification of the actual evolutionary mechanisms leading to the deviation from the Kingman coalescent.
\cite{EBBF15} investigated the ability of a single-locus SFS to distinguish two different deviations from the Kingman coalescent:
\begin{enumerate} 
\item population growth, in particular exponential or algebraic population growth, and 
\item gene genealogies described by so-called $\Lambda$-coalescents \citep{S1999, P1999, DK1999} featuring multiple mergers. There is growing evidence that such coalescents are an appropriate model for organisms with high fecundity coupled with a skewed offspring distribution \citep{B1994,A2004,EW2006,SW2008,HP2011, BBS2011, SBB2013, TL14}.
\end{enumerate}
Both scenarios lead to an excess of singletons in the SFS compared to the Kingman coalescent, leading e.g.~to a negative value of Tajima's $D$ \citep{DS05}.
\citet{EBBF15} showed that a single-locus SFS can distinguish between scenarios 1 and 2 with moderate statistical power, at least when the rate of mutation was sufficiently high and the deviation from the Kingman coalescent, the population growth rate in the scenario 1 and the prevalence of multiple mergers in scenario 2, was sufficiently large.
 
The present work builds on the result of \cite{EBBF15} by proposing a new statistical test to distinguish between scenarios 1 and 2.
A key development is that the proposed tests are based on multiple site frequency spectra corresponding to multiple unlinked loci.
Modern sequencing technologies have made sequence data from multiple linkage groups commonplace, and we show that making use of multi-locus data in this way greatly improves the power of statistical tests.
We assume that no recombination takes place within loci.

In addition, we extend the work of \cite{EBBF15} by explicitly including diploidy into the models under consideration.
Specifically, we consider a diploid, biparental population with symmetric mating, and with each parent contributing a single chromosome to their offspring.
This results in ancestries modeled as time-changed Kingman coalescents in scenario 1, and $\Xi$-coalescents, incorporating up to four simultaneous mergers of groups of multiple lineages, in scenario 2 \citep{BBE13a}.
In brief, the presence of up to four simultaneous mergers arises because genetic material at given locus can coalesce at either copy in either diploid parent.

While our new statistical tests contain the population rescaled mutation rate, $\theta / 2$, as a nuisance parameter, we show empirically that the statistical power achieved by our method is extremely robust to misspecification of this parameter.
Hence, in practice relying on a generalised Watterson estimator for the mutation rate can be expected to produce reliable inferences.

In addition to selecting the model class which best explains the data, it is also of interest to infer those parameters within that class which result in the best fit.
Hence, parameter inference for $\Lambda$- and $\Xi$-coalescents has been a growing area of research in recent years \citep{EW2006, BB2008, SW2008, EW2009, BBS2011, E2011, BBE13b, SBB2013, KJS2015, ZDGE15, BCEH16, KJS2017}.
We contribute to this body of work by demonstrating that the statistic used in our hypothesis test is also able to distinguish between different parameter values within the alternative hypothesis multiple merger class.
Our method is scalable to large data sets in contrast to many of those cited above, and we demonstrate its robustness and unbiasedness empirically via simulations.

\section{Summary statistic and approximate likelihood}\label{methods}

Consider a sample of $n$ DNA sequences taken at a given genetic locus, and assume that derived mutations can be distinguished from ancestral states.
For $n \in \N$ let $[ n ]:=\{ 1, \dots, n \}$, and let $\xi_i^{(n)}$ denote the total number of sites at which the mutant base appears $i \in [ n - 1 ]$ times. 
Then
\begin{equation*}
\bm{\xi}^{(n)} := \left(\xi_1^{(n)}, \ldots , \xi_{n-1}^{(n)} \right)
\end{equation*}
is  referred to as the \emph{unfolded} site-frequency spectrum based on the $n$ DNA sequences.  
If mutant and ancestral type cannot be distinguished, the \emph{folded} spectrum $\bm{ \eta }^{ ( n ) } := ( \eta_1^{ ( n ) }, \ldots, \eta_{ \lfloor n / 2 \rfloor }^{ ( n ) } )$ \citep{Fu1995} is often considered instead, where
\begin{equation*}
\eta_i^{(n)} :=  \frac{\xi_i^{(n)} + \xi_{n-i}^{(n)}}{1  +  \delta_{i,n-i}}, \quad 1 \le i \le \lfloor n/2 \rfloor,
\end{equation*}
and $\delta_{i,j} = 1$ if $i = j$, and is zero otherwise. 
Define $\bm{\zeta}^{(n)} = \left(\zeta_1^{(n)}, \ldots , \zeta_{n-1}^{(n)} \right)$ as the normalised unfolded SFS, whose entries are given by $\zeta_i^{(n)} := \xi_i^{(n)}/|\xi^{(n)}|$, where $|\xi^{(n)}| := \xi_1^{(n)} + \cdots + \xi_{n-1}^{(n)}$ is the total number of segregating sites. 
We adopt the convention that $\bm{\zeta}^{(n)} = \bm{0}$ if there are no segregating sites.

Even though both scenarios 1 and 2 predict an excess of singletons, \cite{EBBF15} showed that the expected tail of the normalised SFS varies between the two when the singletons have been matched \cite[Figure 1]{EBBF15}.
Hence we define the lumped tail
\begin{equation}\label{lumping}
\overline{ \zeta }^{ (n ) }_k := \sum_{ j = k }^{ n - 1 } \zeta_j^{ ( n ) }, \quad 3 \leq k \le n - 1
\end{equation}
and consider the summary statistic $( \zeta_1^{ ( n ) }, \overline{ \zeta }_k^{ ( n ) } )$ for some fixed $k$, to be specified.
This two-dimensional summary of the SFS can be expected to distinguish scenarios 1 and 2, while exploiting a lower dimensionality to reduce sampling variance (see the discussion on the effect of lumping in \citep{EBBF15} for an account of the same phenomenon in the context of approximate Bayesian computation) and reduce the number of Monte Carlo simulations needed to robustly characterise its sampling distribution.

We make use of multi-locus data by computing an SFS independently for each locus, and averaging over all available loci to reduce variance.
To this end, let $( \zeta_1^{ ( n ) }( j ), \overline{ \zeta }_k^{ ( n ) }( j ) )$ denote the singleton class and lumped tail corresponding to the $j^{\text{th}}$ locus, and 
\begin{equation}\label{mean_sfs}
( \zeta_{1, L}^{ ( n ) }, \overline{ \zeta }_{k, L}^{ ( n ) } ) := \frac{ 1 }{ L }\sum_{ j = 1 }^L ( \zeta_1^{ ( n ) }( j ), \overline{ \zeta }_k^{ ( n ) }( j ) )
\end{equation}
denote the mean singletons and lumped tail when there are $L$ loci.
A description of how \eqref{mean_sfs} can be computed from observed data, as well as from simulated trees, is provided in the appendix.
\vskip 11pt
\begin{rmk}
In scenario 1, and for binary, Kingman-type coalescents in general, \eqref{mean_sfs} converges to its expected value as $L \rightarrow \infty$ due to the strong law of large numbers.
In contrast, in scenario 2, and for multiple merger coalescents in general, unlinked loci have positively correlated coalescence times and hence a law of large numbers does not hold.
This is easily verified by considering a haploid sample of size 2 from 2 unlinked loci.
Let 
\begin{equation*}
\lambda_{ n, k } := \Lambda( \{ 0 \} ) \mathds{ 1 }_{ \{ k = 2 \} } + \int_{ ( 0, 1 ] } x^{ k - 2 }( 1 - x )^{ n - k } \Lambda( dx )
\end{equation*}
be the usual merger rate of any $2 \leq k \leq n$ lineages under a $\Lambda$-coalescent.
Let $T_i$ be the coalescence time at locus $i \in \{ 1, 2 \}$.
Note that $\E[ T_i ] = 1$, and that the total rate of effective mergers among the four lineages is 
\begin{equation*}
2 \lambda_{ 4, 2 } + 4 \lambda_{ 4, 3 } + \lambda_{ 4, 4 }.
\end{equation*}
By conditioning on whether the first merger involves four lineages (an event of probability $\lambda_{ 4, 4 } / ( 2 \lambda_{ 4, 2 } + 4 \lambda_{ 4, 3 } + \lambda_{ 4, 4 } )$) or strictly fewer than four lineages (an event of the complementary probability), it is straightforward to verify that
\begin{align*}
\E[ T_1 T_2 ] &= \frac{ 1 }{ 2 \lambda_{ 4, 2 } + 4 \lambda_{ 4, 3 } + \lambda_{ 4, 4 } } \left( 1 + \frac{ 2 - \lambda_{ 4, 4 } }{ 2 \lambda_{ 4, 2 } + 4 \lambda_{ 4, 3 } + \lambda_{ 4, 4 } } \right) \\
&= \frac{ 1 }{ 2 - \lambda_{ 4, 4 } } \left( 1 + \frac{ 2 - \lambda_{ 4, 4 } }{ 2 - \lambda_{ 4, 4 } } \right) = \frac{ 2 }{ 2 - \lambda_{ 4, 4 } } > 1 \text{ whenever } \Lambda \neq \delta_0,
\end{align*}
where the second line follows from the expansions
\begin{align*}
\lambda_{ 4, 3 } &= \lambda_{ 3, 3 } - \lambda_{ 4, 4 }, \\
\lambda_{ 4, 2 } &= 1 - 2 \lambda_{ 3, 3 } + \lambda_{ 4, 4 }.
\end{align*}
Hence, 
\begin{equation*}
\operatorname{Cov}( T_1, T_2 ) = 0 \Leftrightarrow \Lambda = \delta_0,
\end{equation*}
and a slightly more cumbersome version of this calculation verifies the same statement for $\Xi$-coalescents.
Sampling variance will still be reduced by averaging across unlinked loci, but by less than in scenario 1, and an $L \rightarrow \infty$ limit will be random if it exists.
\end{rmk}

Let $\Pi$ denote an arbitrary coalescent process, $\theta / 2$ denote a population-rescaled mutation rate, and $P^{ \Pi, \theta }( \zeta_{1, L}^{ ( n ) }, \overline{ \zeta }_{k, L}^{ ( n ) } )$ denote the sampling distribution of \eqref{mean_sfs} under the coalescent $\Pi$ and mutation rate $\theta / 2$.
This distribution is intractable, but if it were available then the likelihood ratio test statistic
\begin{equation}\label{exact_likelihood}
\frac{ \sup_{ \Pi \in \Theta_1, \theta > 0 } P^{ \Pi, \theta }( \zeta_{1, L}^{ ( n ) }, \overline{ \zeta }_{k, L}^{ ( n ) } ) }{ \sup_{ \Pi \in \Theta_0, \theta > 0 } P^{ \Pi, \theta }( \zeta_{1, L}^{ ( n ) }, \overline{ \zeta }_{k, L}^{ ( n ) } ) }
\end{equation}
would be a desirable tool for distinguishing between coalescent models in the class $\Theta_0$, the null hypothesis, from an alternative class $\Theta_1$, based on an observed value of \eqref{mean_sfs}.

\citet{EBBF15} proposed a Poisson approximation to a test akin to \eqref{exact_likelihood}, based on the whole unnormalised SFS, by assuming that the hidden coalescent ancestry of their sample coincided with the expected coalescent branch lengths.
These expected branch lengths were then computed by recursion.
The number of mutations was also assumed to always coincide exactly with the number of observed segregating sites, in what is known as the ``fixed-$s$" method \citep[see e.g.][]{DV1998, ROR2002}.
In this paper we approximate $P^{ \Pi, \theta }$ by simulation directly.
This avoids the unquantifiable bias introduced by the Poisson and fixed-$s$ approximations, and is scalable because only a two-dimensional sampling distribution needs to be approximated, regardless of the sample size or the number of alleles.
Of course replacing the sampling distributions in \eqref{exact_likelihood} with estimators will itself introduce bias, but provided that the estimators are close to the truth then it is reasonable to expect the bias to be small in comparison to an ad-hoc Poisson approximation.
In particular, Figure \ref{mles} in Section \ref{parameters} shows no evidence of bias when using our approximate sampling distributions to infer parameters.
\vskip 11pt
\begin{rmk}
A similar method based on kernel density estimation of an intractable likelihood function from simulated data was developed by \citet{Diggle84}, albeit with the goal of obtaining point estimators by maximising their approximation of the likelihood.
Our method is also reminiscent of the approximate Bayesian computation paradigm \citep{Beau10}, in that  likelihood evaluations are replaced by simulations from a family of models which is assumed to contain the model generating the data.
\end{rmk}

In order to produce an approximate sampling distribution $\hat{ P }^{ \Pi, \theta }$, a simulation algorithm was developed for producing samples from a specified coalescent $\Pi$ with mutation rate $\theta / 2$.
The algorithm takes as inputs a sample size and a desired number of unlinked loci per sample, within which we assume no recombination takes place.
The method is described in detail in the appendix, and a C++ implementation is available at \href{https://github.com/JereKoskela/Beta-Xi-Sim}{https://github.com/JereKoskela/Beta-Xi-Sim}.
Three diploid model classes have been implemented:
\begin{enumerate}
\item $\operatorname{Beta}( 2 - \alpha, \alpha )$-$\Xi$-coalescents for $\alpha \in [1, 2]$ derived from $\Lambda$-coalescents with $\Lambda = \operatorname{Beta}( 2 - \alpha, \alpha )$ by randomly splitting all merging lineages into four groups, and coalescing each group separately as in \citep[equation (29)]{BBE13a} and \citep[equation (15)]{BCEH16},
\item the Kingman coalescent with exponential population growth at rate $\gamma > 0$,
\item the Kingman coalescent with algebraic population growth with exponent $\gamma > 0$.
\end{enumerate}
Implementations for further model classes, such as $\Xi$-coalescents driven by measures other than the $\operatorname{Beta}(2 - \alpha, \alpha )$ class or derived from more general models of mating patterns, such as those described in \citep{BLS17}, could be easily added.
Further evolutionary forces such as natural selection could also be incorporated, though naturally both of these generalisation come at increased computational cost.

Once a sample of simulated realisations of \eqref{mean_sfs} has been produced under the desired null and alternative hypotheses (or suitable discretisations in the case of interval hypotheses), kernel density estimators were fitted to these samples using the \texttt{kde} function in the \texttt{ks} package (version 1.10.4) in \textsf{R} under default settings.
For a comprehensive introduction to kernel density estimation, see e.g.~\citep{Scott92}.
In particular, the bandwitdh is determined using the SAMSE estimator of \citep[equation (6)]{Duong03} as implemented in the \texttt{Hpi} function of the aforementioned \texttt{ks} package.
This package allows kernel density estimation in up to six dimensions, so that higher dimensional statistics could be used to distinguish model classes that are not well separated by just singletons and the lumped tail.
However, this increase in dimensionality would come at the cost of increasingly intensive simulation time as more samples are needed to accurately represent a higher dimensional sampling distribution, and was not necessary for our purposes.

Once these kernel density estimators have been produced, they can be substituted into \eqref{exact_likelihood} to obtain an approximation
\begin{equation*}
\frac{ \sup_{ \Pi \in \Theta_1, \theta > 0 } \hat{ P }^{ \Pi, \theta }( \zeta_{1, L}^{ ( n ) }, \overline{ \zeta }_{k, L}^{ ( n ) } ) }{ \sup_{ \Pi \in \Theta_0, \theta > 0 } \hat{ P }^{ \Pi, \theta }( \zeta_{1, L}^{ ( n ) }, \overline{ \zeta }_{k, L}^{ ( n ) } ) }.
\end{equation*}
However, simulating samples and producing kernel density estimators for combinations of coalescent processes and mutation rates also incurs a significant computational cost.
To alleviate it, we assume that a computationally cheap estimator $\hat{\theta}$ is available, e.g.~the generalised Watterson estimator 
\begin{equation*}
\hat{ \theta } = | \xi^{ ( n ) } | / \E^{ \Pi }[ T^{ ( n ) } ],
\end{equation*}
where $\E^{ \Pi }[ T^{ ( n ) } ]$ denotes the expected tree length from $n$ leaves under coalescent mechanism $\Pi$.
We then consider the test statistic
\begin{equation}\label{approx_test}
\frac{ \sup_{ \Pi \in \Theta_1 } \hat{ P }^{ \Pi, \hat{\theta} }( \zeta_{1, L}^{ ( n ) }, \overline{ \zeta }_{k, L}^{ ( n ) } ) }{ \sup_{ \Pi \in \Theta_0 } \hat{ P }^{ \Pi, \hat{\theta} }( \zeta_{1, L}^{ ( n ) }, \overline{ \zeta }_{k, L}^{ ( n ) } ) }
\end{equation}
instead.
Because our method makes use of the normalised SFS, it is highly insensitive to misspecification of the mutation rate, unless the mutation rate is small enough that normalisation of the SFS itself becomes a source of variance.
This robustness will be confirmed by simulation in the next section.
See also \cite[Supporting Information, pages SI 12--SI 13]{EBBF15} for a theoretical result in this direction.

The statistic \eqref{approx_test} is straightforward to evaluate pointwise, and further simulated data can be used to obtain an empirical quantile $\rho_{ \varepsilon }$ such that
\begin{equation*}
\sup_{ \Pi \in \Theta_0 } \Prb^{ \Pi, \hat{ \theta } }\left( \frac{ \sup_{ \Pi \in \Theta_1 } \hat{ P }^{ \Pi, \hat{\theta} }( \zeta_{1, L}^{ ( n ) }, \overline{ \zeta }_{k, L}^{ ( n ) } ) }{ \sup_{ \Pi \in \Theta_0 } \hat{ P }^{ \Pi, \hat{\theta} }( \zeta_{1, L}^{ ( n ) }, \overline{ \zeta }_{k, L}^{ ( n ) } ) }  \geq \rho_{ \varepsilon } \right) \leq \varepsilon,
\end{equation*}
where $\Prb^{ \Pi, \hat{ \theta } }$ denotes the law of the coalescent process under coalescent mechanism $\Pi$ and mutation rate $\hat{\theta} / 2$.
Given such a quantile, a hypothesis test of approximate size $\varepsilon$ for an observed pair $( \zeta_{1, L}^{ ( n ) }, \overline{ \zeta }_{k, L}^{ ( n ) } )$ is given by
\begin{equation}\label{test}
\Phi( \zeta_{1, L}^{ ( n ) }, \overline{ \zeta }_{k, L}^{ ( n ) } ) = \begin{cases}
0 &\text{if } \frac{ \sup_{ \Pi \in \Theta_1 } \hat{ P }^{ \Pi, \hat{\theta} }( \zeta_{1, L}^{ ( n ) }, \overline{ \zeta }_{k, L}^{ ( n ) } ) }{ \sup_{ \Pi \in \Theta_0 } \hat{ P }^{ \Pi, \hat{\theta} }( \zeta_{1, L}^{ ( n ) }, \overline{ \zeta }_{k, L}^{ ( n ) } ) } \leq \rho_{ \varepsilon } \\
1 &\text{if } \frac{ \sup_{ \Pi \in \Theta_1 } \hat{ P }^{ \Pi, \hat{\theta} }( \zeta_{1, L}^{ ( n ) }, \overline{ \zeta }_{k, L}^{ ( n ) } ) }{ \sup_{ \Pi \in \Theta_0 } \hat{ P }^{ \Pi, \hat{\theta} }( \zeta_{1, L}^{ ( n ) }, \overline{ \zeta }_{k, L}^{ ( n ) } ) } > \rho_{ \varepsilon }
\end{cases},
\end{equation}
where $\Phi( \zeta_{1, L}^{ ( n ) }, \overline{ \zeta }_{k, L}^{ ( n ) } ) = 1$ corresponds to rejecting the null hypothesis.
The size of the test is only approximate because the quantile $\rho_{ \varepsilon }$ is computed from finitely many simulated realisations, and because the sampling distribution $\hat{P}^{ \Pi, \hat{\theta} }$ is itself a kernel density approximation.
In the next section we demonstrate the performance and robustness of this test, and compare it to the Poisson-fixed-$s$ test of \citet{EBBF15}.
\vskip 11pt
\begin{rmk}
Hypothesis tests only provide information on the model fit of the alternative hypothesis relative to the null.
Hence, if the null hypothesis is a poor fit to the data, a high value of \eqref{approx_test} does not necessarily indicate that the alternative hypothesis is a good fit, but only a better one than the null hypothesis.
The empirical sampling distributions constructed in order to evaluate \eqref{approx_test} can be used to address this ambiguity to some extent, e.g.~by plotting observed data along with simulated realisations in scatter plots akin to Figure \ref{scatter} in the next section.
This ability to overcome a limitation of hypothesis tests arises out of the low dimensionality of the statistic \eqref{mean_sfs}, and the ease with which its sampling distribution can hence be visualised.
\end{rmk}

\section{Results: model selection}\label{results}

\begin{figure}[!ht]
\centering
\includegraphics[width = 0.49 \linewidth]{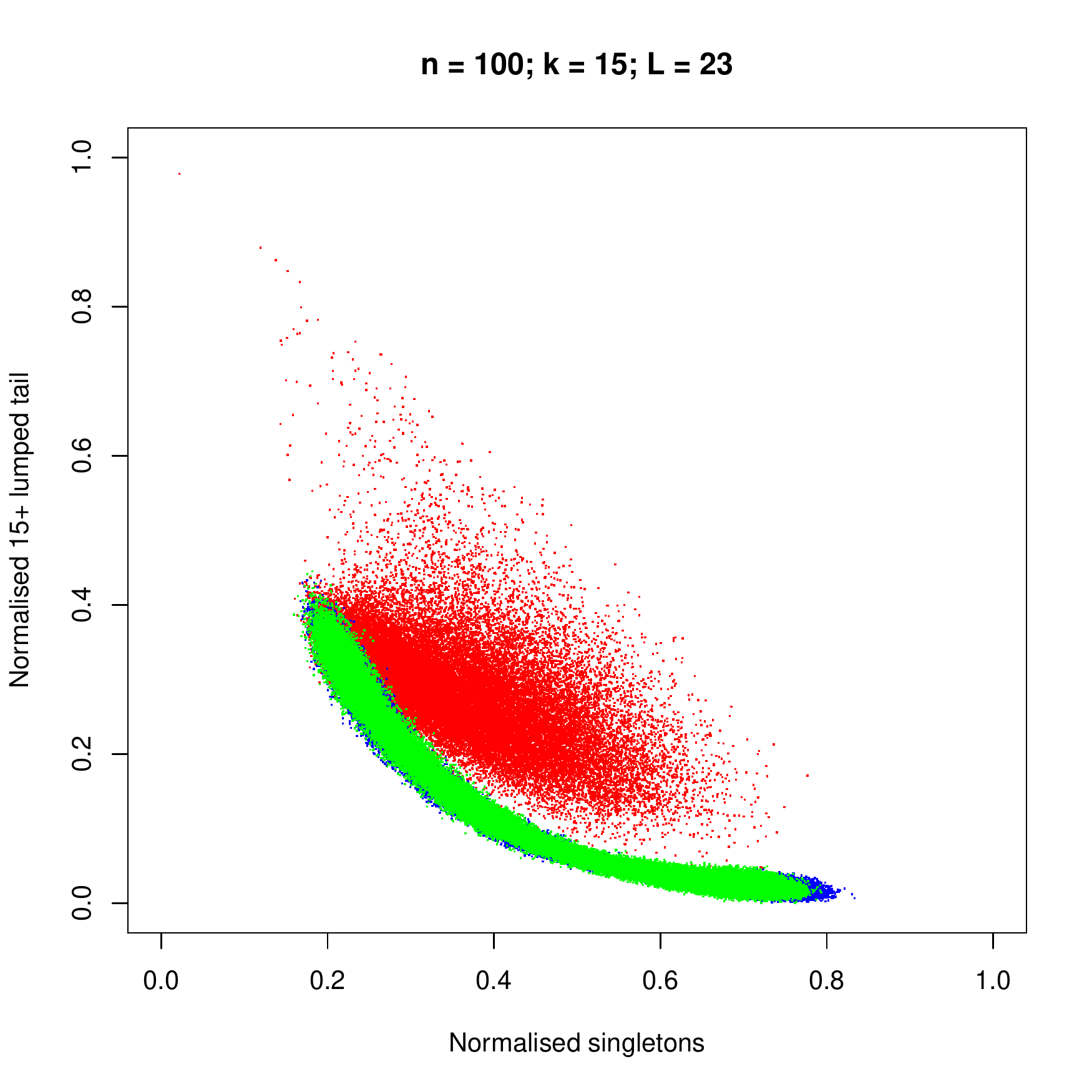}
\includegraphics[width = 0.49 \linewidth]{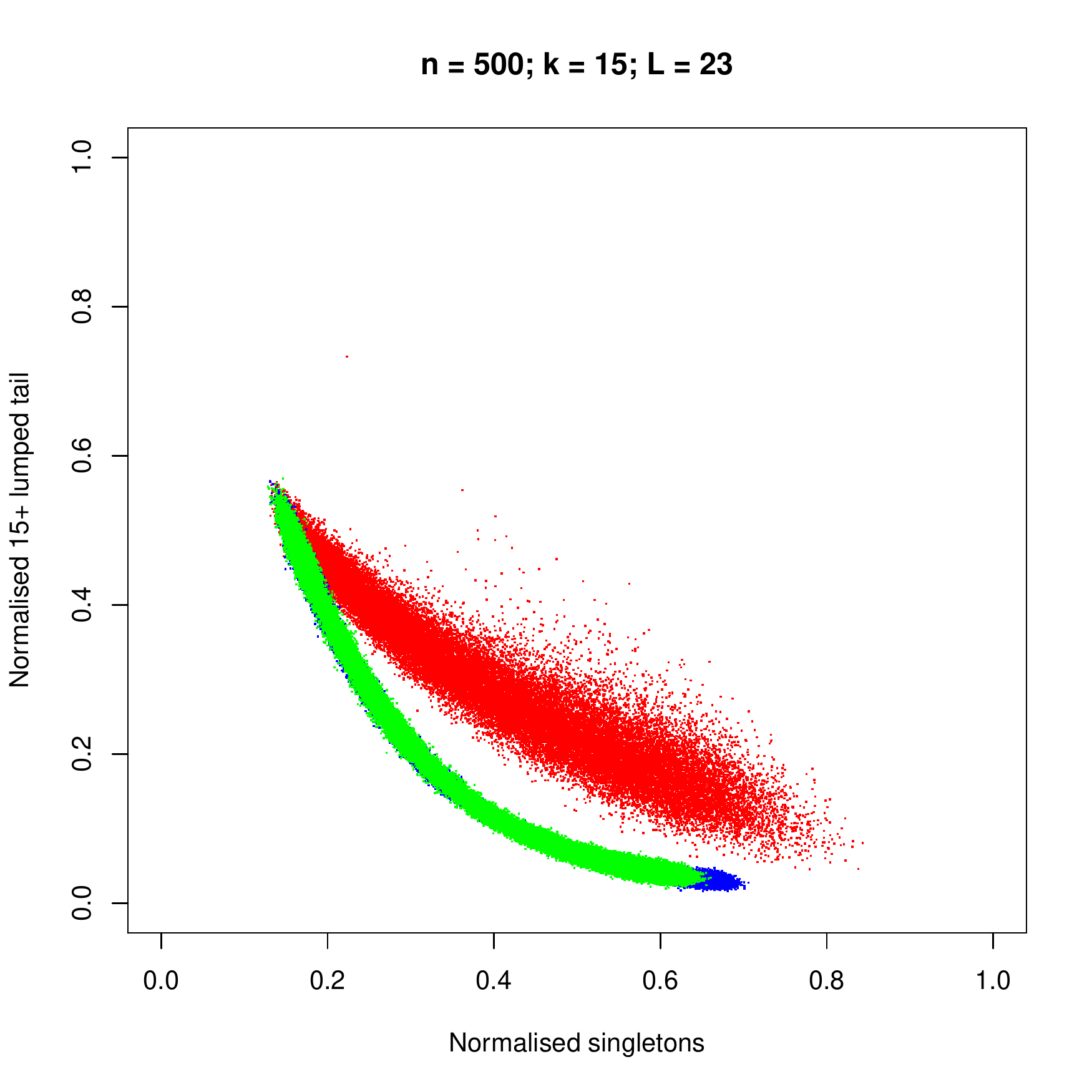}
\caption{Scatter plot of the joint distribution of the summary statistic \eqref{mean_sfs} with $L = 23$ loci, lumping from $k = 15$ and sample size $n = 100$ on the left and $n = 500$ on the right.
Red dots denote realisations from a $\operatorname{Beta}( 2 - \alpha, \alpha )$-$\Xi$-coalescent, green dots from an algebraic growth coalescent, and blue dots from an exponential growth coalescent.
The runtimes to generate these samples on a single Intel i5-2520M 2.5 GHz processor were 6.5 hours for $n = 100$, and 59 hours for $n = 500$.}
\label{scatter}
\end{figure}

In this section we present simulation studies demonstrating the statistical power and robustness  of \eqref{test}.
For comparison, the same simulation study will also be undertaken using the Poisson-fixed-$s$ approximate likelihood ratio test of \citep{EBBF15}.
For ease of computation we discretise the null and alternative hypotheses, and take them to be
\begin{align*}
\Theta_0 := \{& \text{Kingman coalescent with exponential or algebraic growth, each at rates } \gamma \in \{ 0, 0.1, \\
&0.2, \ldots, 0.9, 1, 1.5, 2, 2.5, 3, 3.5, 4, 5, 6, \ldots, 19, 20, 25, 30, 35, 40, 50, 60, \ldots, 990, 1000 \} \}, \\
\Theta_1 := \{&\text{Beta}(2 - \alpha, \alpha)\text{-}\Xi\text{-coalescents with } \alpha \in \{ 1, 1.025, \ldots, 1.975, 2\} \},
\end{align*}
respectively.
The somewhat irregular discretisation has been chosen to yield comprehensive coverage of the sampling distribution of the continuous hypotheses based on trial runs (see Figure \ref{scatter}).
Note that all three model classes coincide with the Kingman coalescent at the far left end of the cloud of points, which corresponds to $\gamma = 0$ in scenario 1 and $\alpha = 2$ in scenario 2.
We also fix the approximate sizes of all hypothesis tests at $\varepsilon = 0.01$.

Note that \citet{EBBF15} considered the two classes in our null hypothesis as separate hypotheses.
Figure \ref{scatter} indicates that considering either class in isolation is indeed sufficient, as the sampling distributions corresponding to exponential and algebraic growth coincide almost exactly across the considered ranges of rates.
We also emphasize that distinguishing a $\operatorname{Beta}( 2 - \alpha, \alpha )$-$\Xi$-coalescent from a class of binary coalescents is harder than distinguishing the corresponding $\operatorname{Beta}( 2 - \alpha )$-$\Lambda$-coalescent from the same binary class.
This is because randomly assigning mergers into four groups can splits multiple mergers into either outright binary mergers, or smaller multiple mergers.
Thus, the SFS of the resulting $\Xi$-coalescent tree will resemble the SFS of a binary coalescent more closely than the same $\Lambda$-coalescent tree would have done.

For each of the coalescent processes in $\Theta_0$ and $\Theta_1$, we simulated 1000 replicates of samples of size $n = 100$ and $n = 500$, with $L = 23$ unlinked loci and with per-locus mutation rates $\theta_i / 2 = N_e^{ \alpha - 1 } L_i \mu$, with effective population size $N_e = 1000$, per-site-per-generation mutation rate set at either $\mu = 10^{ -8 }$ or $10^{ -7 }$, and the number of sites per locus varying between $L_i = 19 \times 10^6$ and $L_i = 31 \times 10^6$ as recorded for the linkage groups of Atlantic cod in \citep[Supplementary Table 3]{T17}.
Atlantic cod is a species for which multiple mergers have frequently been suggested as an important evolutionary mechanism \citep[see e.g.][and references therein]{SBB2013, TL14}.
The scaling parameter $\alpha$ in the mutation rate denotes the parameter of the $\Theta_1$ class, with $\alpha = 2$ corresponding to the Kingman coalescent as per the scaling results of \citet{S2003}.
For each coalescent, the 1000 replicates were split into two groups: 500 replicates as training data for computing the kernel density estimator $\hat{P}^{ \Pi, \hat{ \theta } }$, and an independent 500 replicates as test data from which to estimate the empirical size and power of the hypothesis test \eqref{test}.

\begin{figure}[!ht]
\centering
\includegraphics[width = 0.49 \linewidth]{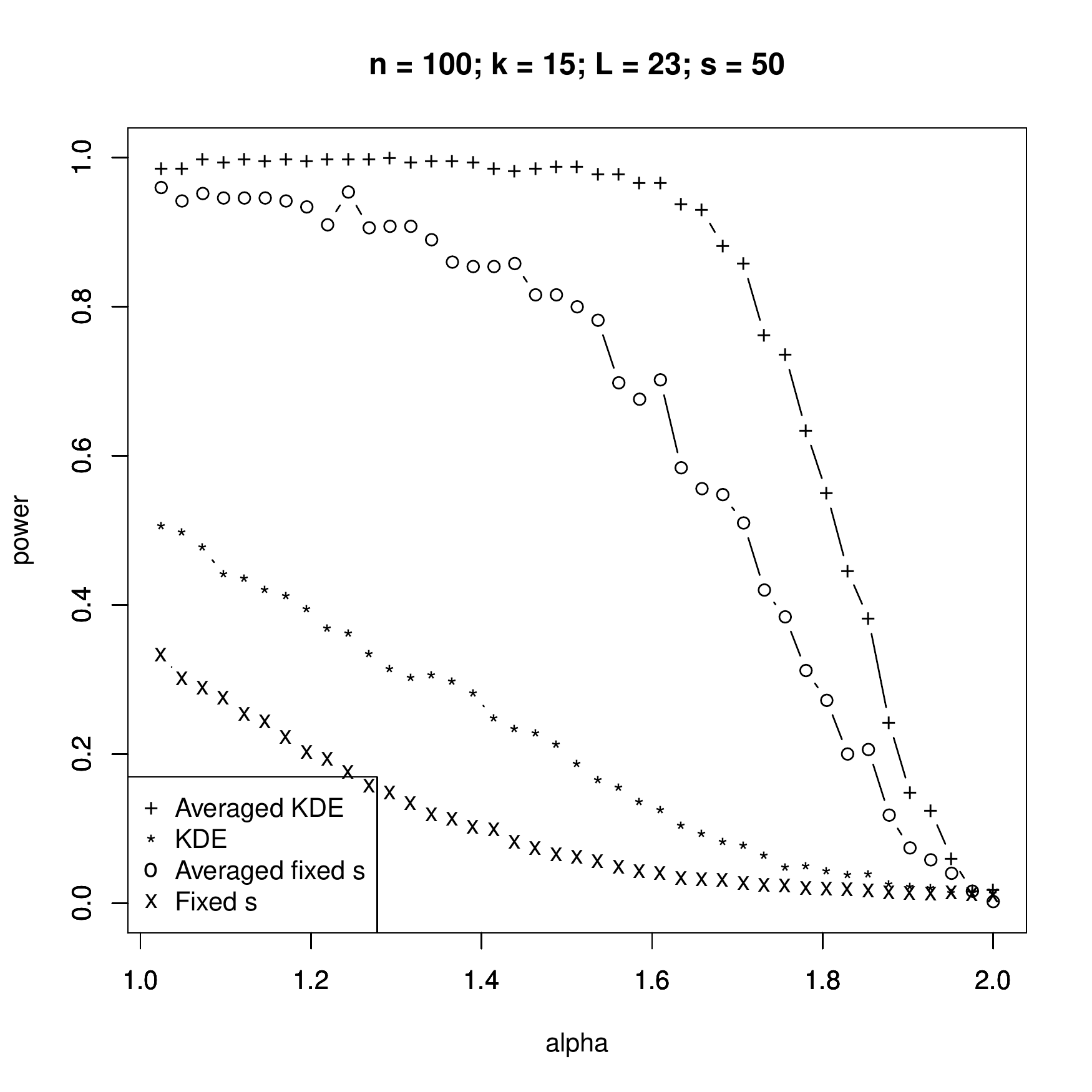}
\includegraphics[width = 0.49 \linewidth]{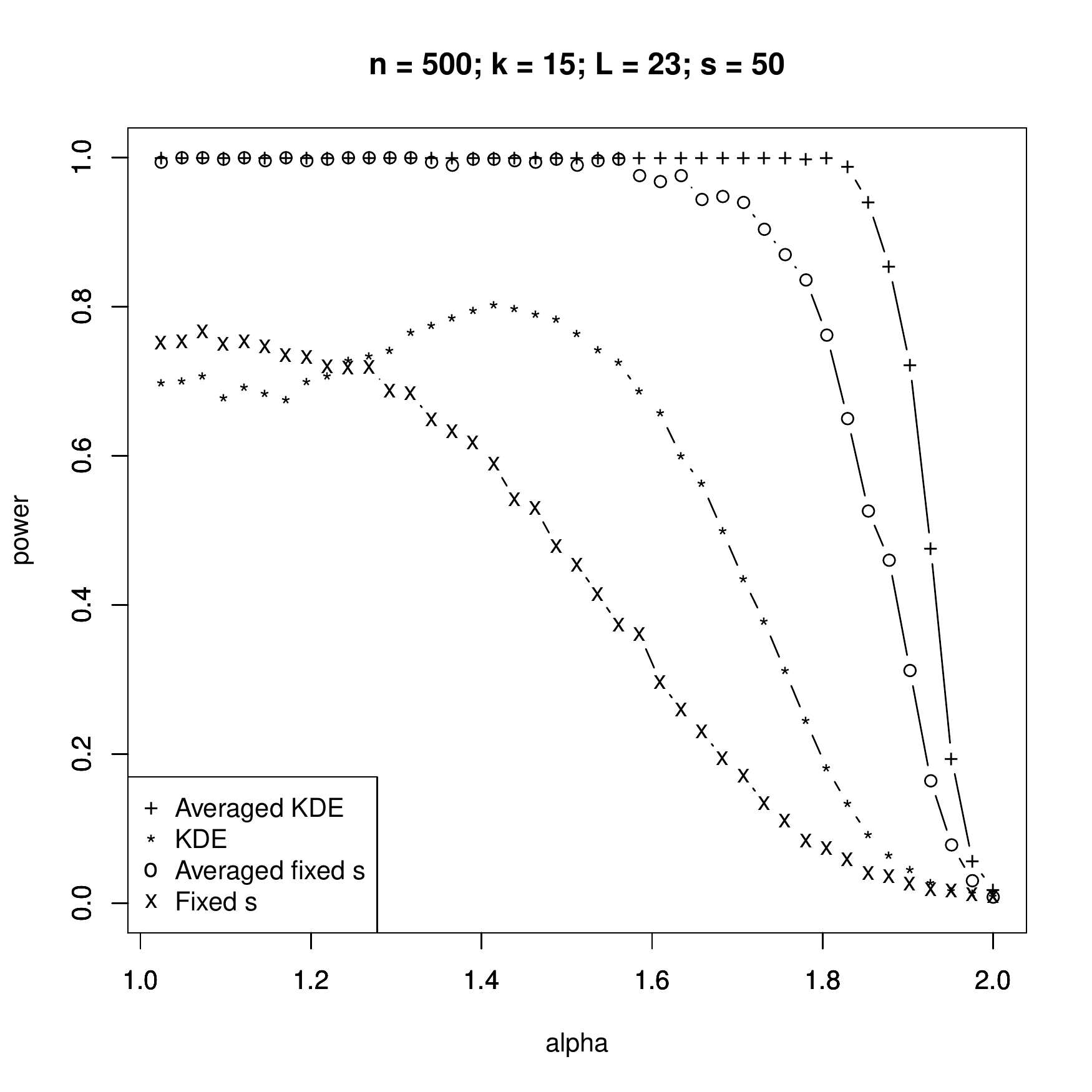}
\includegraphics[width = 0.49 \linewidth]{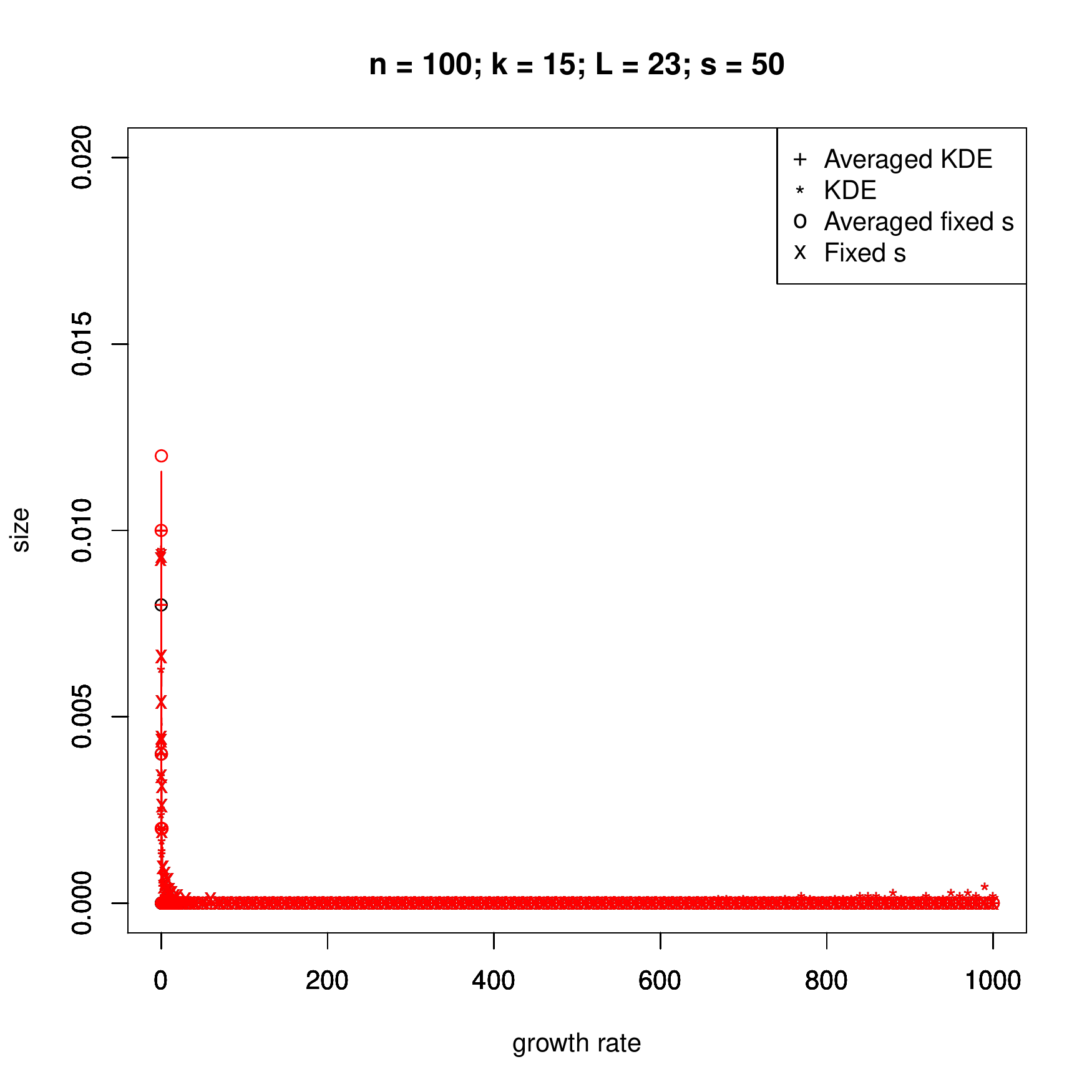}
\includegraphics[width = 0.49 \linewidth]{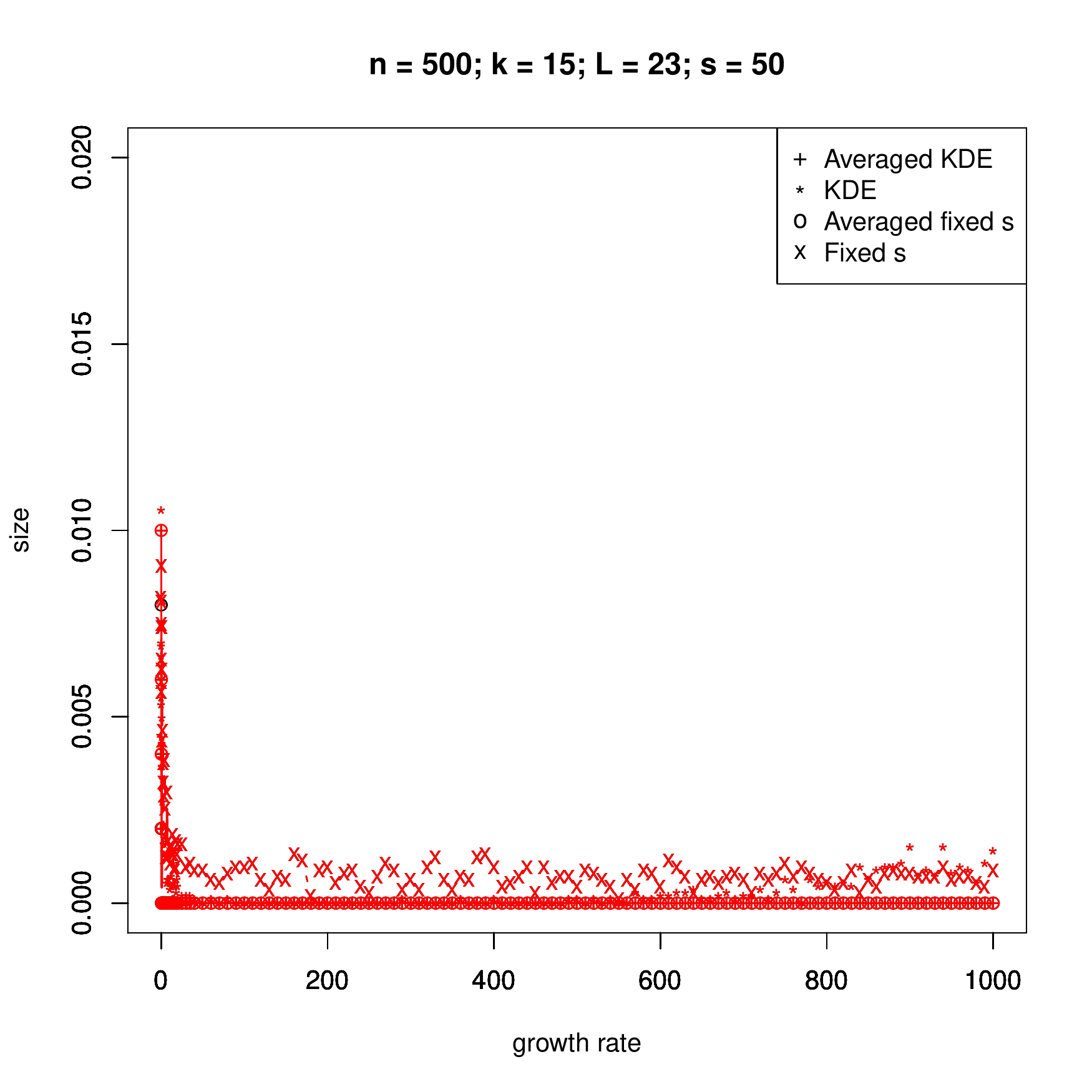}
\caption{Comparison of the Poisson-fixed-$s$ approximate likelihood ratio test with s = 50, and the KDE approximation proposed in this article with $L = 23$ loci and lumping from $k = 15$. 
The first row shows the empirical power of the test, and the second the empirical size with $\varepsilon = 0.01$.
In the second row, red characters denote a true algebraic growth model, and black characters denote a true exponential growth model.
Results are presented under the assumption of independent loci (``KDE" and``fixed $s$") in which case each data set is treated as $1000 \times L$ independent observations, and for a single SFS obtained by averaging all $L$ loci (``Averaged KDE" and ``Averaged fixed $s$").
The sample sizes are $n = 100$ on the left and $n = 500$ on the right. 
Post-processing the simulated data to obtain these graphs in \textsf{R} took 71 minutes for $n = 100$, and 274 minutes for $n = 500$ on a single Intel i5-2520M 2.5 GHz processor.
The expected branch lengths for the Poisson-fixed-$s$ methods were estimated from 100 000 coalescent realisations.}
\label{comparison}
\end{figure}

We also simulated independent realisations of the ``fixed-$s$" data sets of \citet{EBBF15} to facilitate a comparison, keeping the total number of segregating sites at $s = 50$ as in the original article.
We extend their Poisson-fixed-$s$ approximate likelihood test to our multi-locus setting in two ways: by treating multiple loci as independent, or averaging over them as in the KDE case.
Furthermore, we approximate the expected branch lengths required by the ``fixed-$s$" method by averaging over 100 000 coalescent simulations, in contrast with the exact, numerical computations carried out by \citet{EBBF15}.
This makes our ``fixed-$s$" results slightly noisier, but facilitates the straightforward extension of the method from $\Lambda$-coalescents to $\Xi$-coalescents.
While solvable recursions for $\Xi$-coalescent expected branch lengths are known \citep{BCEH16}, they are much more computationally intensive than their $\Lambda$-coalescent counterparts.
Moreover, the exact branch length approach cannot be easily extended to cover more complex scenarios, which are nevertheless easy to simulate.
A comparison of the empirical power as a function of the $\alpha$-parameter in the $\operatorname{Beta}( 2 - \alpha, \alpha )$-distribution is presented in Figure \ref{comparison}.

Figure \ref{comparison} shows that \eqref{test} substantially improves on the test of \citep{EBBF15} in terms of the power to distinguish multiple merger coalescents from population growth.
It yields higher power in the majority of experiments under the incorrect assumption of independent loci, and all of the experiments when loci have been averaged.
The results for $n = 100$ are already promising, and samples of size $n = 500$ yield statistical power estimates of 1 for a majority of parameter values.
The empirical sizes remain bounded by the target value of $\varepsilon = 0.01$ for both methods unless the true model is very close to the constant population size Kingman coalescent, and even then the sizes barely exceed it.

\begin{figure}[!ht]
\centering
\includegraphics[width = 0.49 \linewidth]{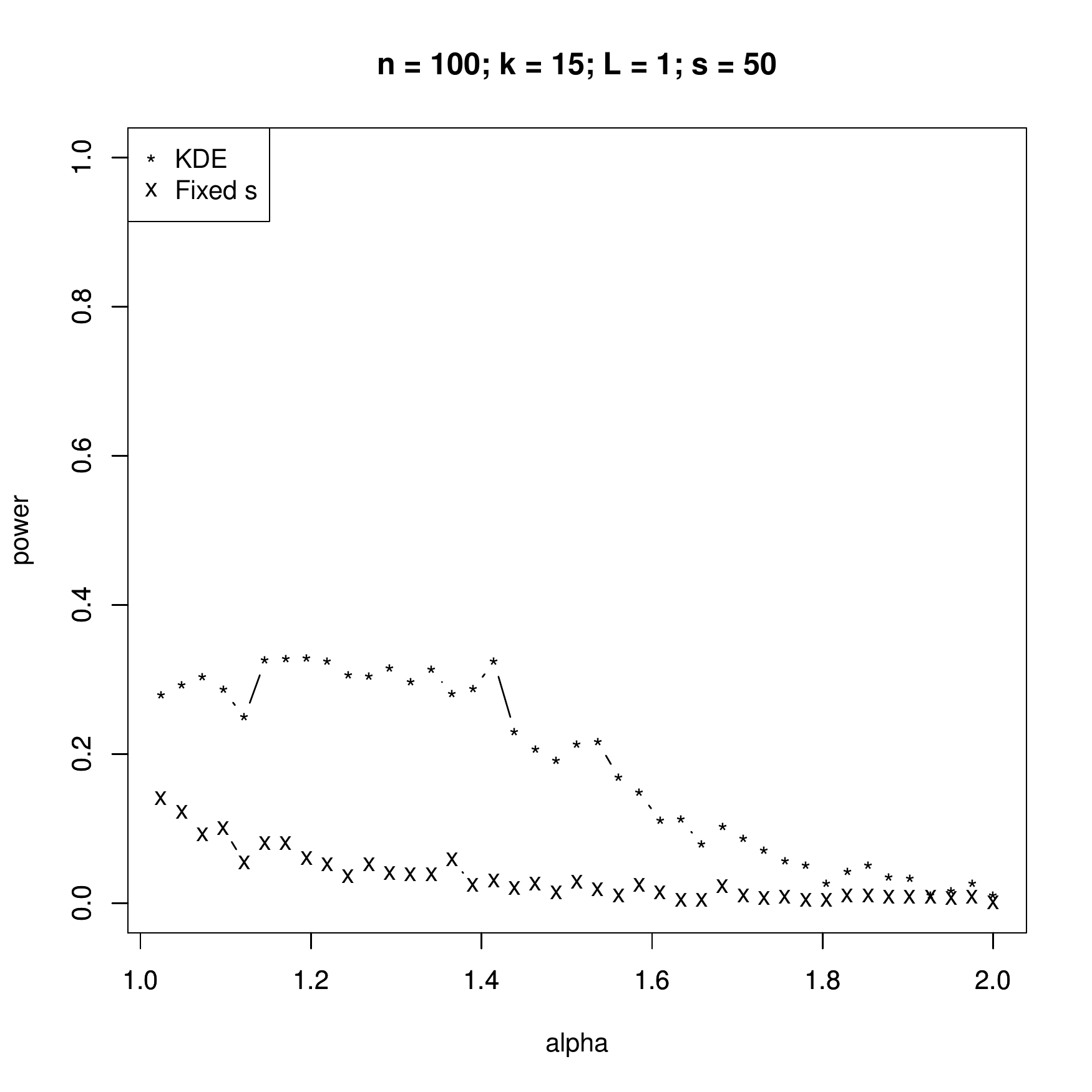}
\includegraphics[width = 0.49 \linewidth]{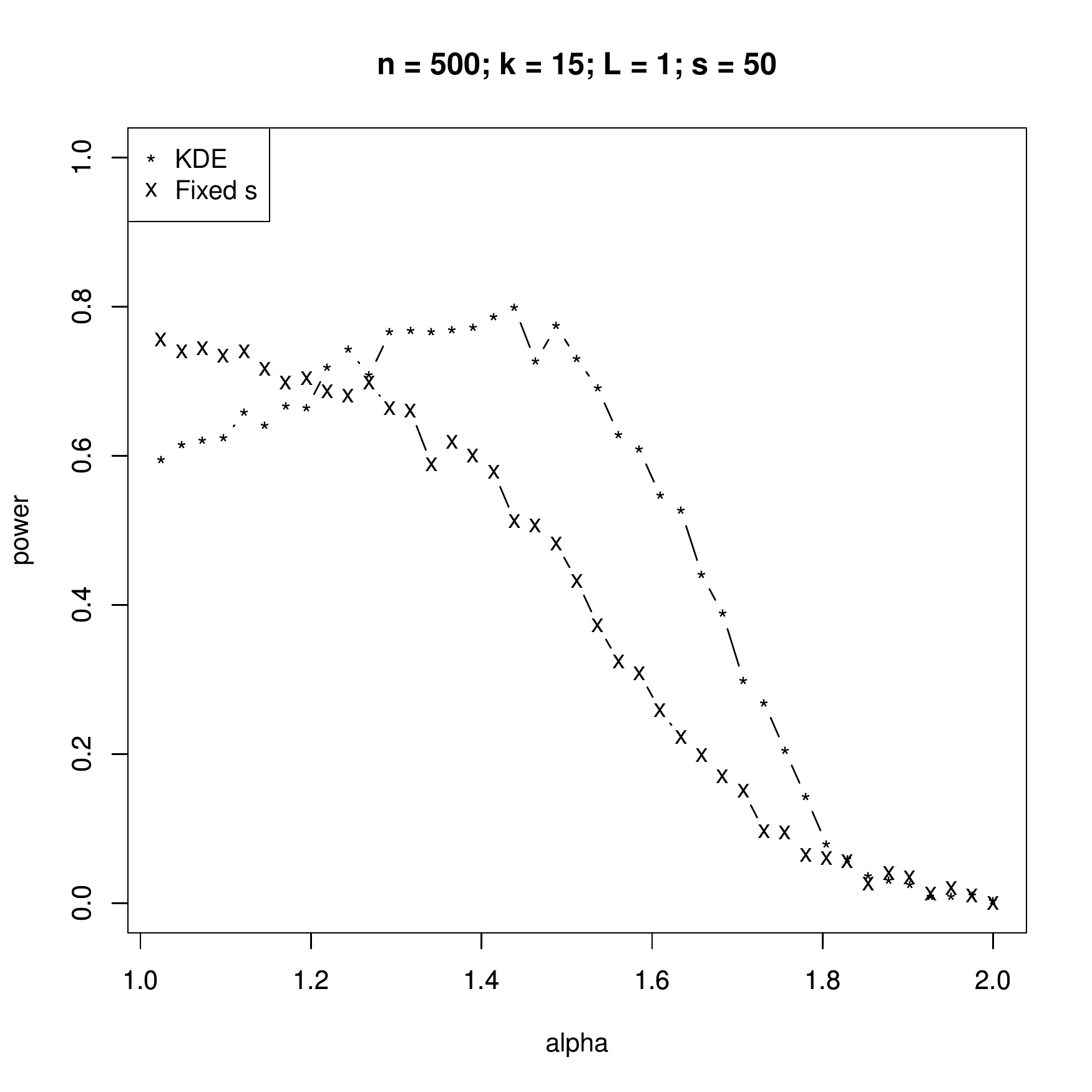}
\includegraphics[width = 0.49 \linewidth]{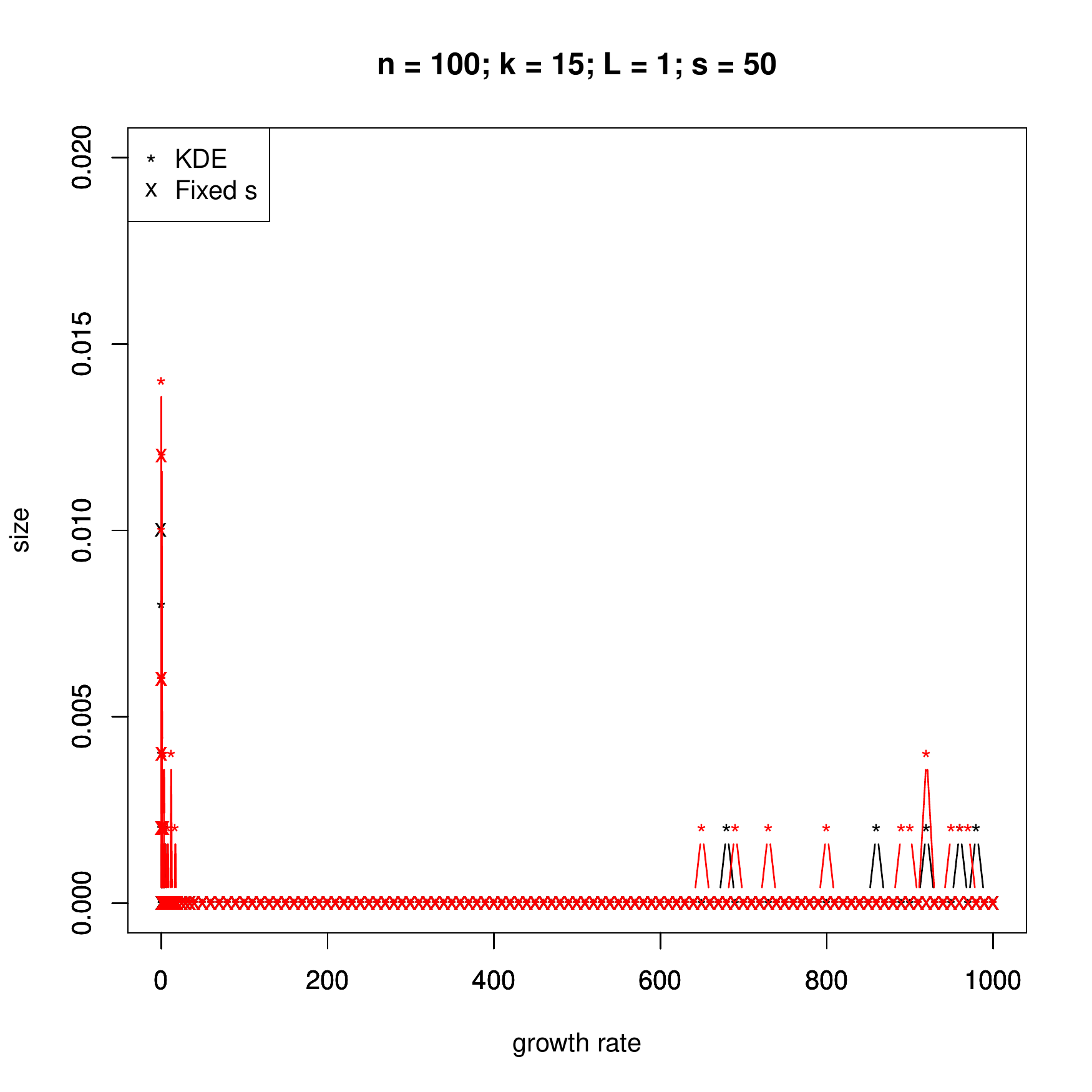}
\includegraphics[width = 0.49 \linewidth]{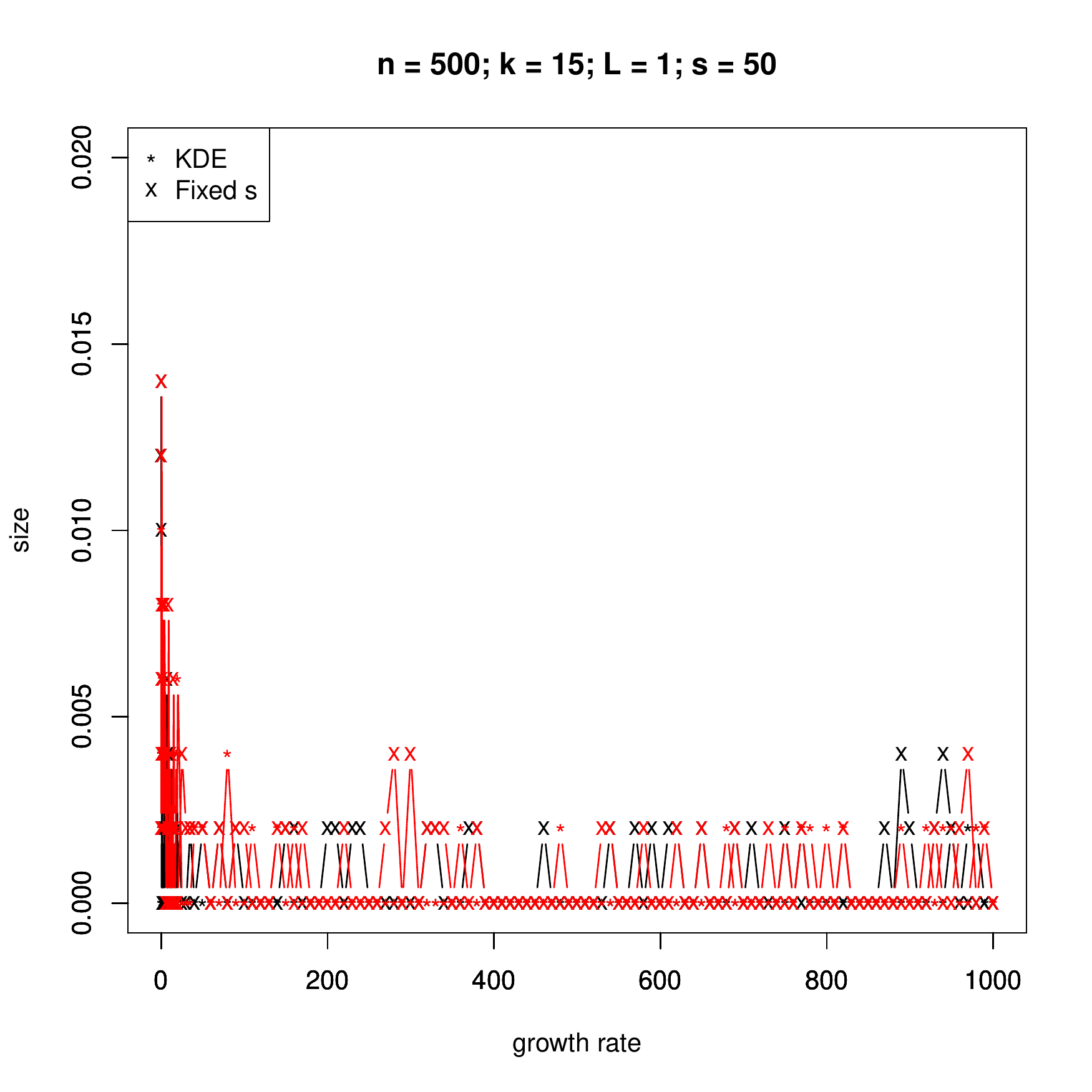}
\caption{Comparison of the Poisson-fixed-$s$ approximate likelihood ratio test with s = 50, and the KDE approximation proposed in this article with a single locus and lumping from $k = 15$. 
The first row shows the empirical power of the test, and the second the empirical size with $\varepsilon = 0.01$.
In the second row, red characters denote a true algebraic growth model, and black characters denote a true exponential growth model.
The sample sizes are $n = 100$ on the left and $n = 500$ on the right. 
Post-processing the simulated data to obtain these graphs in \textsf{R} took 47 minutes for $n = 100$, and 192 minutes for $n = 500$ on a single Intel i5-2520M 2.5 GHz processor.
The expected branch lengths for the Poisson-fixed-$s$ methods were estimated from 100 000 realisations.}
\label{comparison_single_loc}
\end{figure}

Figure \ref{comparison_single_loc} assesses the need for multilocus data by presenting the same comparison as Figure \ref{comparison}, but for $L = 1$ locus.
The results are somewhat noisier than those without averaging in Figure \ref{comparison} because the size of the data set has decreased by a factor of 23, but it is clear that the KDE approximation is superior to the fixed-$s$ method in most cases even for a single locus.
Moreover, multiplying the size of the data set by 23 barely improves statistical power if the assumption of independent (as opposed to merely unlinked) loci is made incorrectly.
In contrast, averaging across loci yields substantial improvements from the larger data set.

\begin{figure}[!ht]
\centering
\includegraphics[width = 0.49 \linewidth]{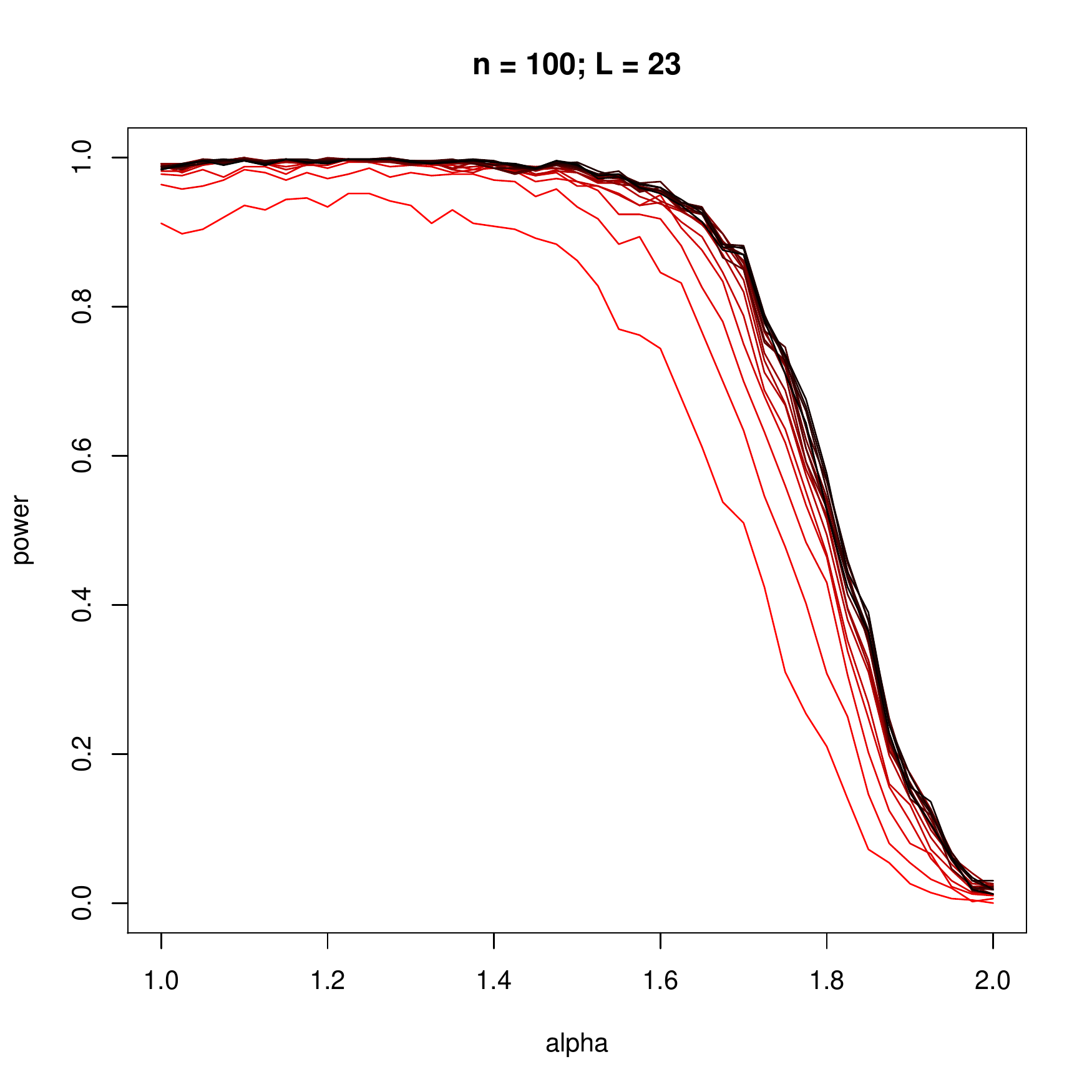}
\includegraphics[width = 0.49 \linewidth]{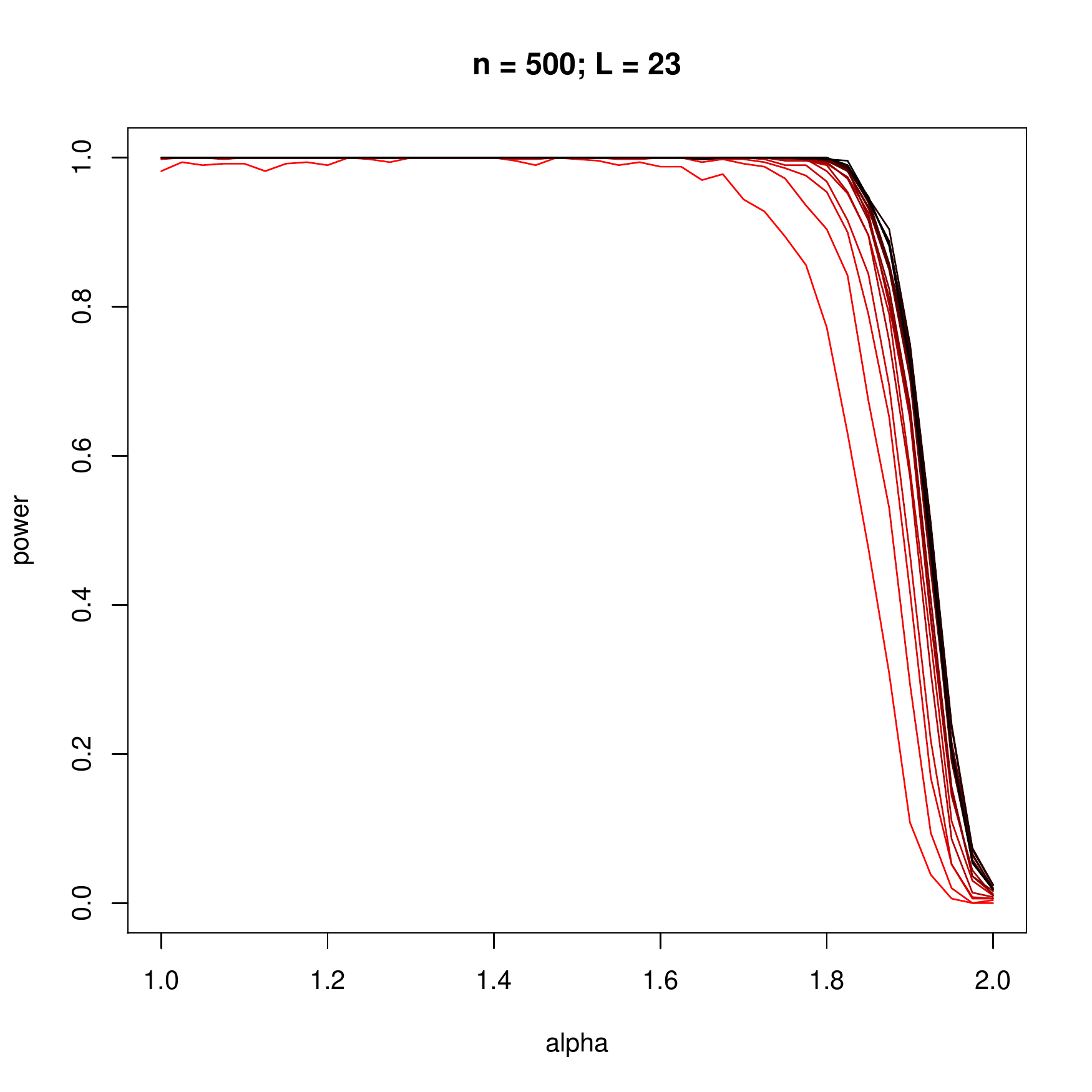}
\includegraphics[width = 0.49 \linewidth]{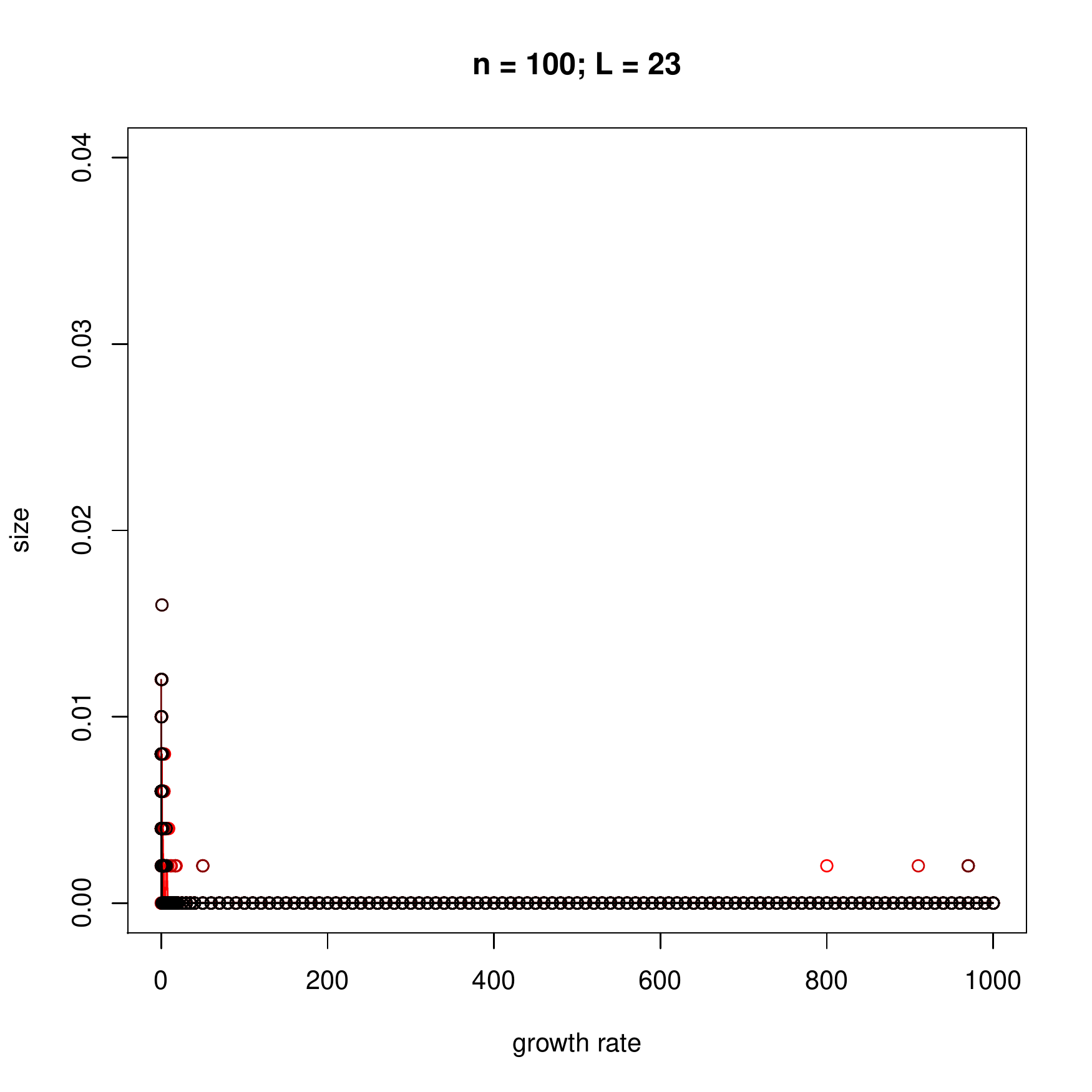}
\includegraphics[width = 0.49 \linewidth]{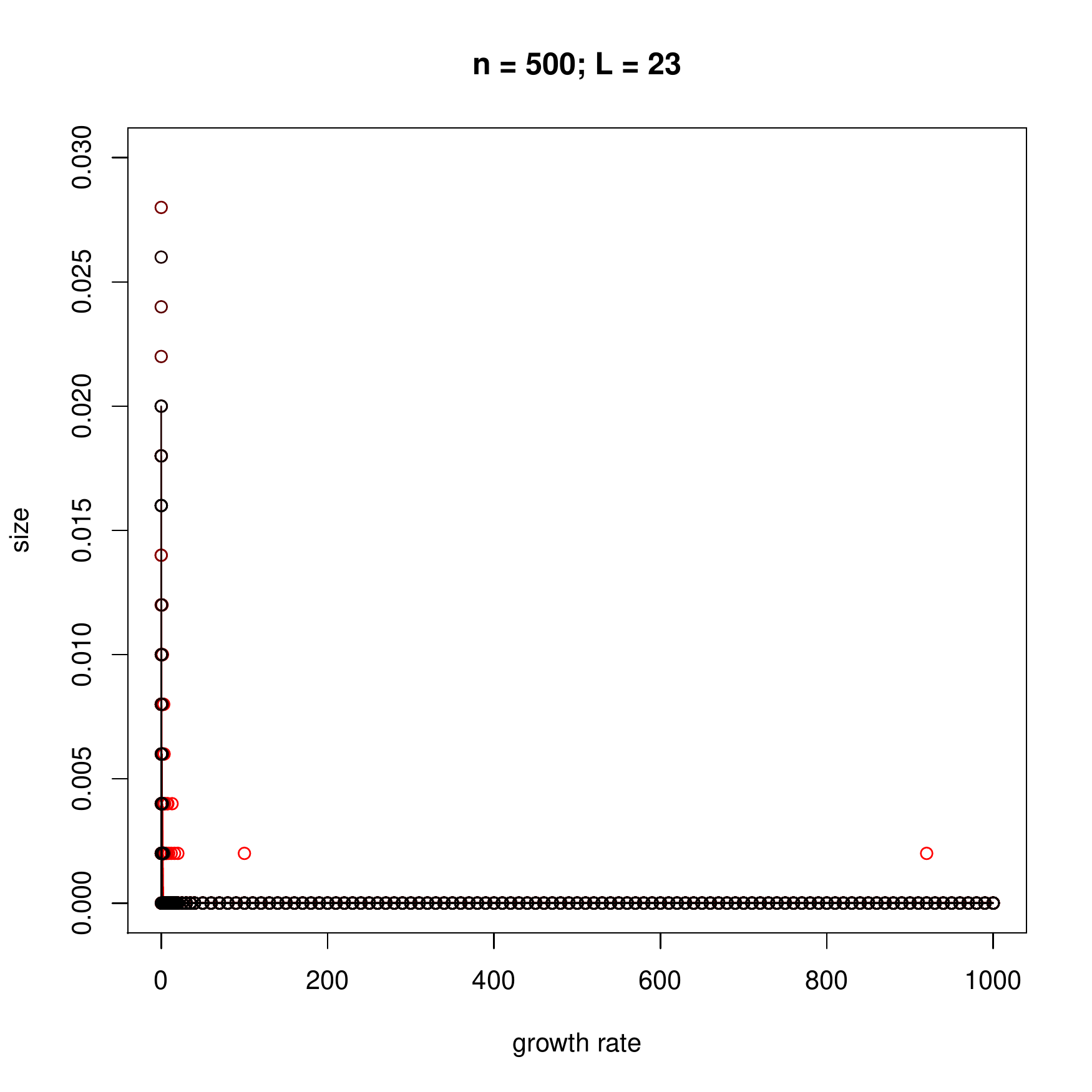}
\includegraphics[width = 0.99 \linewidth]{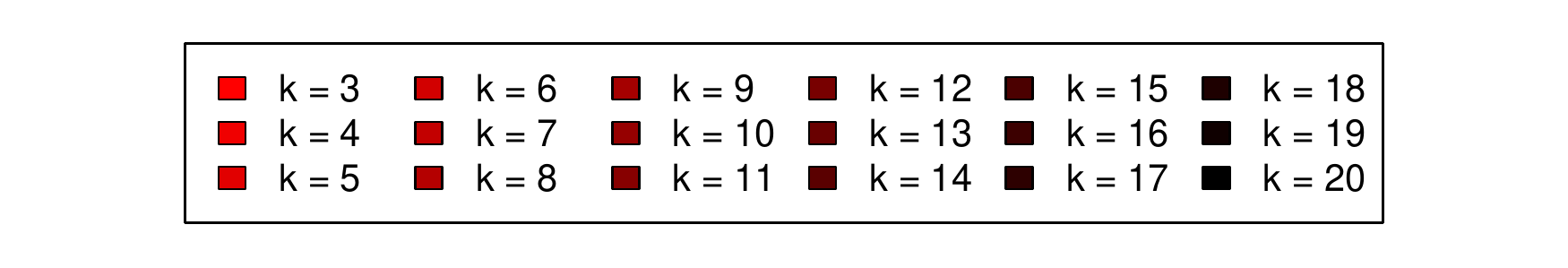}
\caption{Effect of the lumping cutoff in \eqref{mean_sfs}. 
On the second row, solid lines correspond to an exponential growth model, and discrete circles to an algebraic growth model.
Post-processing these graphs from the simulated data in \textsf{R} took 18 minutes for $n = 100$, and 40 minutes for $n = 500$ on a single Intel i5-2520M 2.5 GHz processor.}
\label{cutoff}
\end{figure}

In Figure \ref{cutoff} we investigate the effect of the lumping threshold $k$ in \eqref{lumping}.
Lumping corresponding to $k = 2$ was not included, as it yields an effectively one dimensional statistic corresponding to just the singleton class.
Lumping from a very low level, say $k \in \{ 3, 4, 5 \}$, can diminish statistical power as mid-range entries of the SFS confound the contribution of the lumped tail, but beyond these low numbers the good performance of our method is highly insensitive to this tuning parameter.
The empirical sizes of tests are also bounded by the target $\varepsilon = 0.01$ except when the true model is very close to the Kingman coalescent, the most challenging case, and even then the observed size does not exceed 0.03.
The slight inflation of size near the $\gamma = 0$ Kingman coalescent suggest that sizes smaller than the typical $\varepsilon = 0.05$ should be used.

\begin{figure}[!ht]
\centering
\includegraphics[width = 0.49 \linewidth]{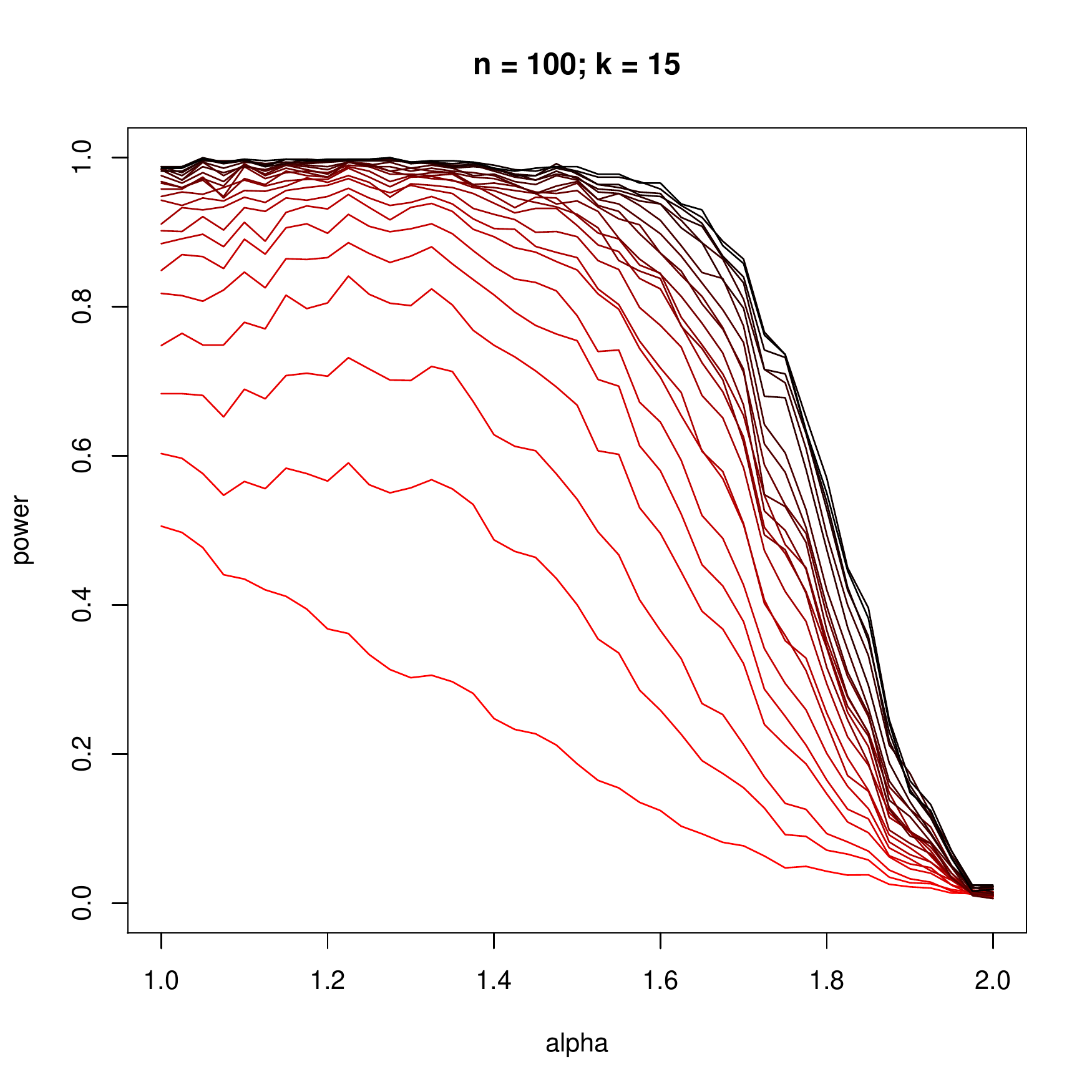}
\includegraphics[width = 0.49 \linewidth]{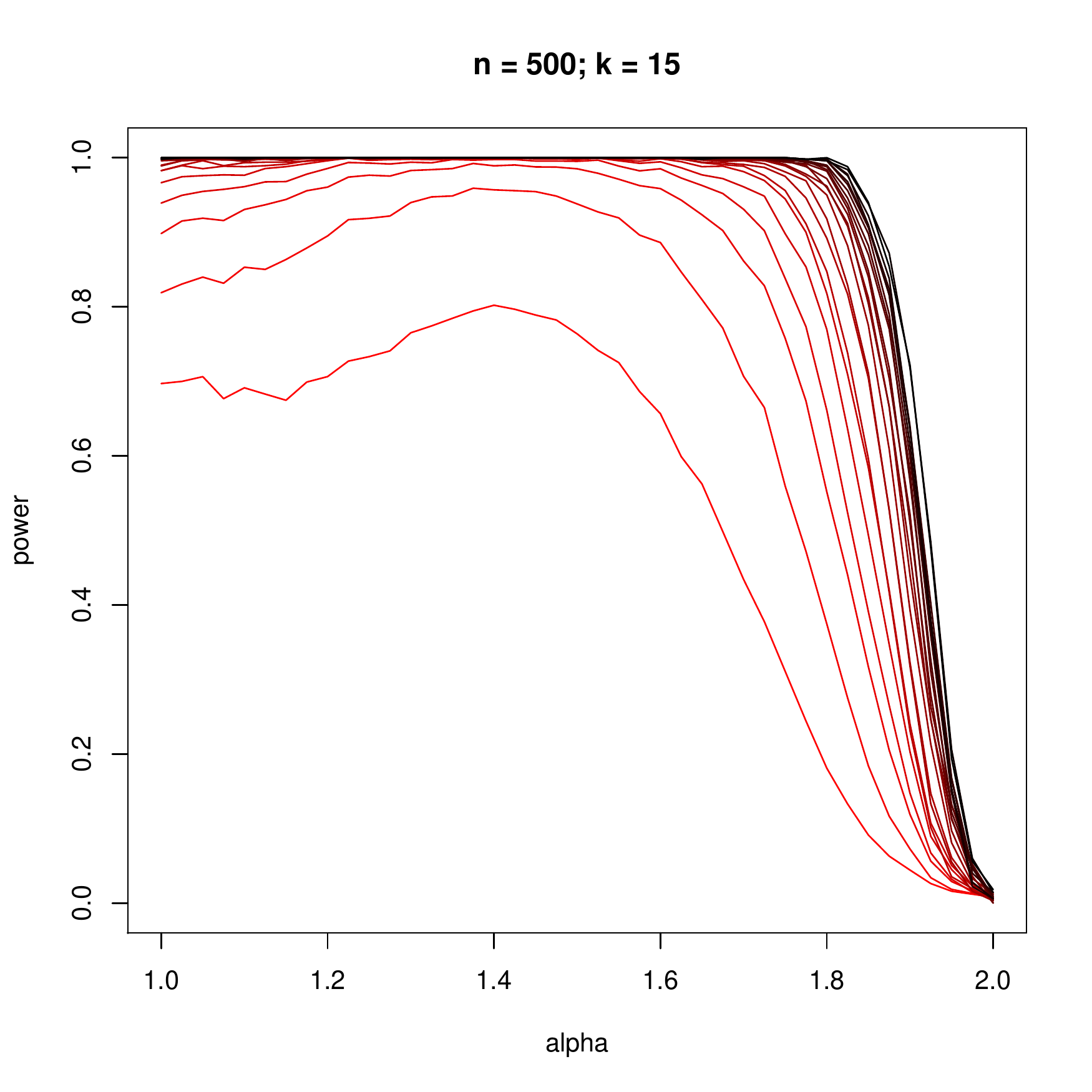}
\includegraphics[width = 0.49 \linewidth]{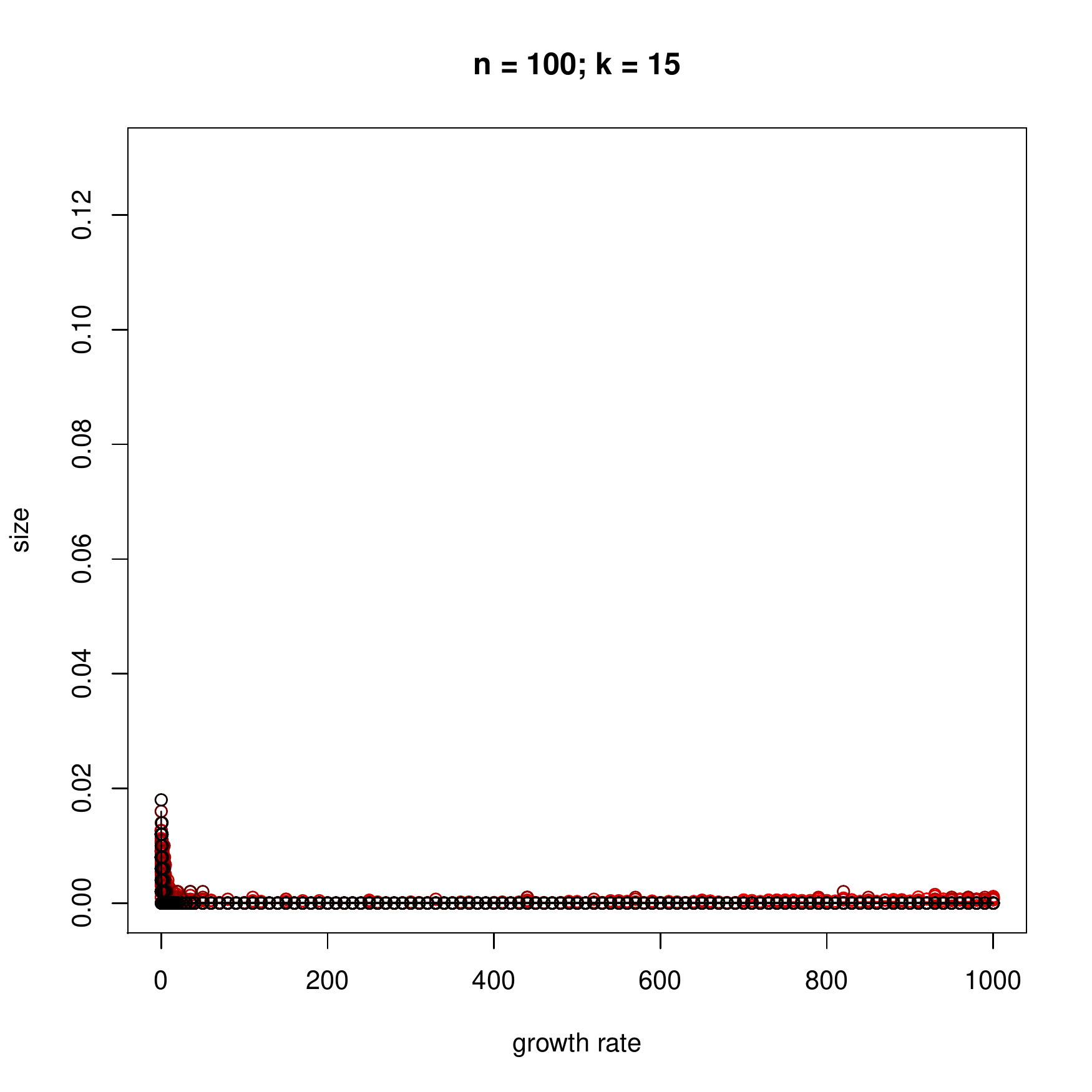}
\includegraphics[width = 0.49 \linewidth]{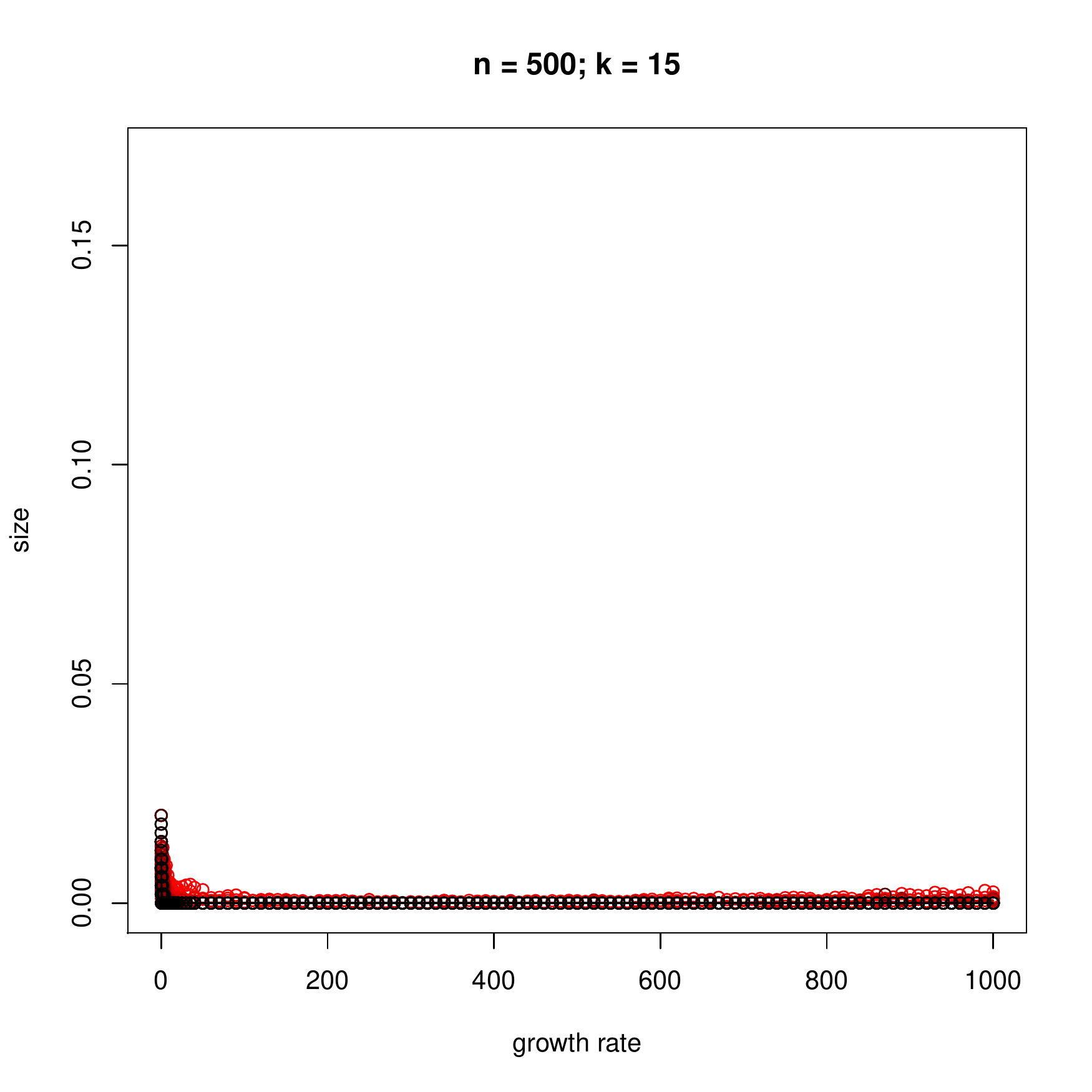}
\includegraphics[width = 0.99 \linewidth]{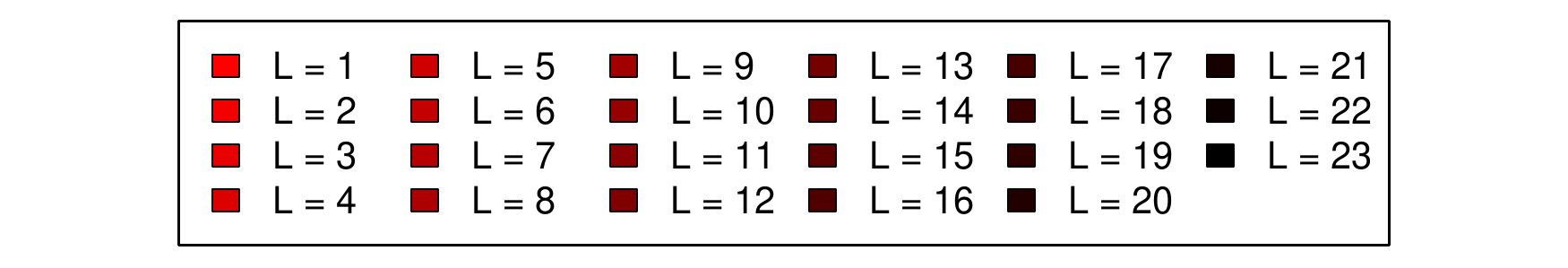}
\caption{Effect of the number of loci over which \eqref{mean_sfs} is averaged.
On the second row, solid lines correspond to an exponential growth model, and discrete circles to an algebraic growth model.
Post-processing these graphs from the simulated data in \textsf{R} took 180 minutes for $n = 100$, and 630 minutes for $n = 500$ on a single Intel i5-2520M 2.5 GHz processor.
The discrepancy in run times in comparison to Figures \ref{cutoff} is due to the necessity of repeatedly computing averages over various numbers of loci.}
\label{loci}
\end{figure}

Figure \ref{loci} investigates the effect of averaging across loci.
When averaging over $l < L = 23$ loci, we treat the the 1000 simulated replicates of $L$ loci as  $1000 \lfloor L / l \rfloor$ replicates to keep the total amount of data as constant as possible.
For example, the $l = 1$ curve corresponds to treating the data as $23 \times 1000$ independent replicates, the $l = 2$ curve corresponds to treating the data as $11 \times 1000$ independent replicates, each being the average of two loci, and so on until the final, $l = 23$ curve corresponds to 1000 replicates, each being the average of all 23 loci.
Figure \ref{loci} shows diminishing returns as the number of loci being averaged increases, but does not reach a saturation point in the same way as Figure \ref{cutoff}; for the hypotheses and parameters considered here, adding more loci always improves power.
The empirical size of the test is not sensitive to the number of loci, as was the case of the cutoff in Figure \ref{cutoff} as well.

\begin{figure}[!ht]
\centering
\includegraphics[width = 0.49 \linewidth]{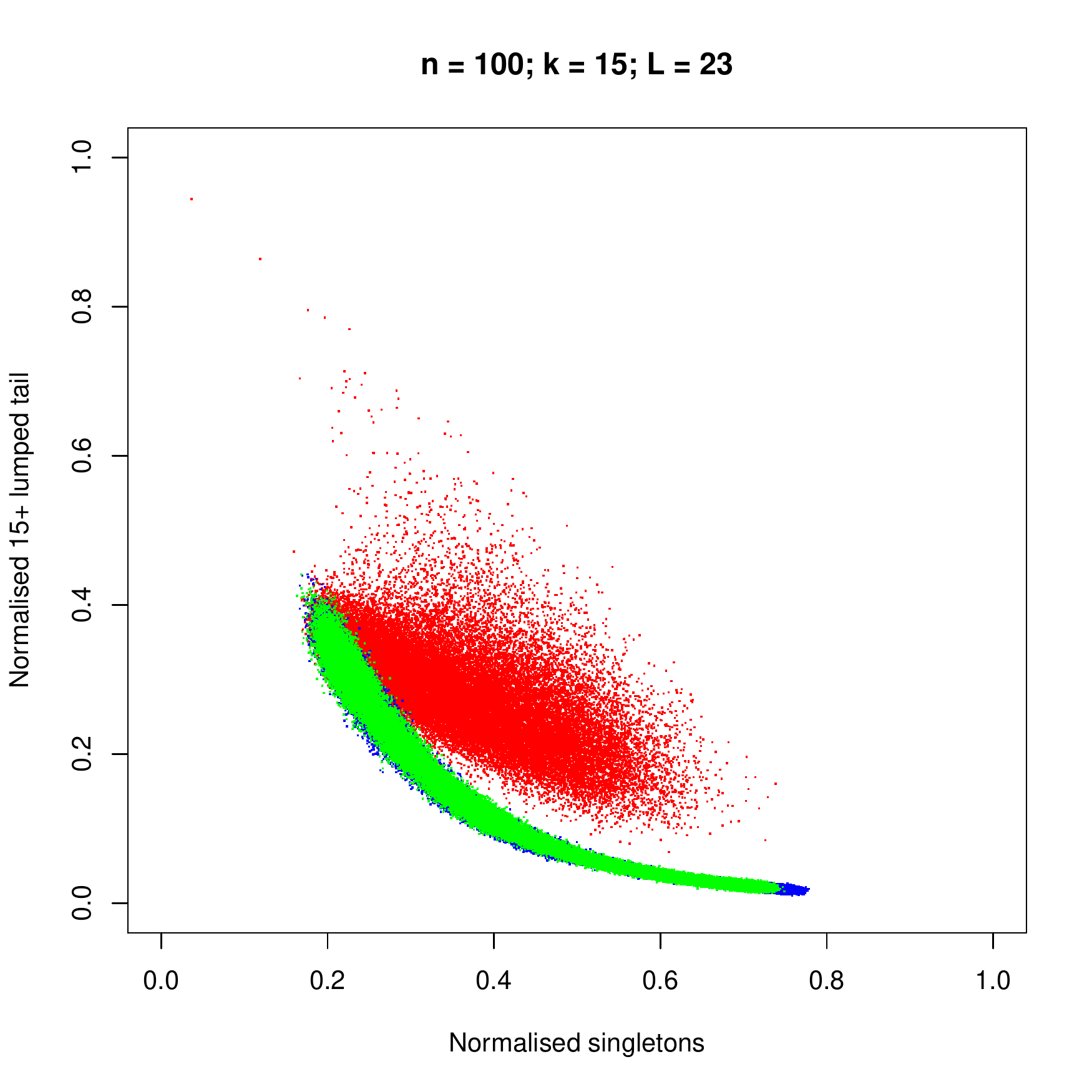}
\includegraphics[width = 0.49 \linewidth]{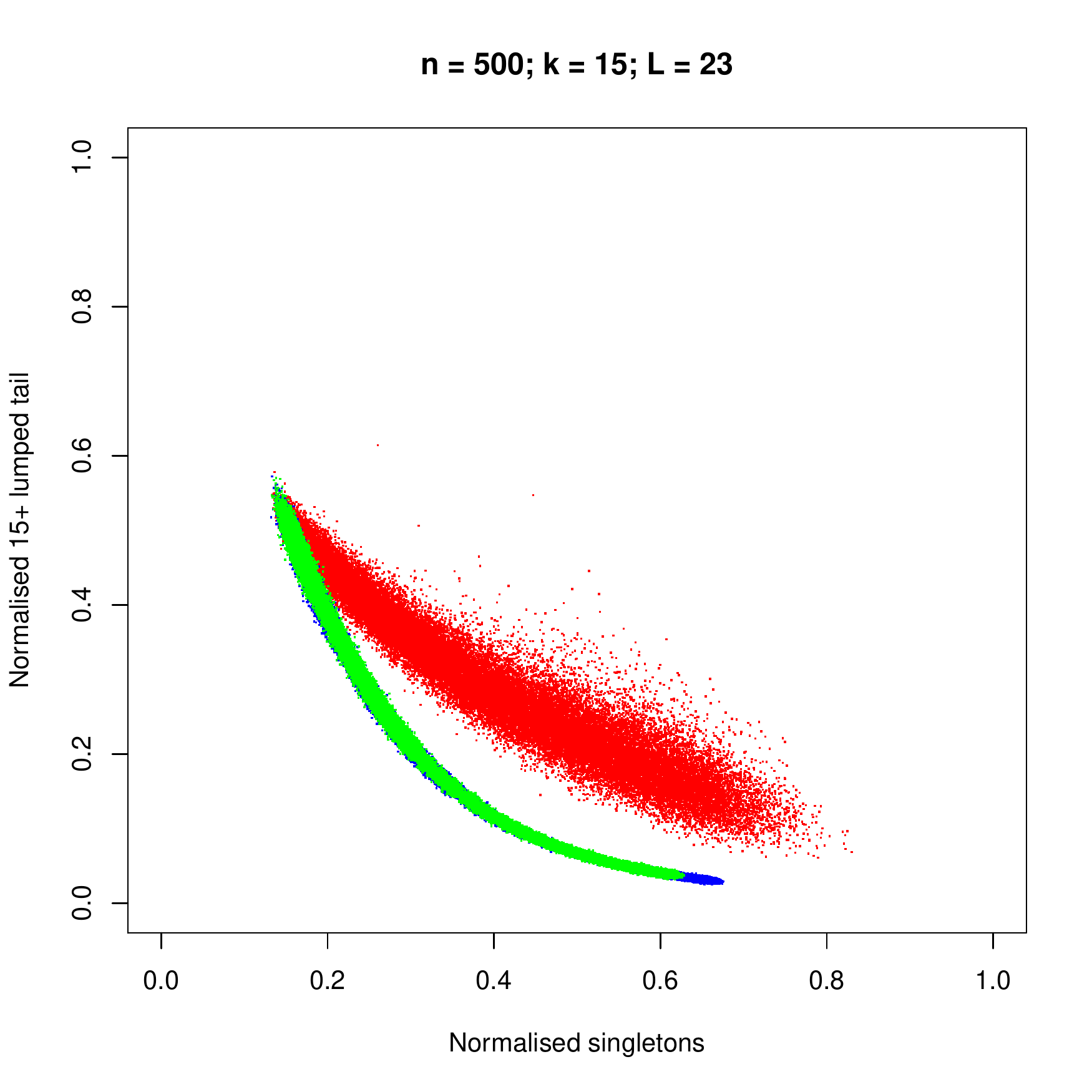}
\caption{A repeat of the scatter plots of Figure \ref{scatter} with mutation rate increased by a factor of 10.}
\label{scatter_hi_mut}
\end{figure}

\begin{figure}[!ht]
\centering
\includegraphics[width = 0.49 \linewidth]{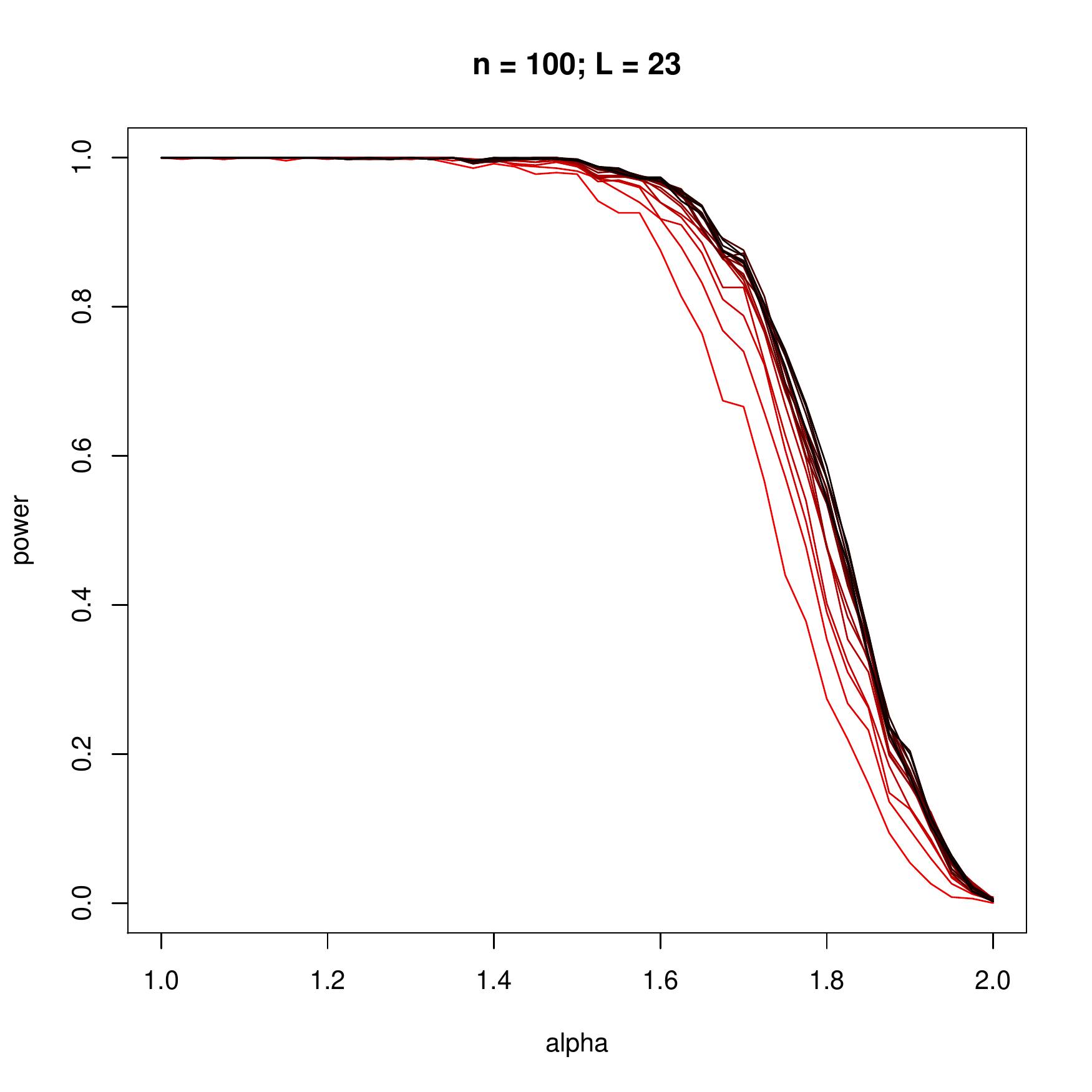}
\includegraphics[width = 0.49 \linewidth]{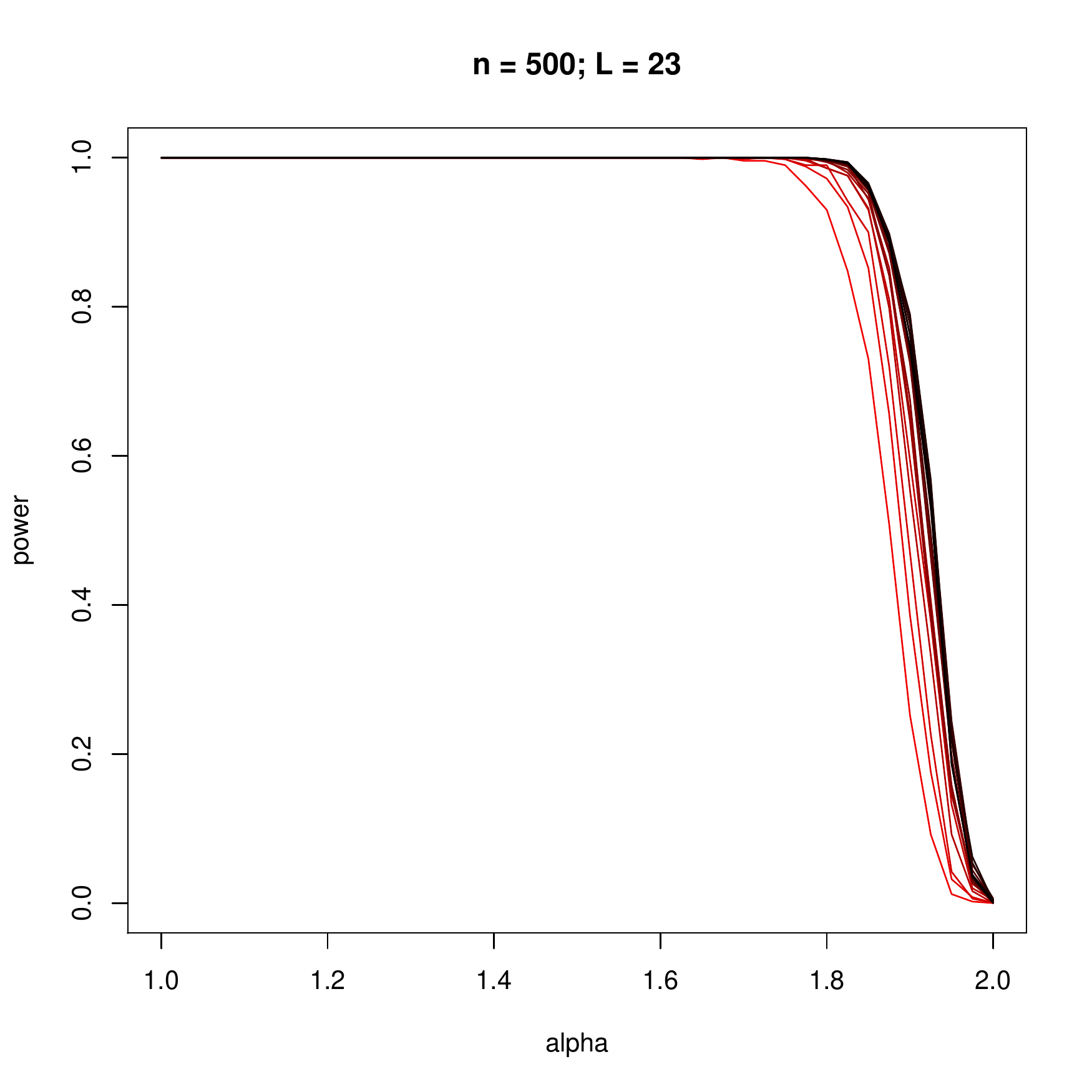}
\includegraphics[width = 0.49 \linewidth]{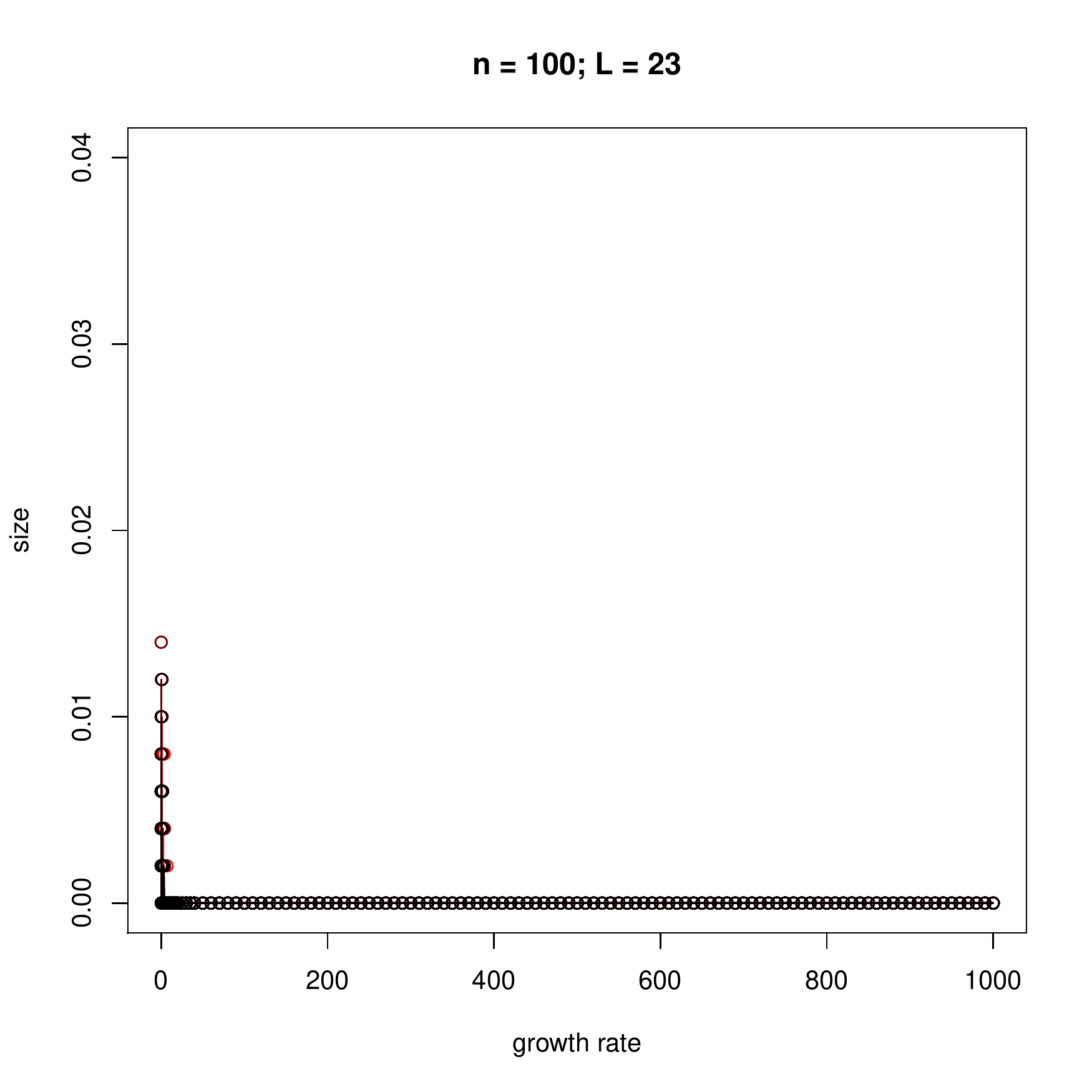}
\includegraphics[width = 0.49 \linewidth]{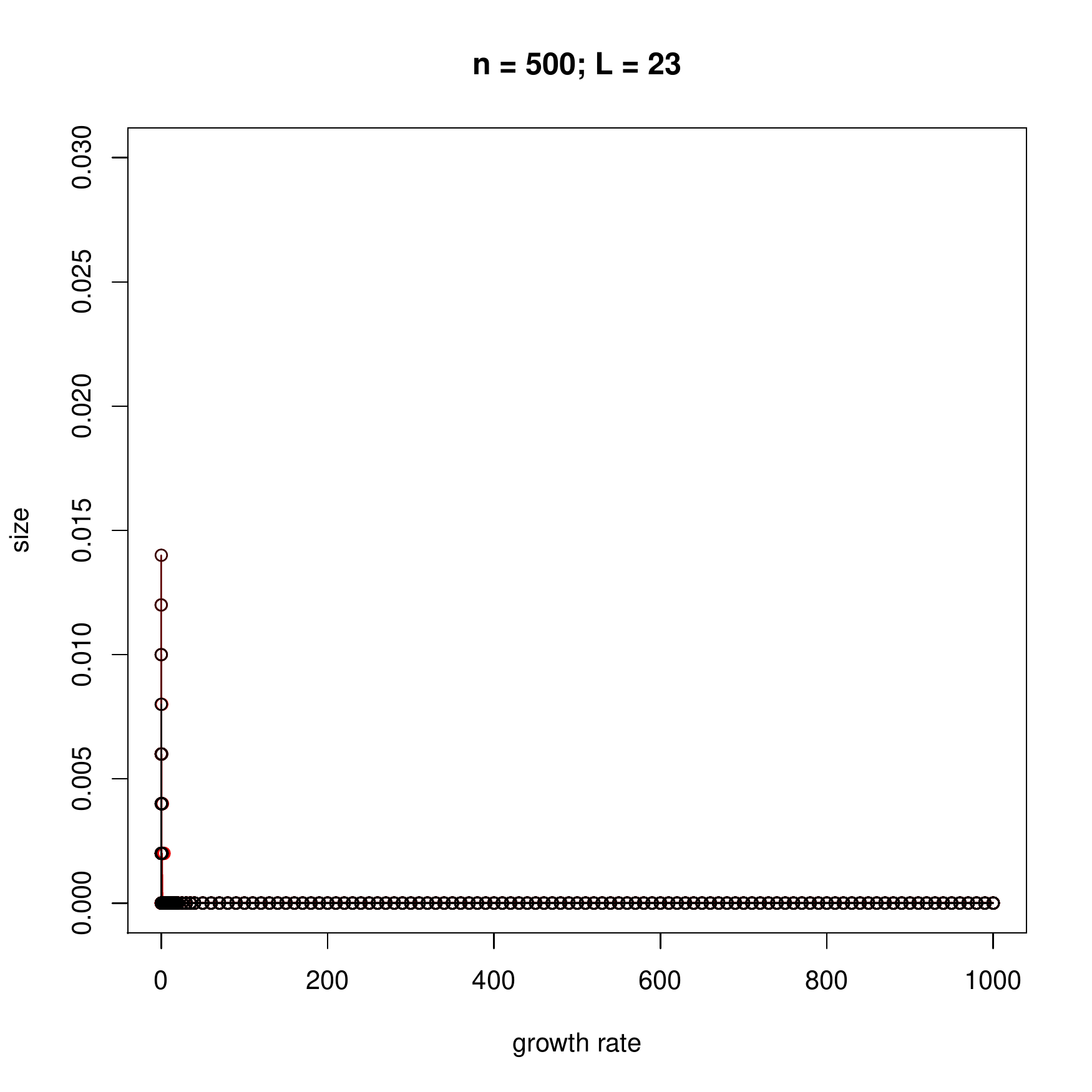}
\includegraphics[width = 0.99 \linewidth]{k_legend.pdf}
\caption{A repeat of the analysis from Figure \ref{cutoff} when the true mutation rate is 10 times higher than that used in the simulation algorithm.}
\label{mutation_hi_cutoff}
\end{figure}

\begin{figure}[!ht]
\centering
\includegraphics[width = 0.49 \linewidth]{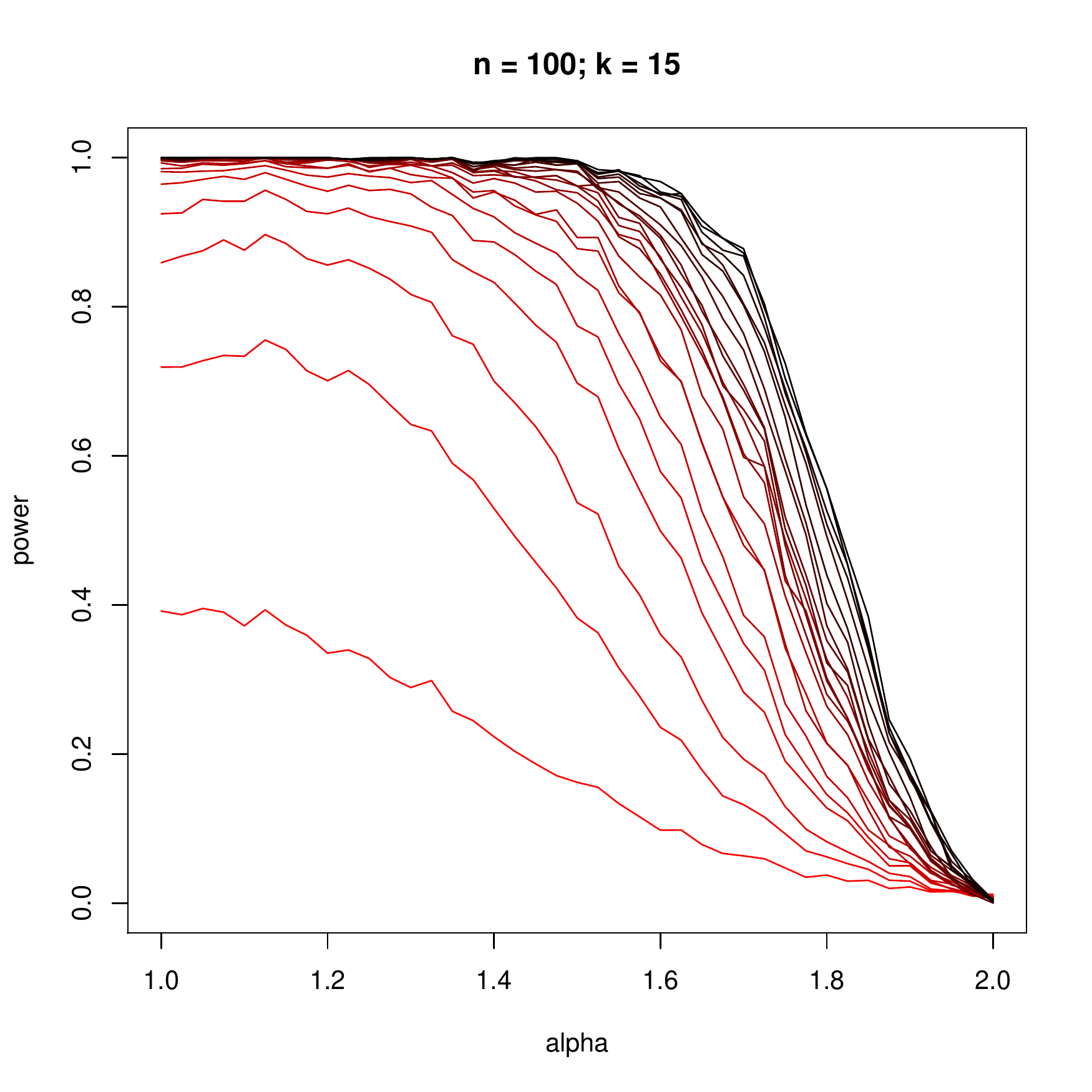}
\includegraphics[width = 0.49 \linewidth]{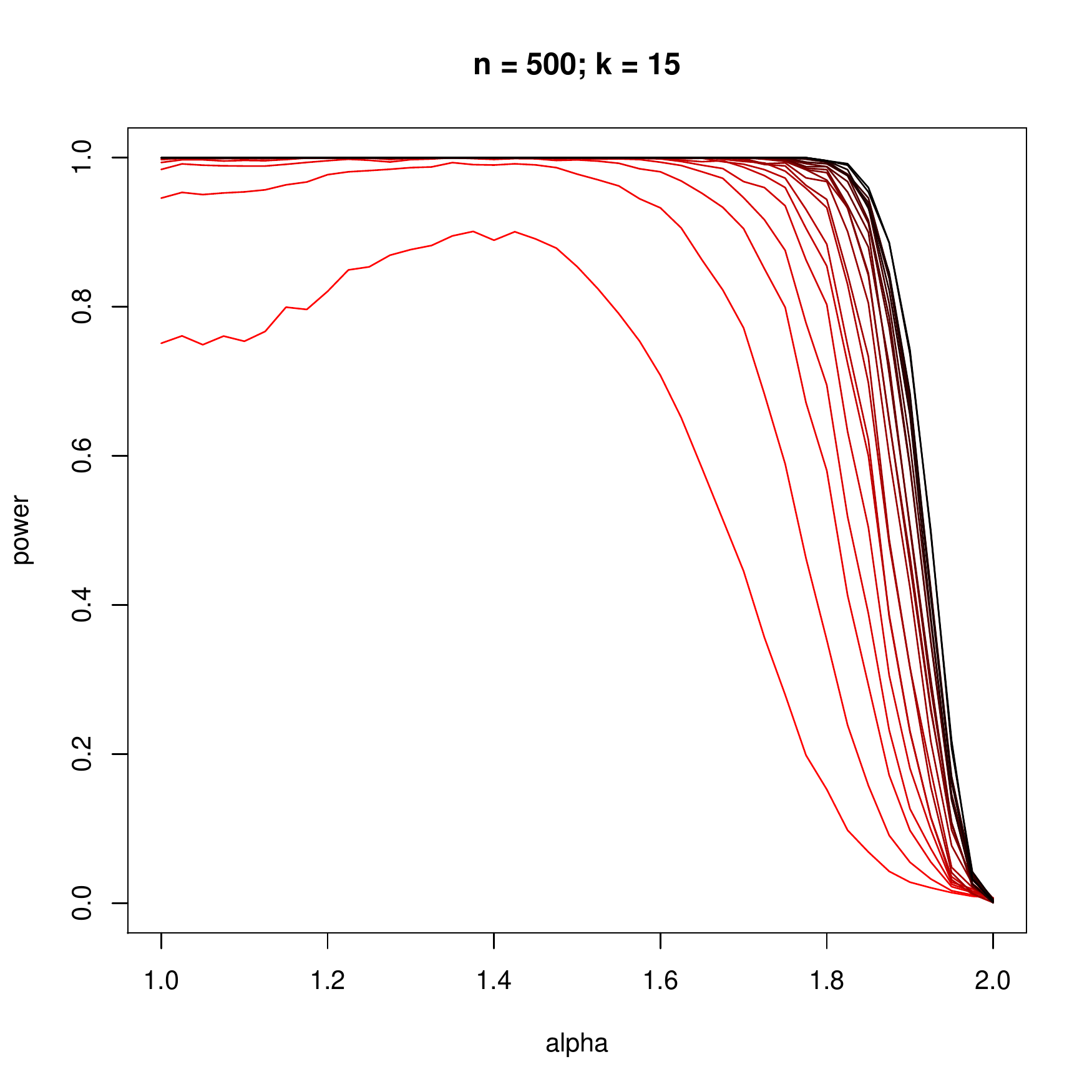}
\includegraphics[width = 0.49 \linewidth]{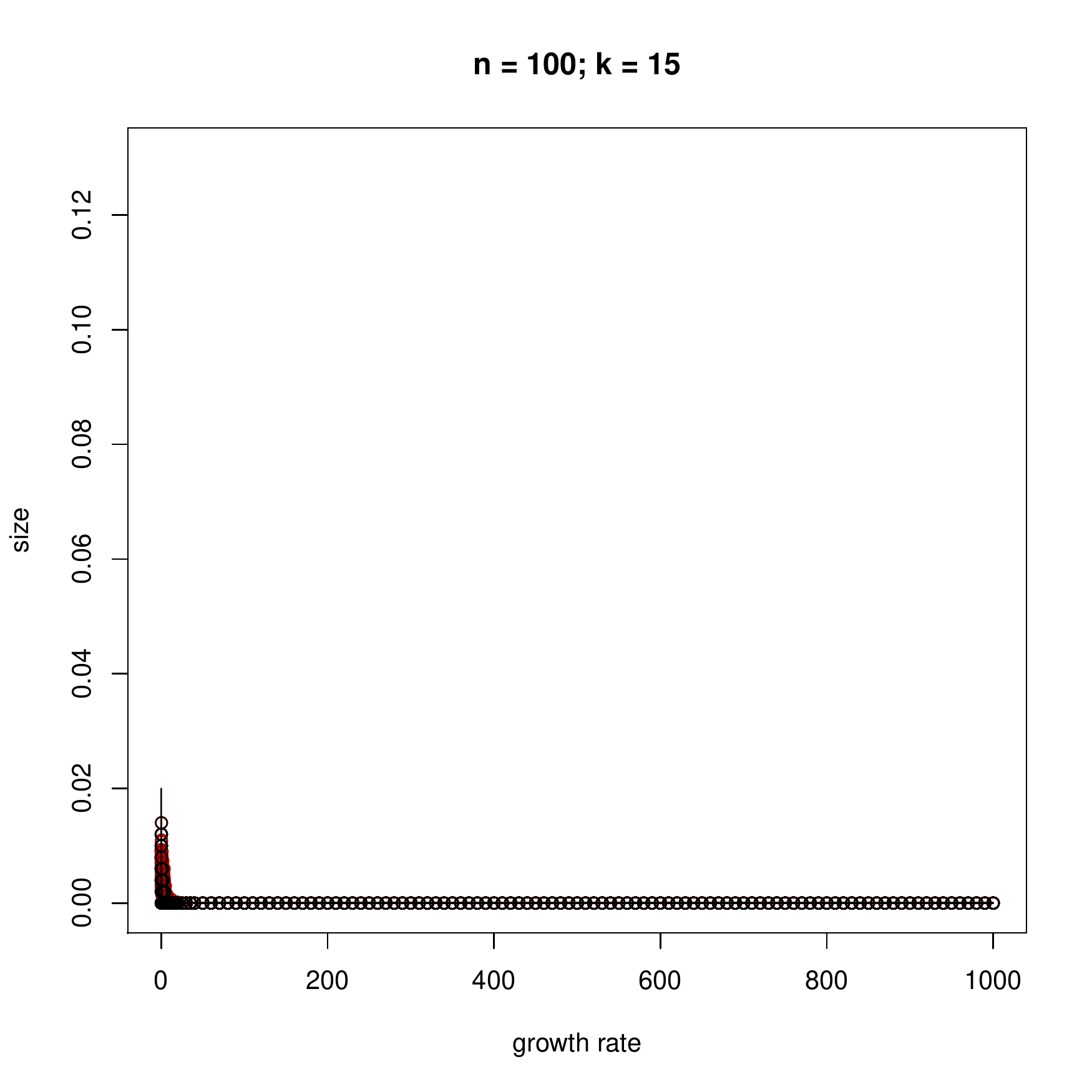}
\includegraphics[width = 0.49 \linewidth]{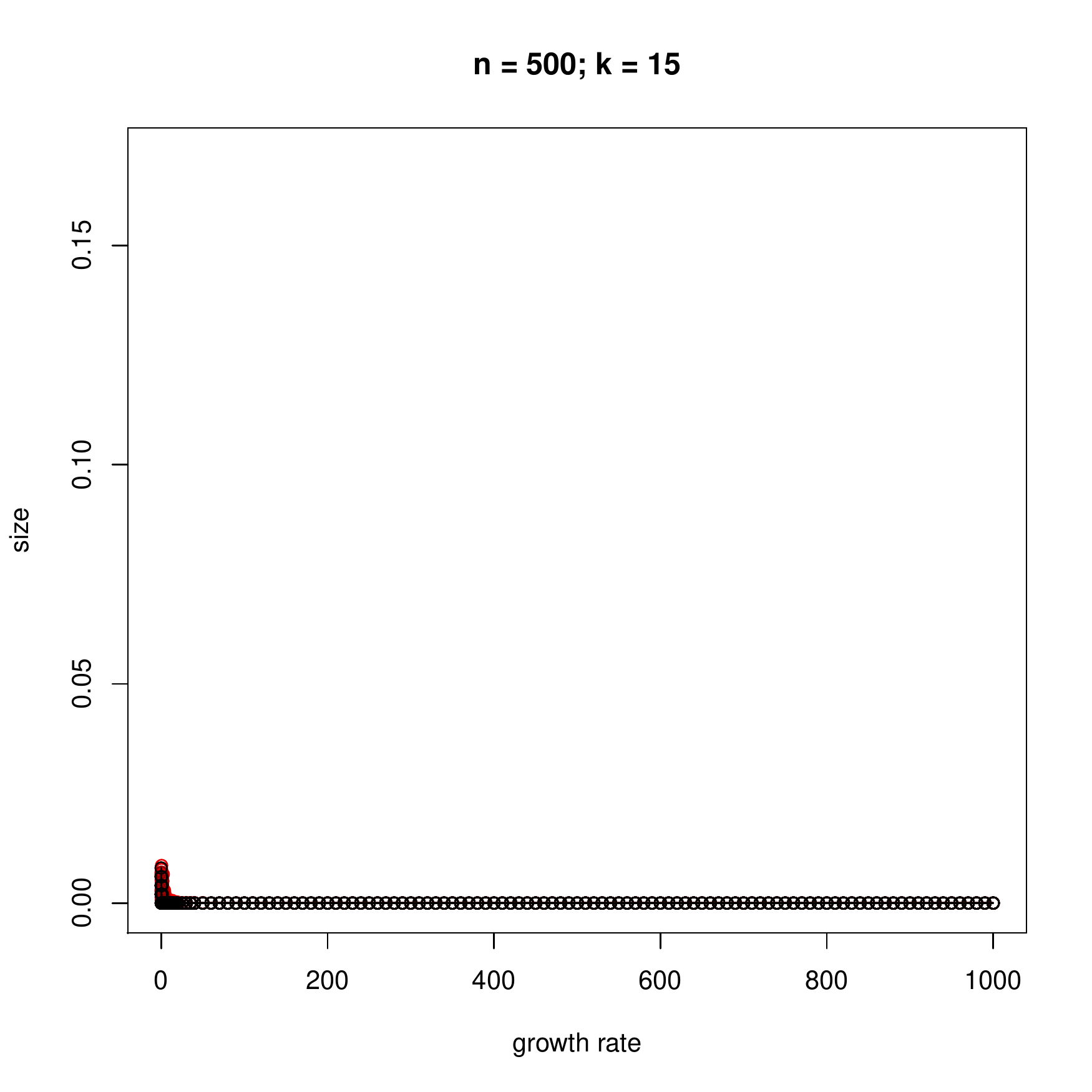}
\includegraphics[width = 0.99 \linewidth]{L_legend.pdf}
\caption{A repeat of the analysis from Figures \ref{loci} when the true mutation rate is 10 times higher than that used in the simulation algorithm.}
\label{mutation_hi_loci}
\end{figure}

\begin{figure}[!ht]
\centering
\includegraphics[width = 0.49 \linewidth]{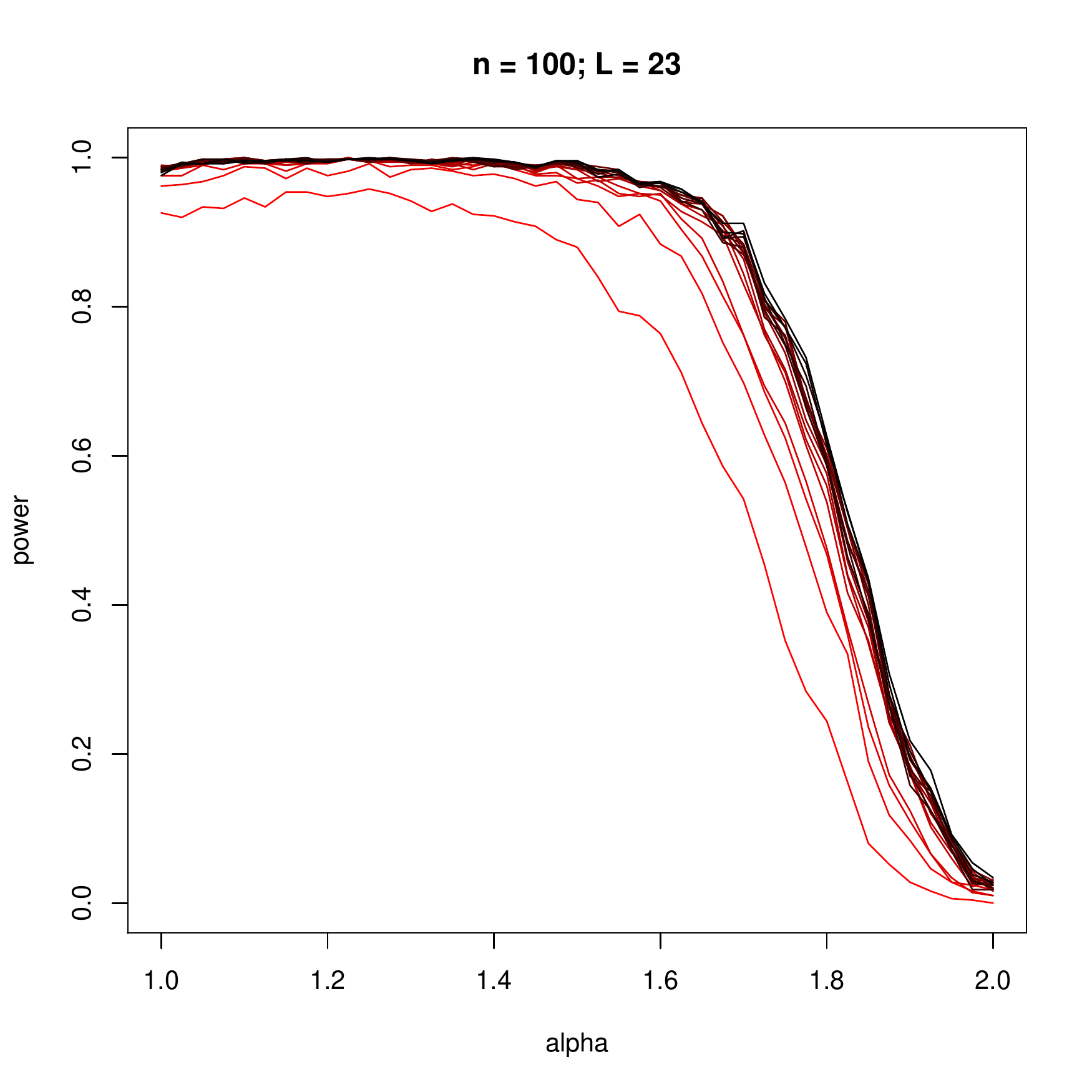}
\includegraphics[width = 0.49 \linewidth]{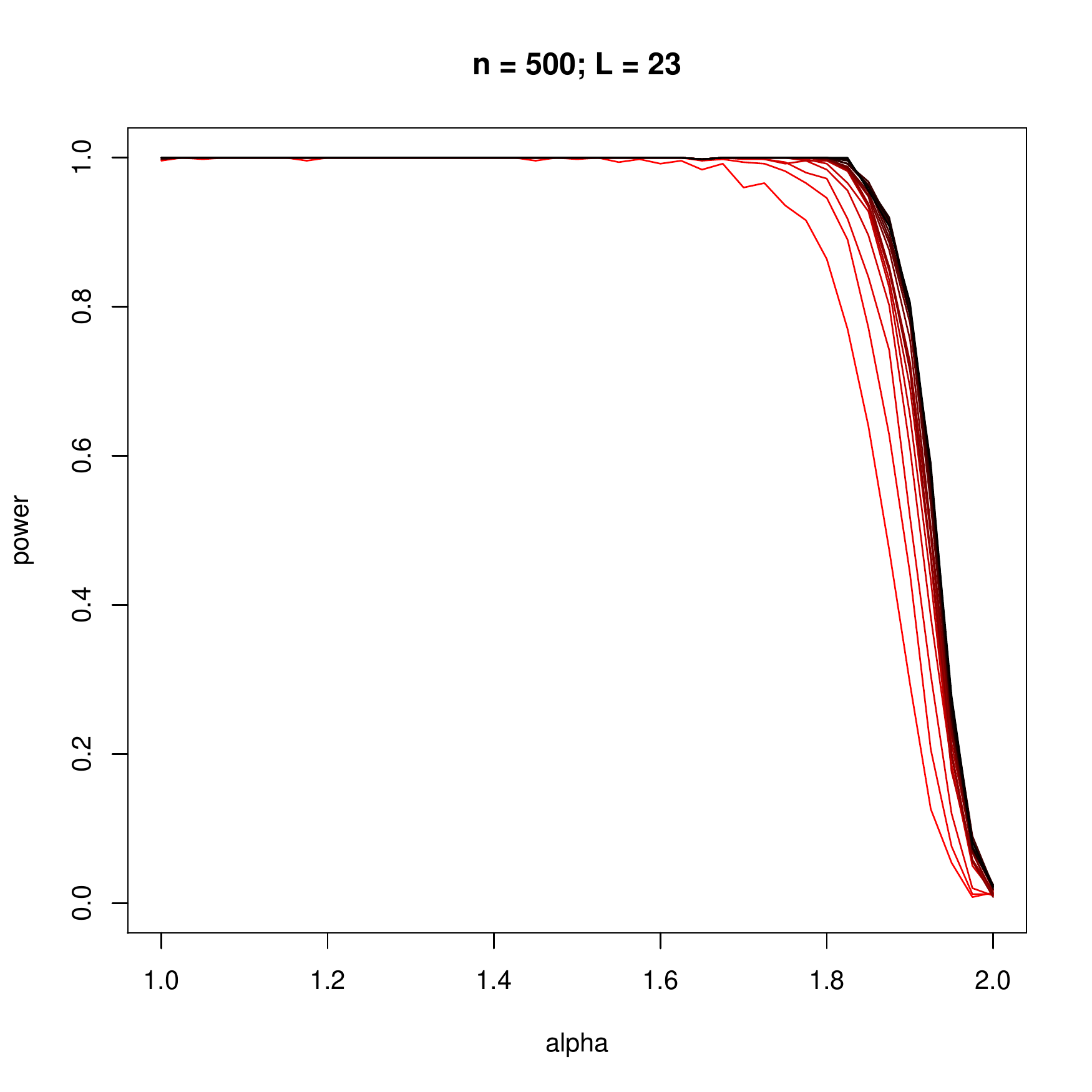}
\includegraphics[width = 0.49 \linewidth]{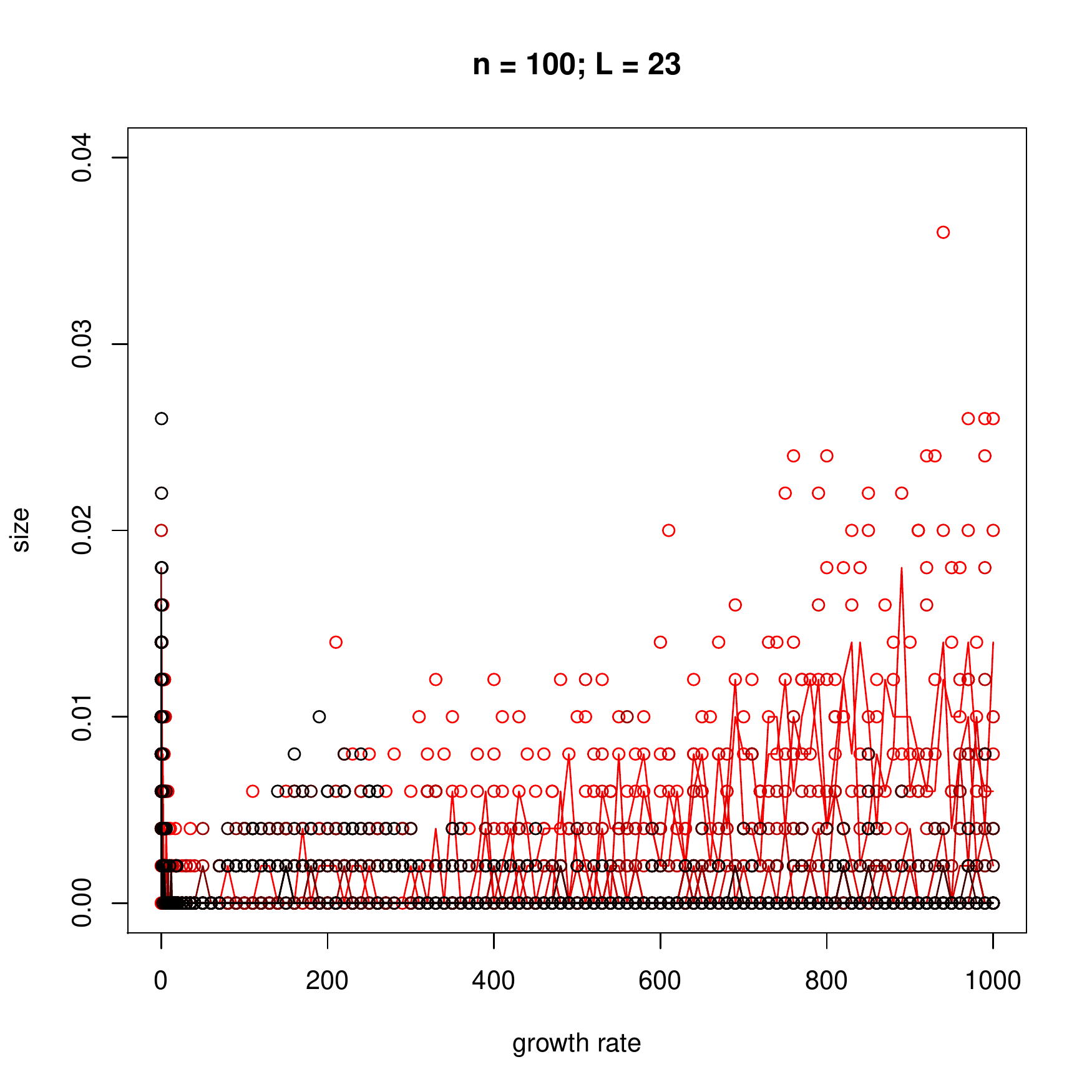}
\includegraphics[width = 0.49 \linewidth]{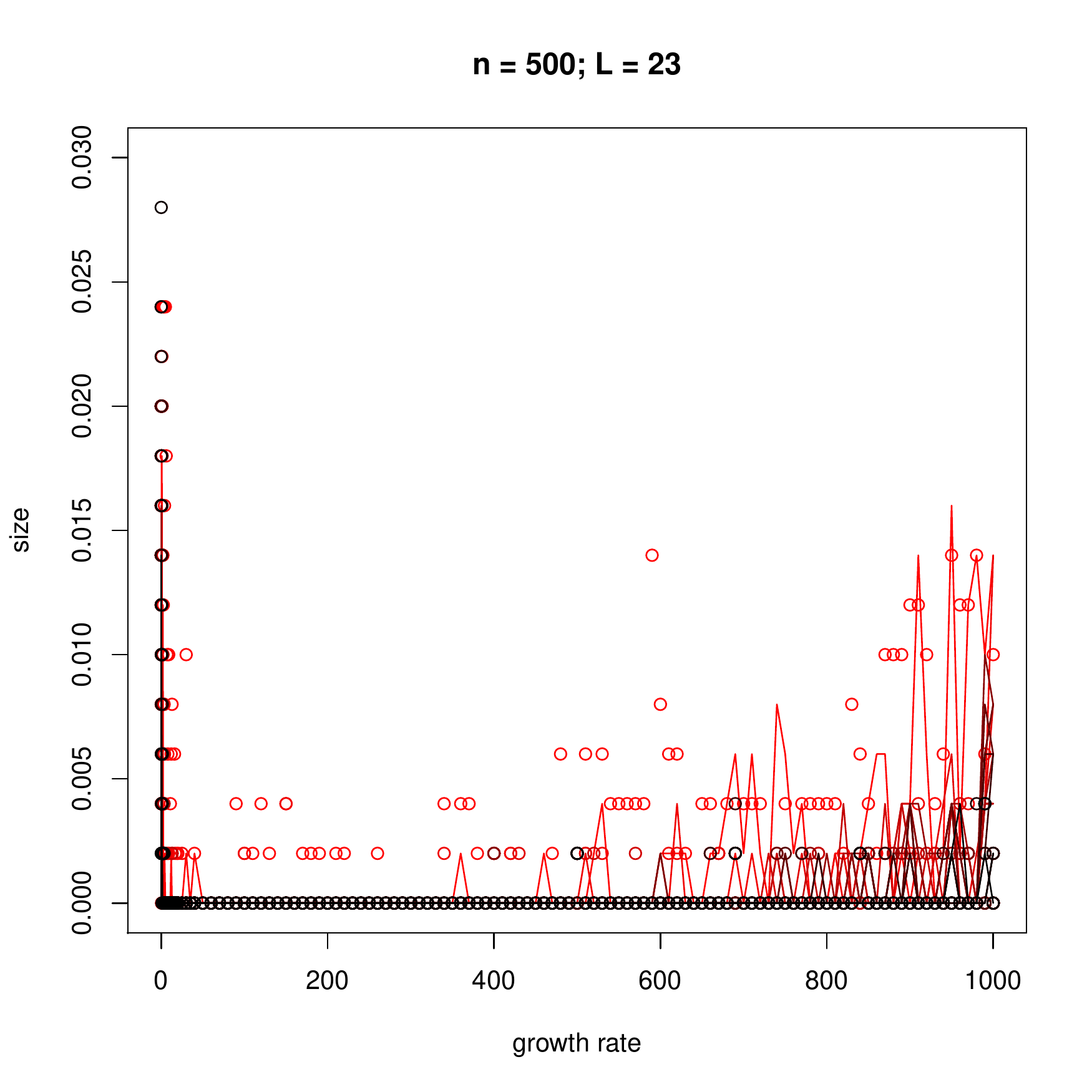}
\includegraphics[width = 0.99 \linewidth]{k_legend.pdf}
\caption{A repeat of the analysis from Figure \ref{cutoff} when the true mutation rate is 10 times lower than that used in the simulation algorithm.}
\label{mutation_lo_cutoff}
\end{figure}

\begin{figure}[!ht]
\centering
\includegraphics[width = 0.49 \linewidth]{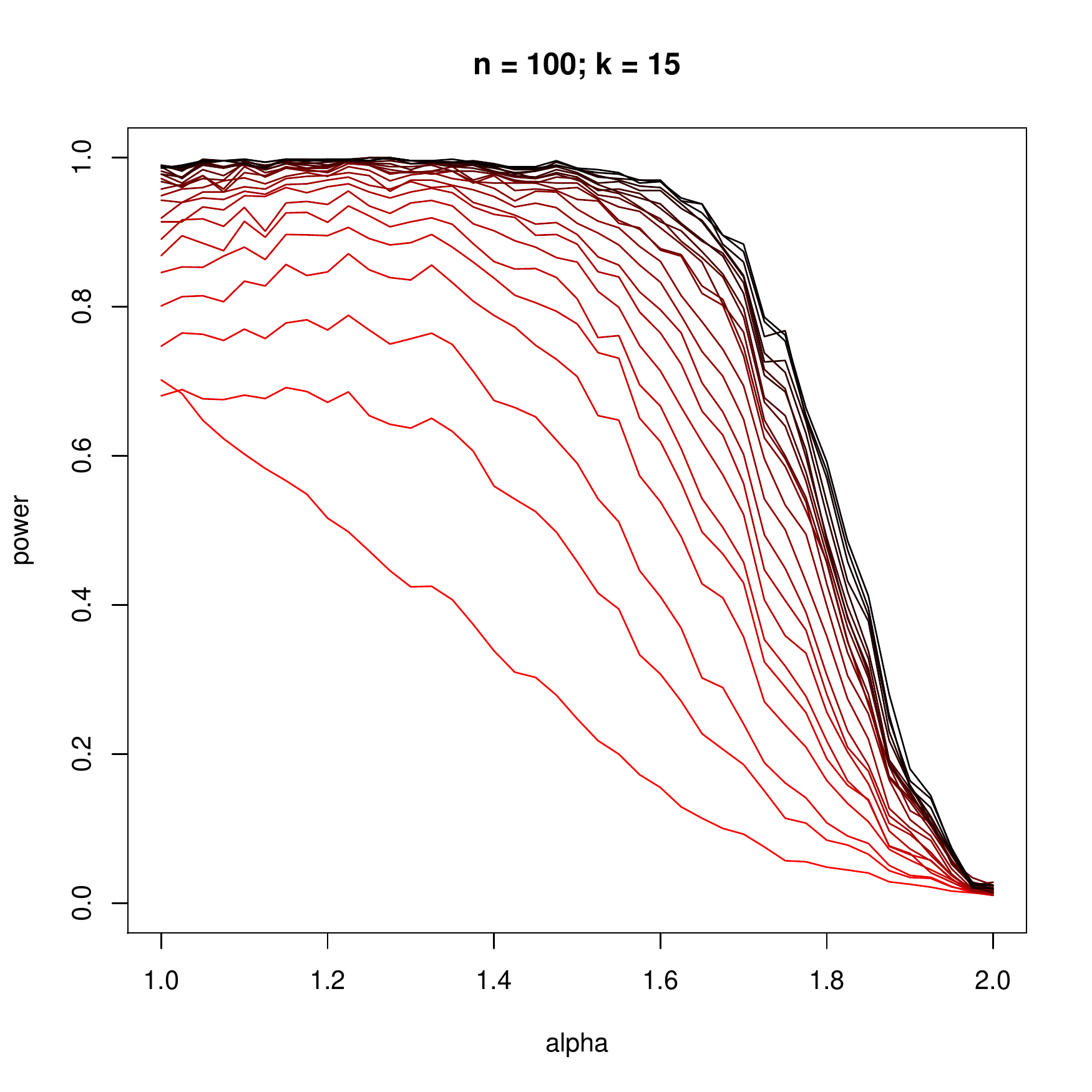}
\includegraphics[width = 0.49 \linewidth]{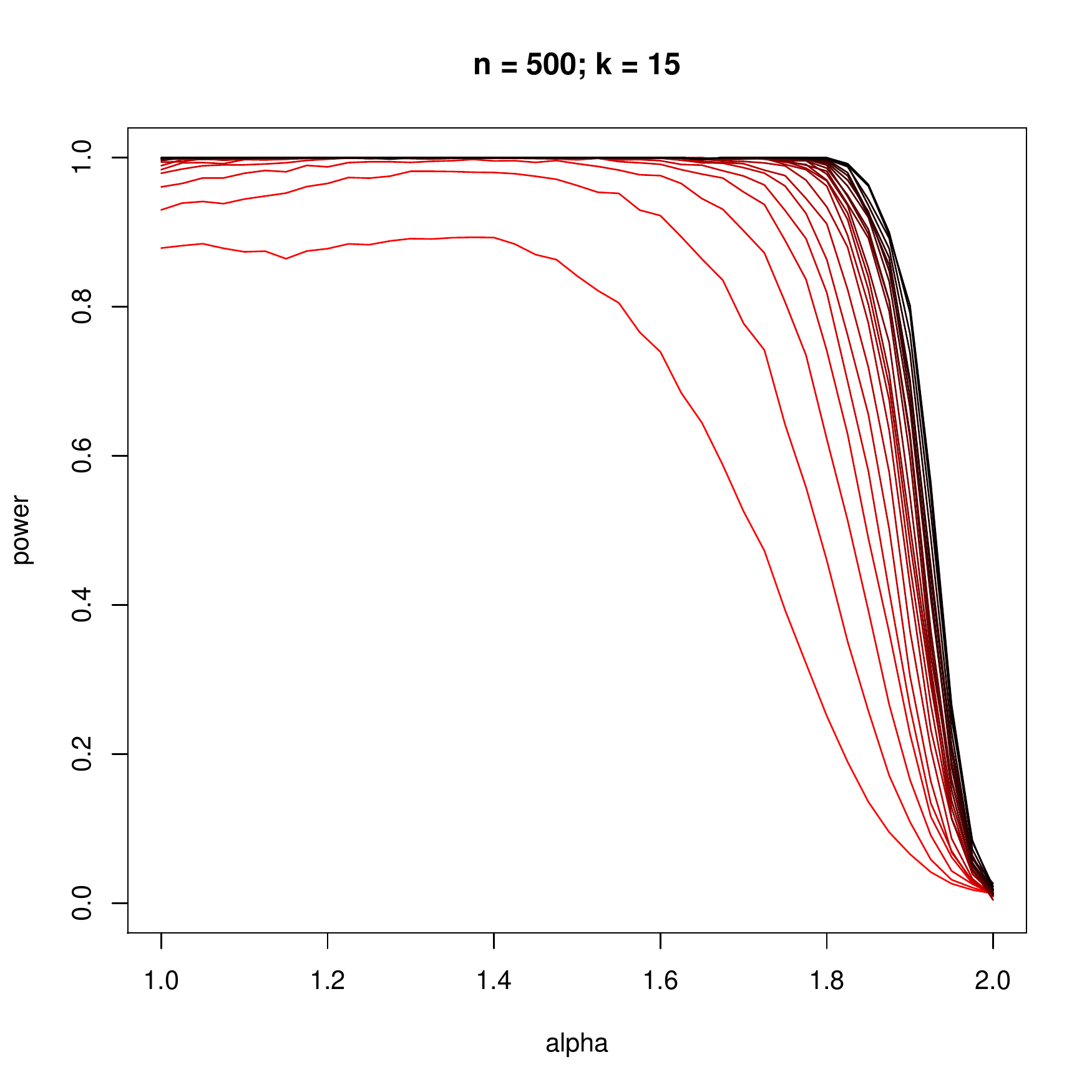}
\includegraphics[width = 0.49 \linewidth]{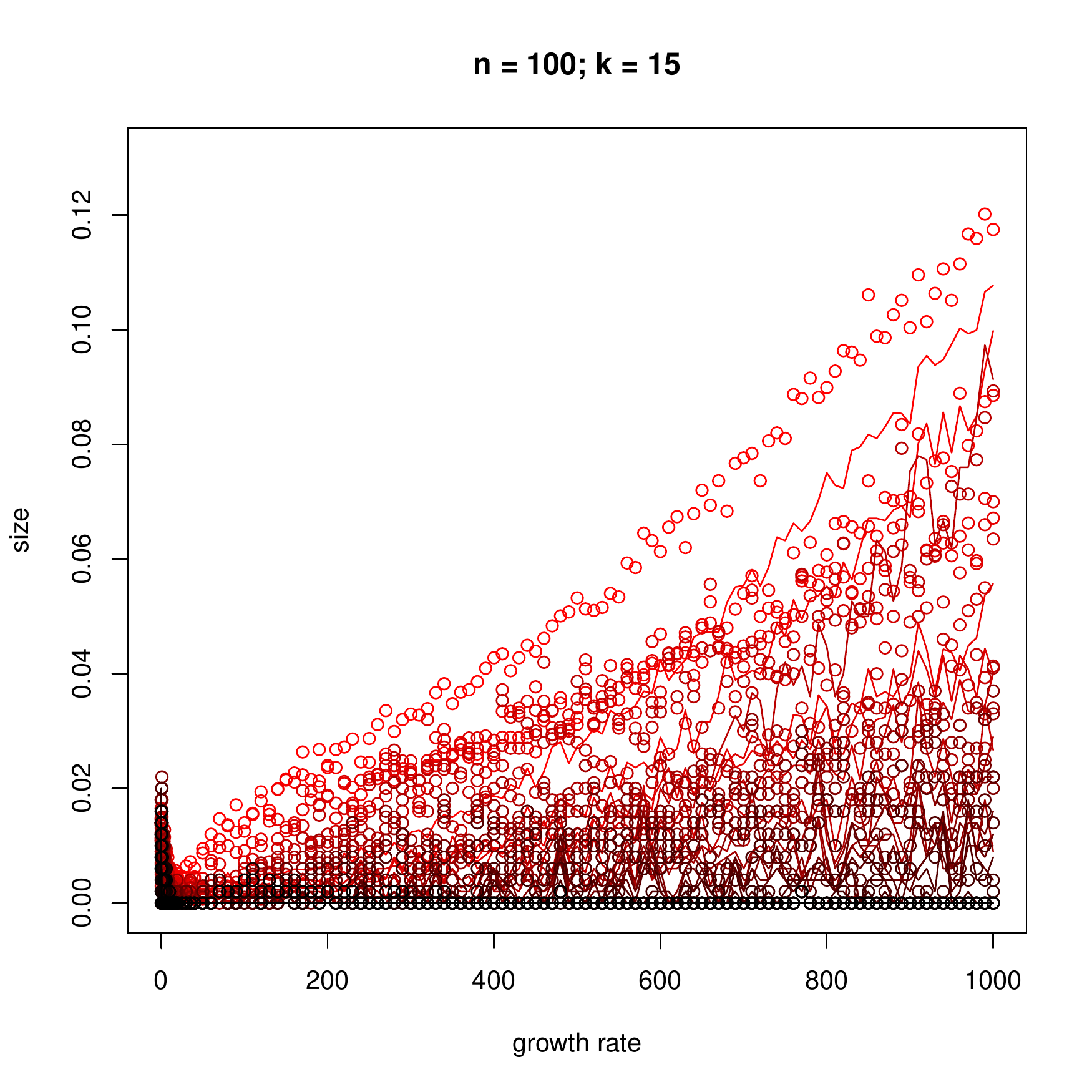}
\includegraphics[width = 0.49 \linewidth]{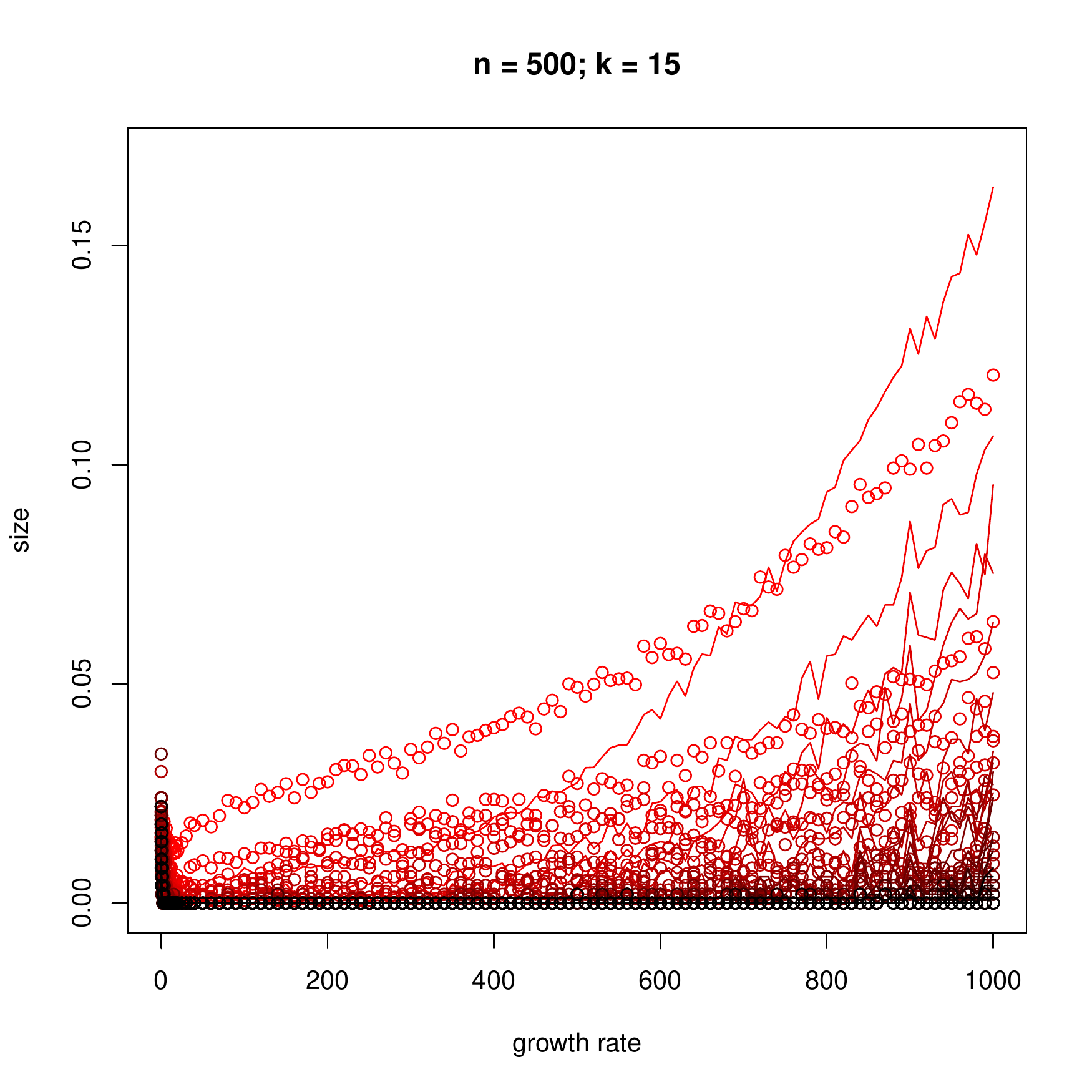}
\includegraphics[width = 0.99 \linewidth]{L_legend.pdf}
\caption{A repeat of the analysis from Figure \ref{loci} when the true mutation rate is 10 times lower than that used in the simulation algorithm.}
\label{mutation_lo_loci}
\end{figure}

Finally, we investigate the effect of misspecification of the mutation rate under which inference is conducted.
Figure \ref{scatter_hi_mut} shows the empirical sampling distribution of the singletons and lumped tail when the mutation rate is 10 times higher than it was in the simulations used to produce Figure \ref{scatter}.
It can be seen that this increase in mutation rate causes a slight reduction in sampling variance, but the difference is small enough to be difficult to detect by eye.
Figures \ref{mutation_hi_cutoff} through \ref{mutation_lo_loci} depict the statistical power resulting from our test when the mutation rate is either too small or too large by a factor of 10.
It can be seen that the power curves are nearly indistinguishable from those obtained using the true mutation rate in earlier figures, at least for sample size $n = 500$ and when both the lumped tail cut off and the number of loci being averaged are larger than 3 or 4.
In contrast, the empirical sizes of tests deteriorate substantially if the mutation rate for the simulated data is too high, and either the cutoff for the lumped tail or the number of loci is too small (c.f.~Figures \ref{mutation_lo_cutoff} and \ref{mutation_lo_loci}).
However, even in these cases the empirical sizes remain close to or below the reference value of $\varepsilon = 0.01$ when both the cutoff for the lumped tail and the number of loci being averaged are at least 10.
These results demonstrate that it is best to use a low value of the mutation rate in simulations when a range of plausible values is available.
The reason is intuitively clear from Figures \ref{scatter} and \ref{scatter_hi_mut}: a higher mutation rate decreases the variance of simulation output, which results in unstable hypothesis tests if the true mutation rate is lower because the observed statistic will lie far in the tails of the simulated distributions.

\section{Results: parameter inference}\label{parameters}

The kernel density estimators produced in order to approximate the hypothesis test \eqref{test} can also be used to perform parameter inference.
In this section we investigate the ability of the statistic \eqref{mean_sfs} to distinguish between $\operatorname{Beta}(2 - \alpha, \alpha)$-$\Xi$-coalescents by treating the kernel density estimators $\hat{ P }^{ \Pi, \hat{ \theta } }$ as true likelihood functions.

\begin{figure}[!ht]
\centering
\includegraphics[width = 0.24 \linewidth]{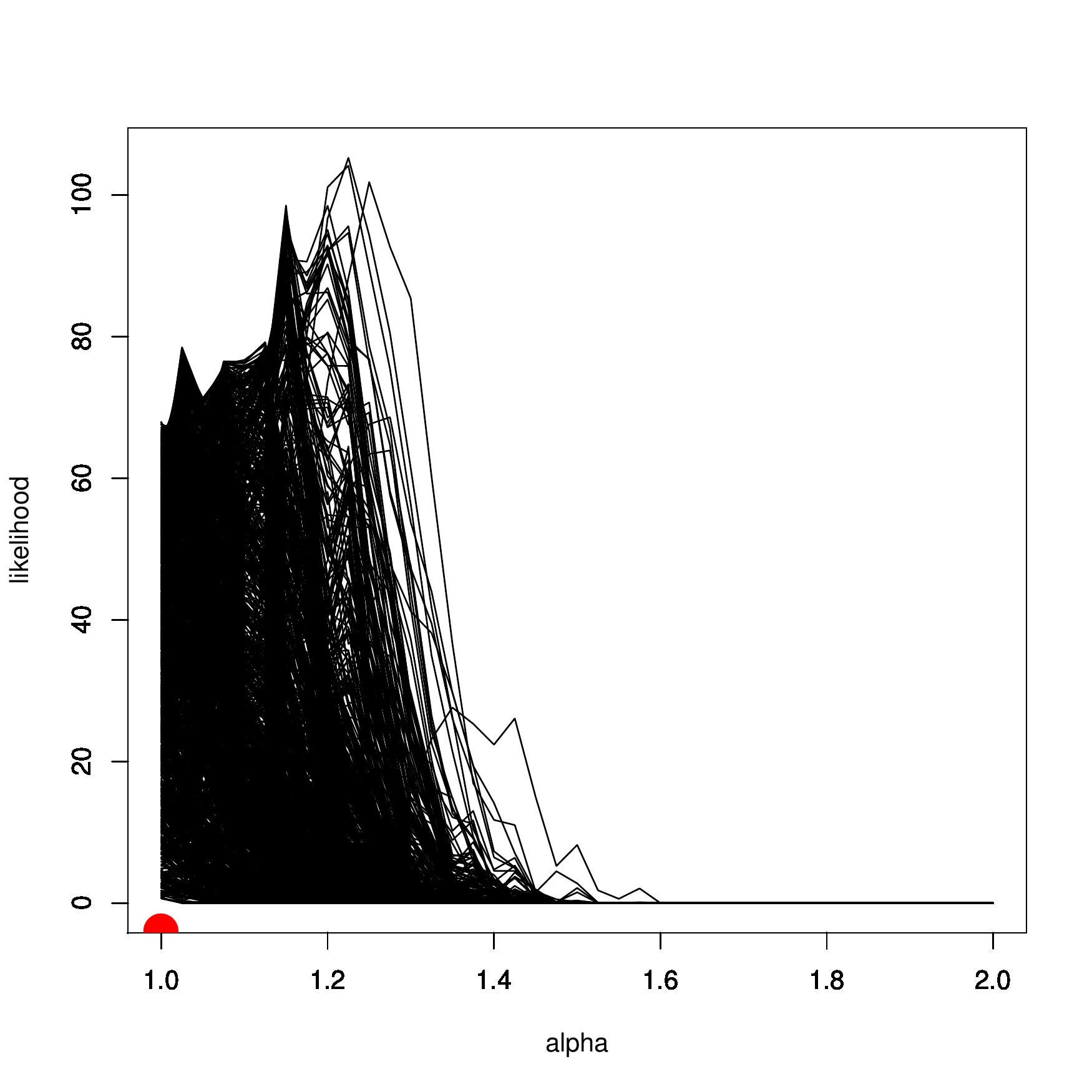}
\includegraphics[width = 0.24 \linewidth]{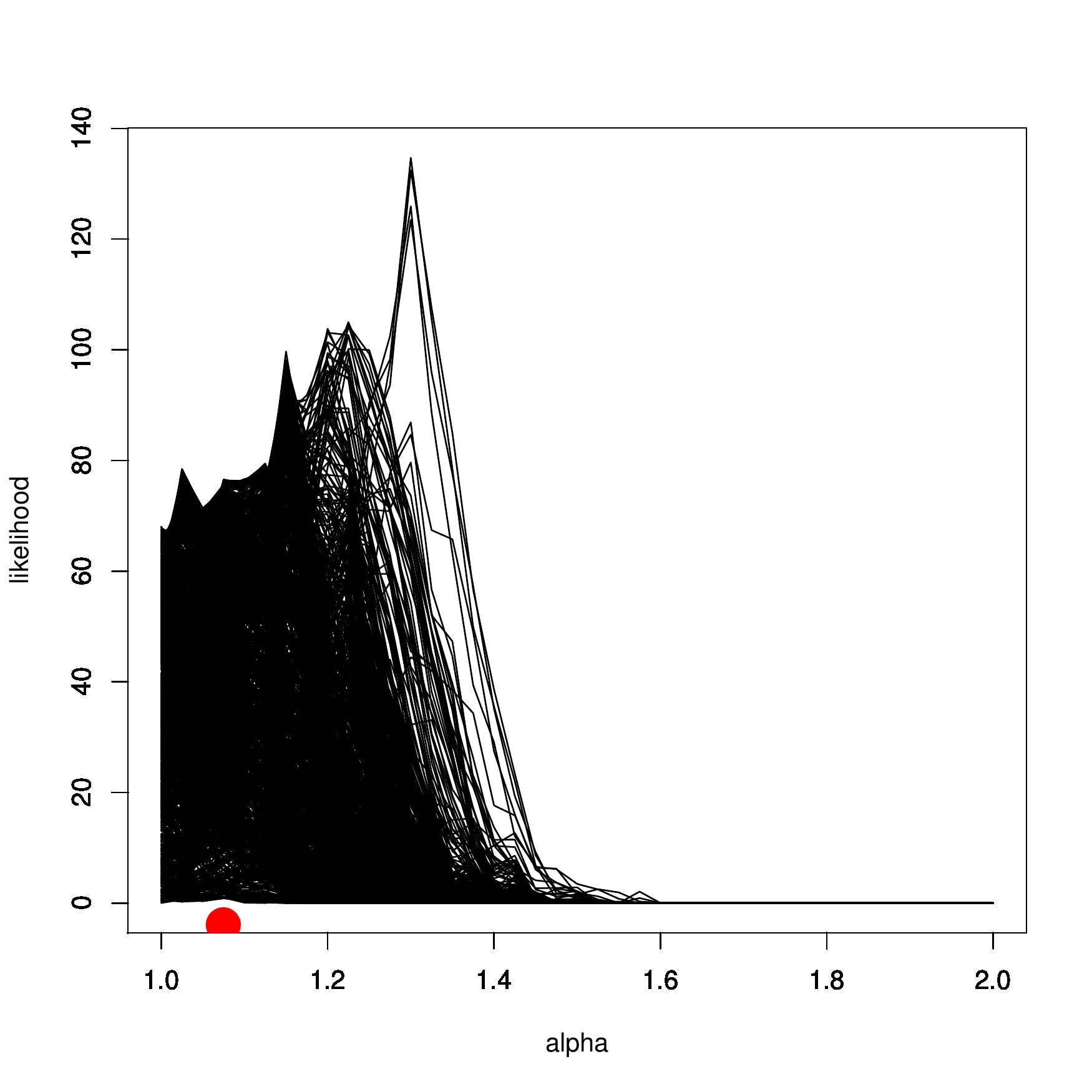}
\includegraphics[width = 0.24 \linewidth]{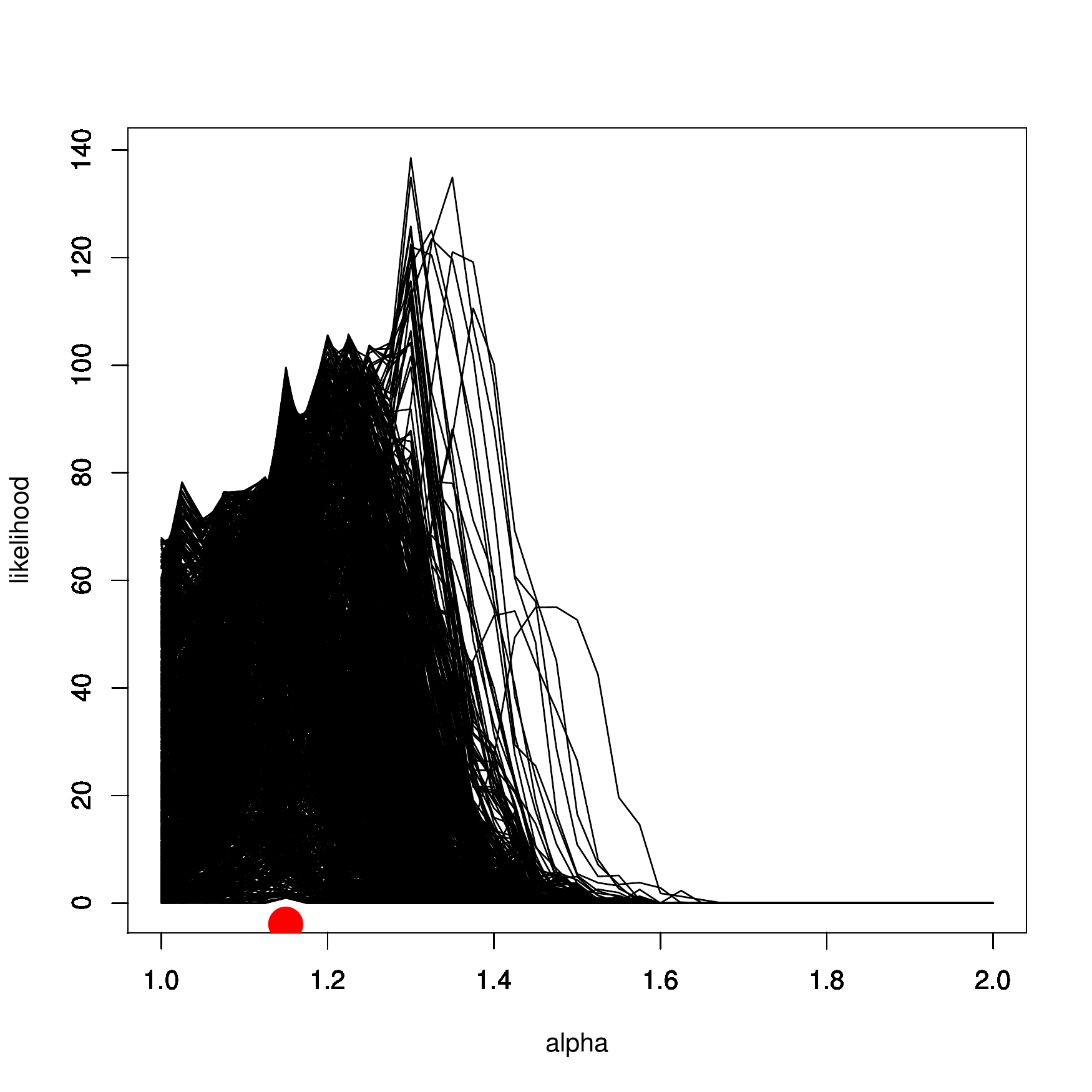}
\includegraphics[width = 0.24 \linewidth]{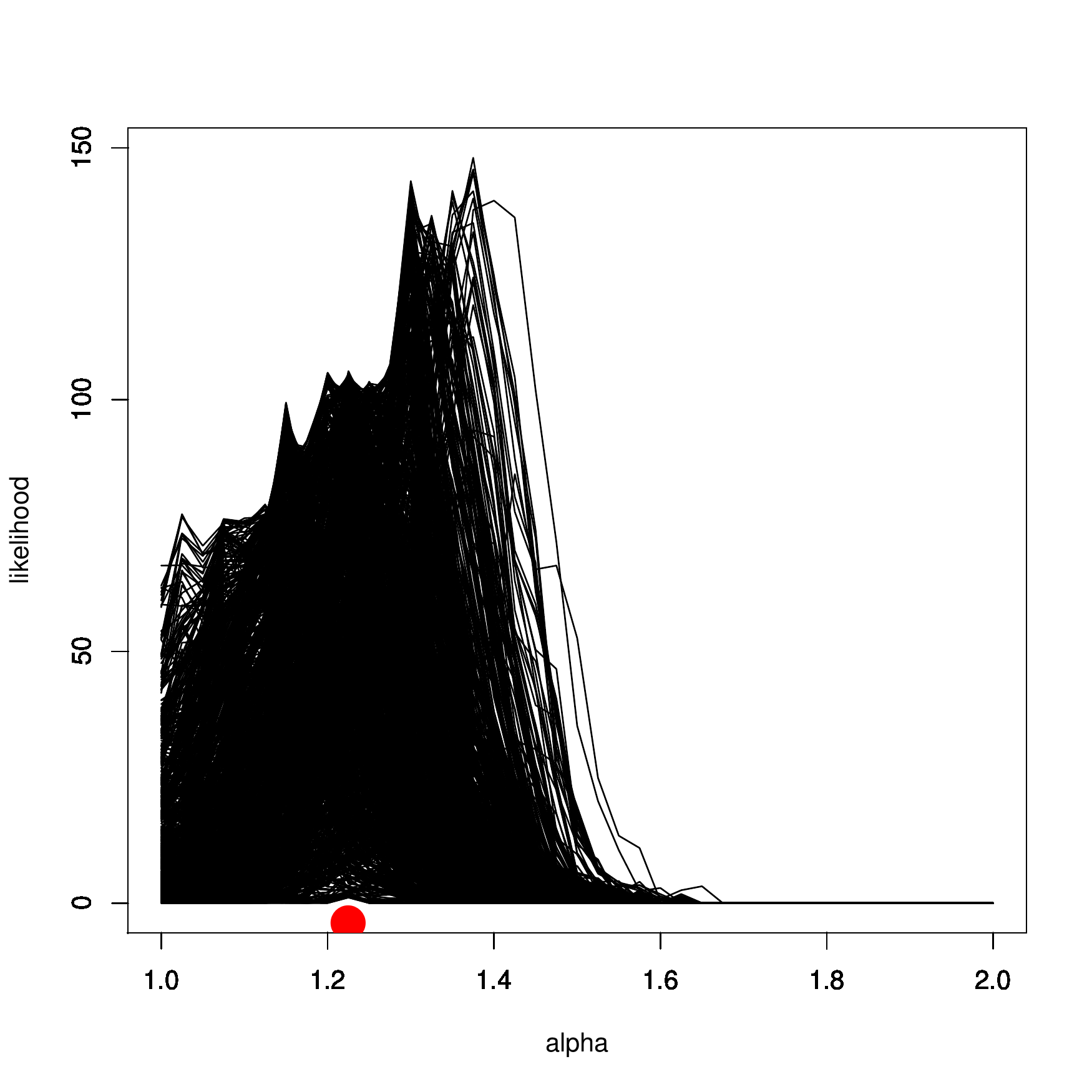}
\includegraphics[width = 0.24 \linewidth]{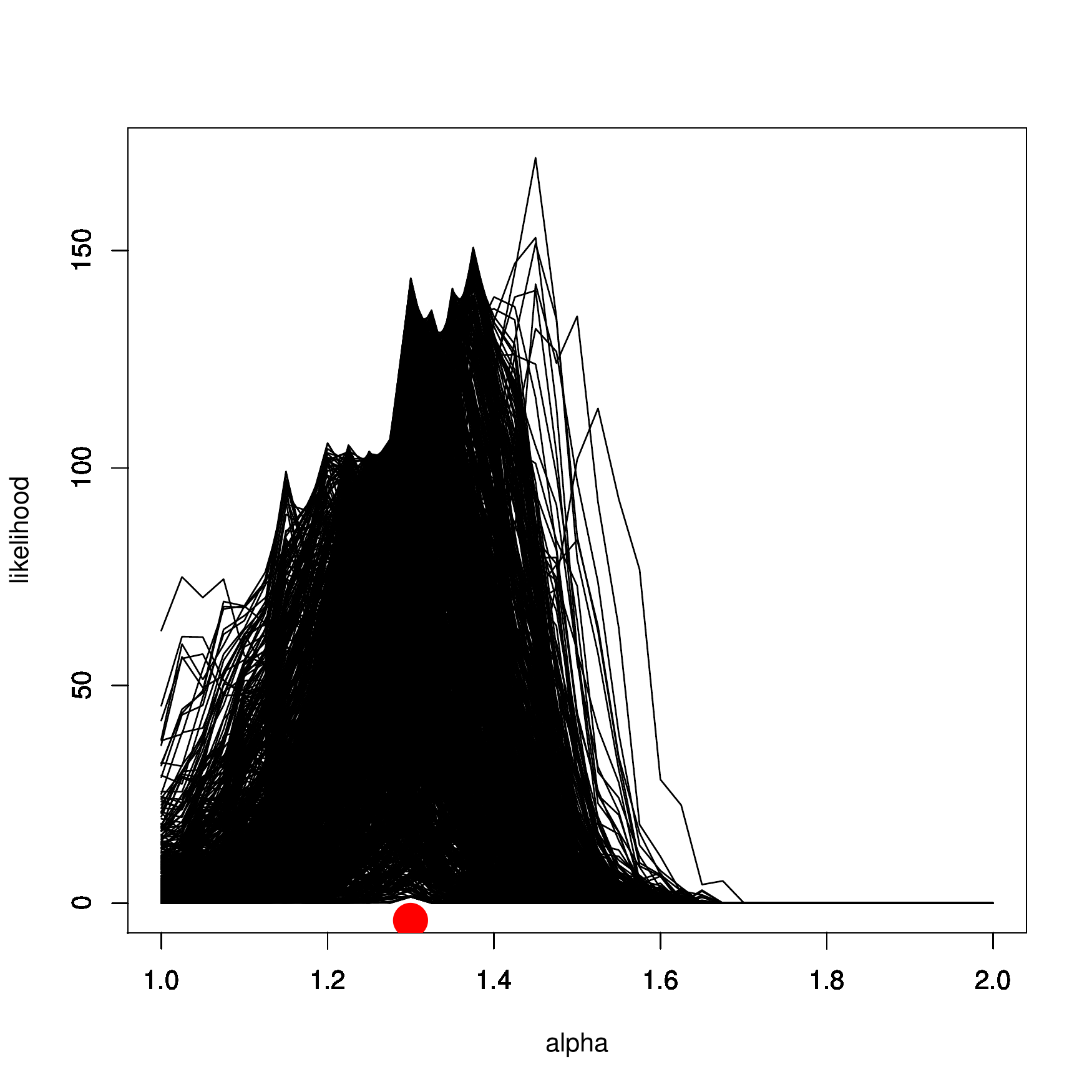}
\includegraphics[width = 0.24 \linewidth]{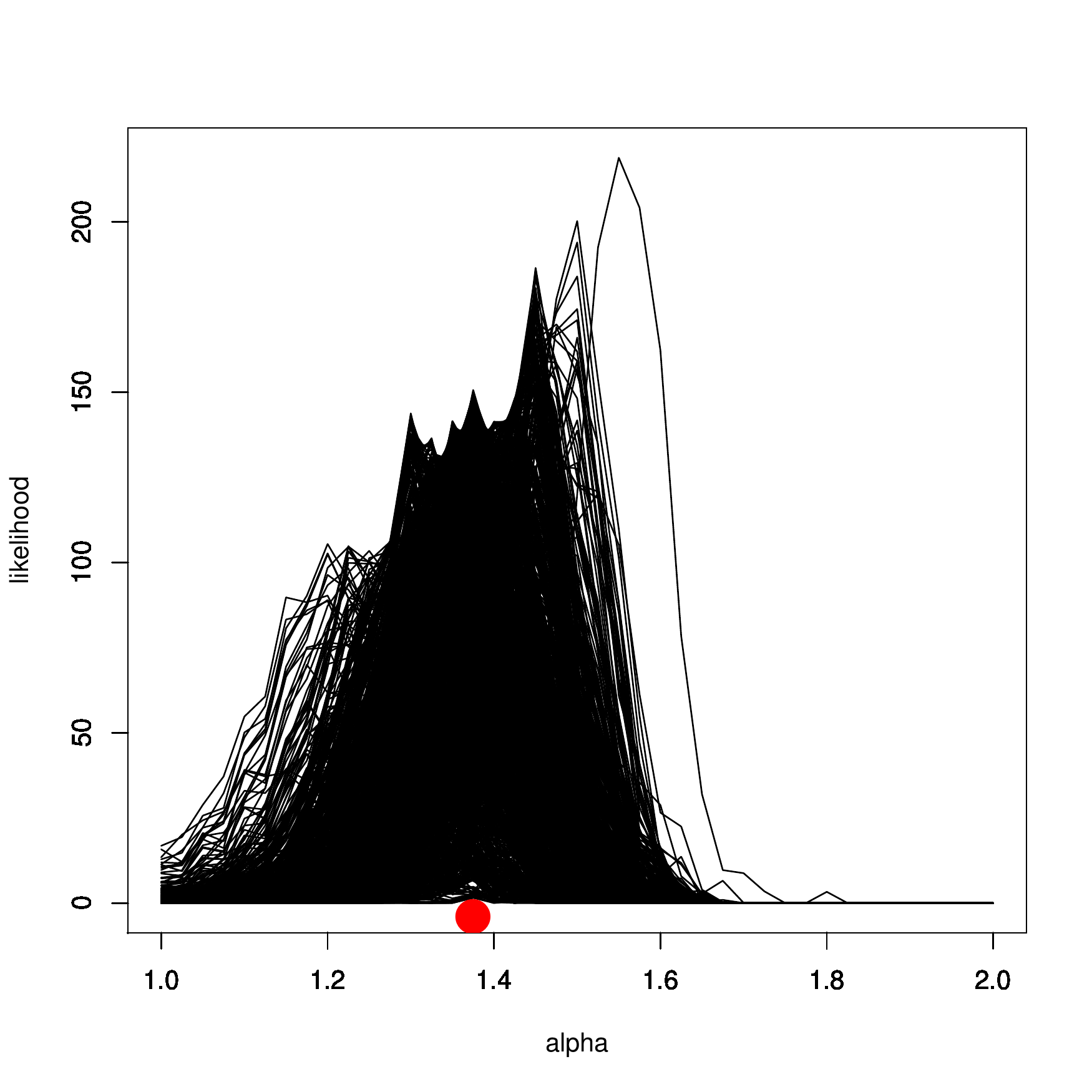}
\includegraphics[width = 0.24 \linewidth]{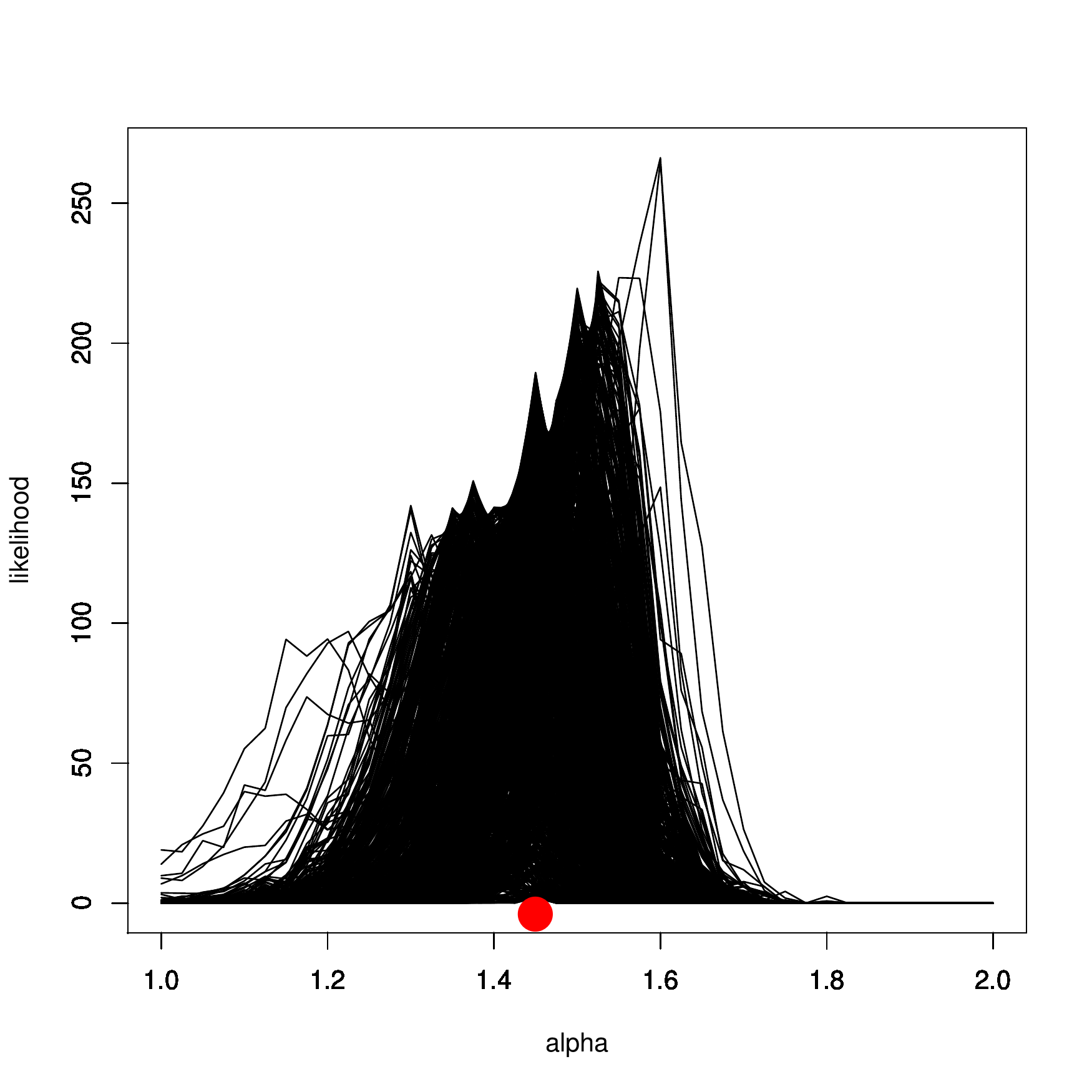}
\includegraphics[width = 0.24 \linewidth]{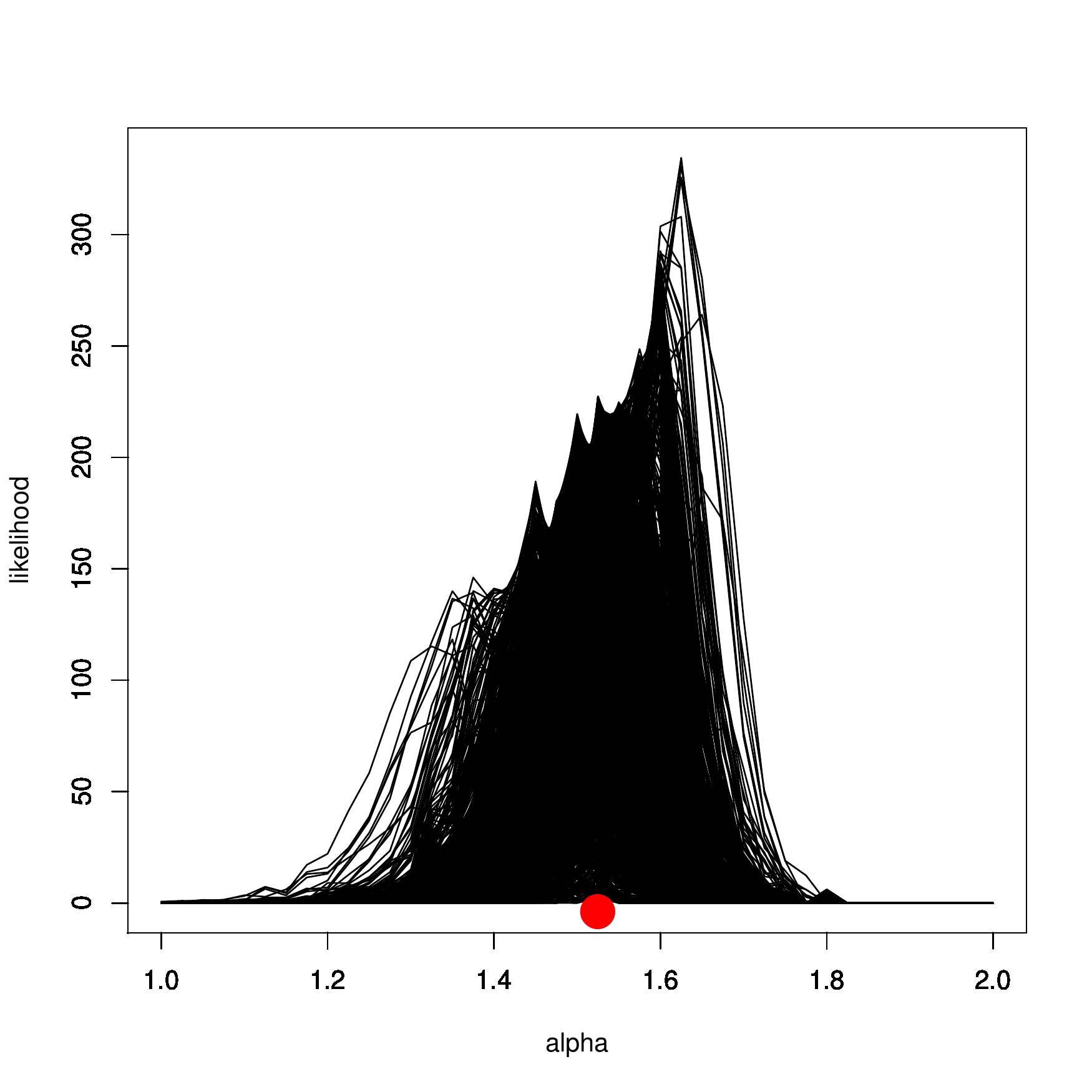}
\includegraphics[width = 0.24 \linewidth]{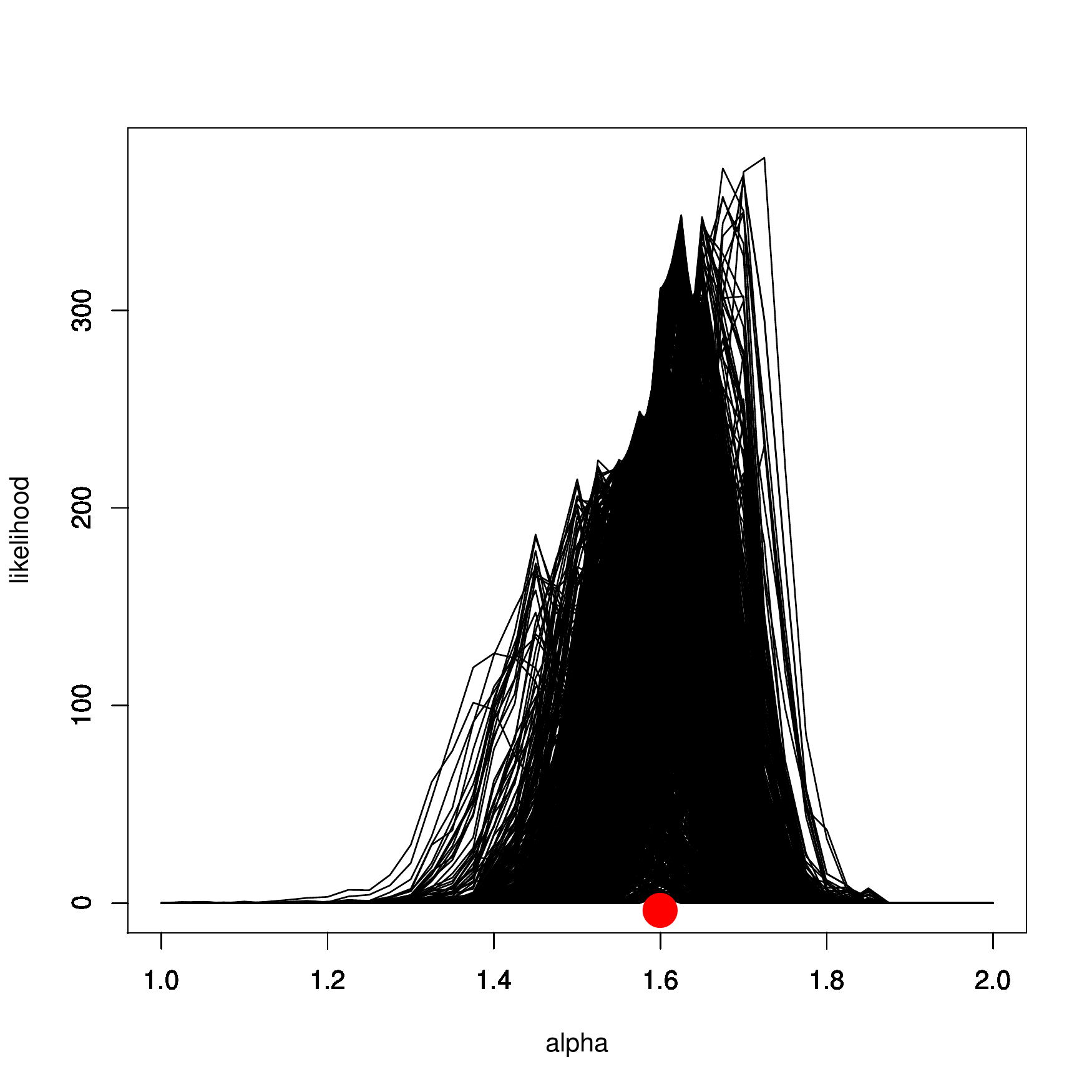}
\includegraphics[width = 0.24 \linewidth]{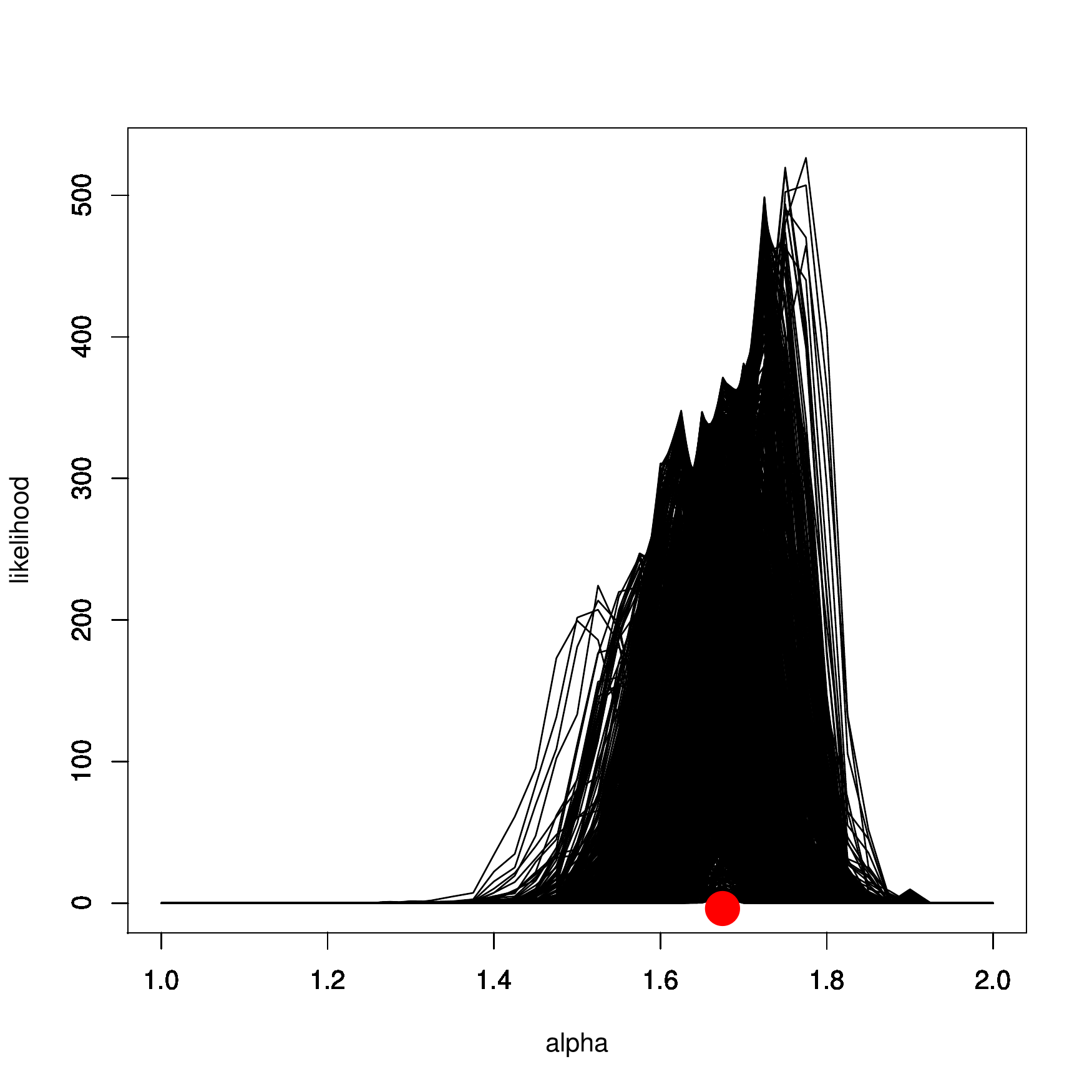}
\includegraphics[width = 0.24 \linewidth]{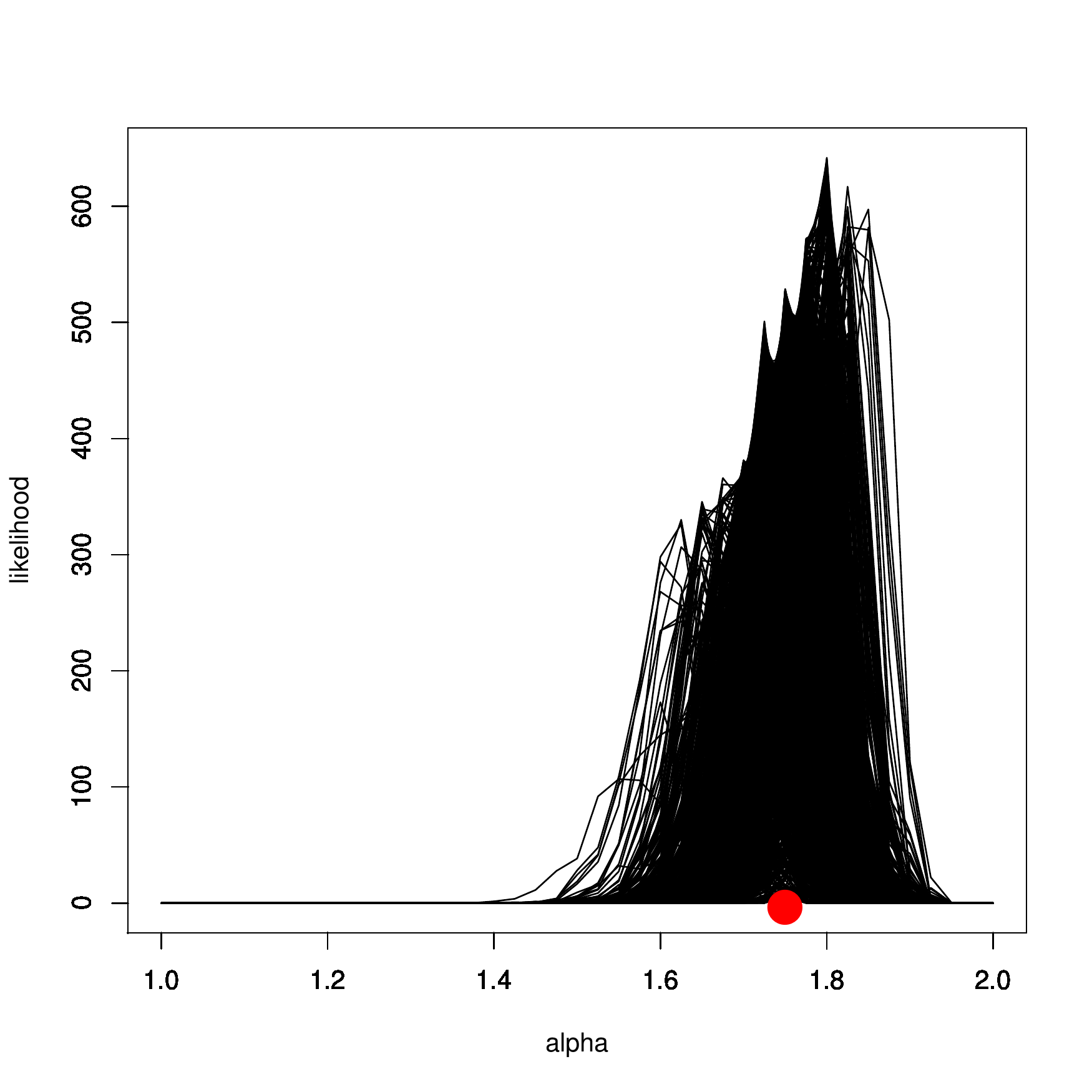}
\includegraphics[width = 0.24 \linewidth]{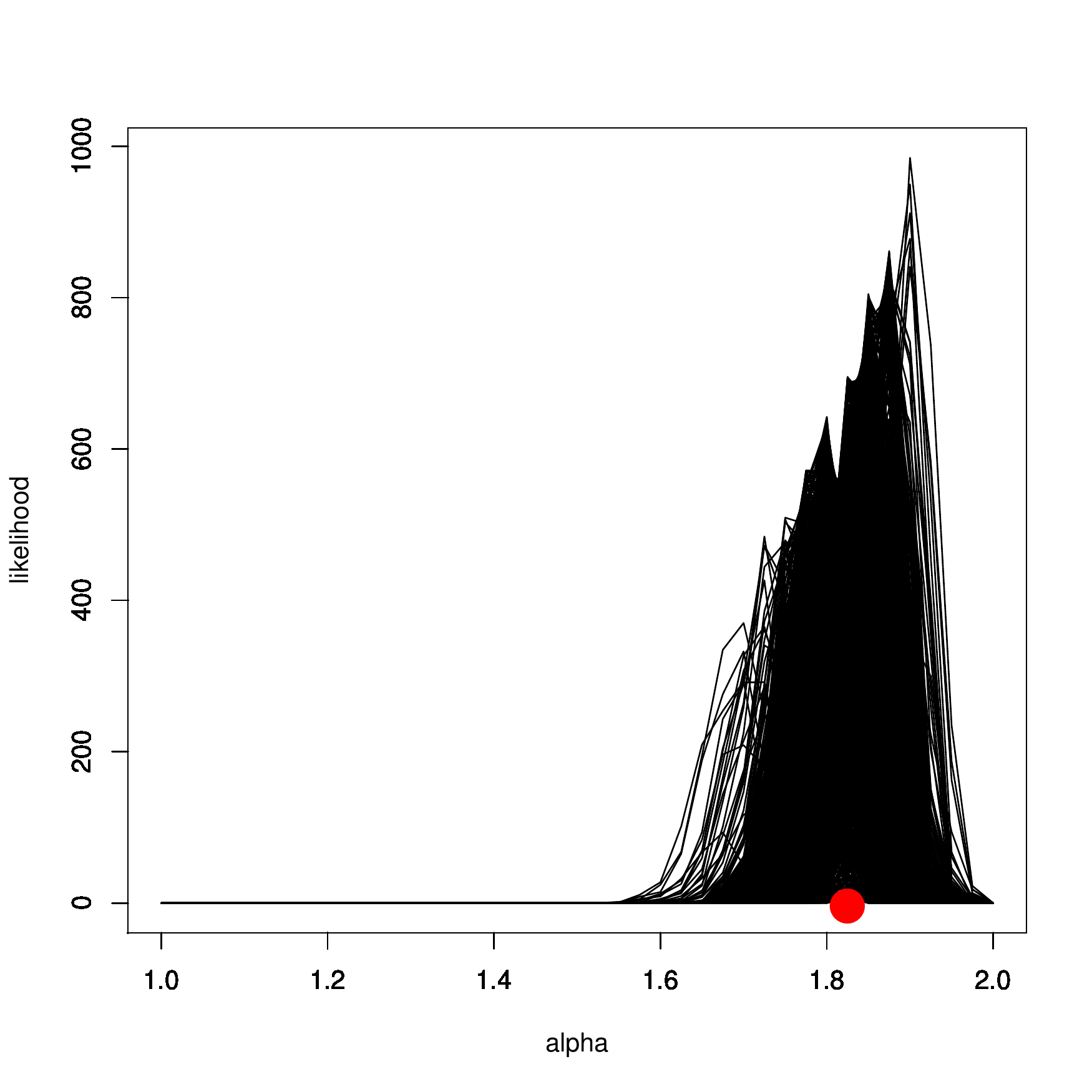}
\includegraphics[width = 0.24 \linewidth]{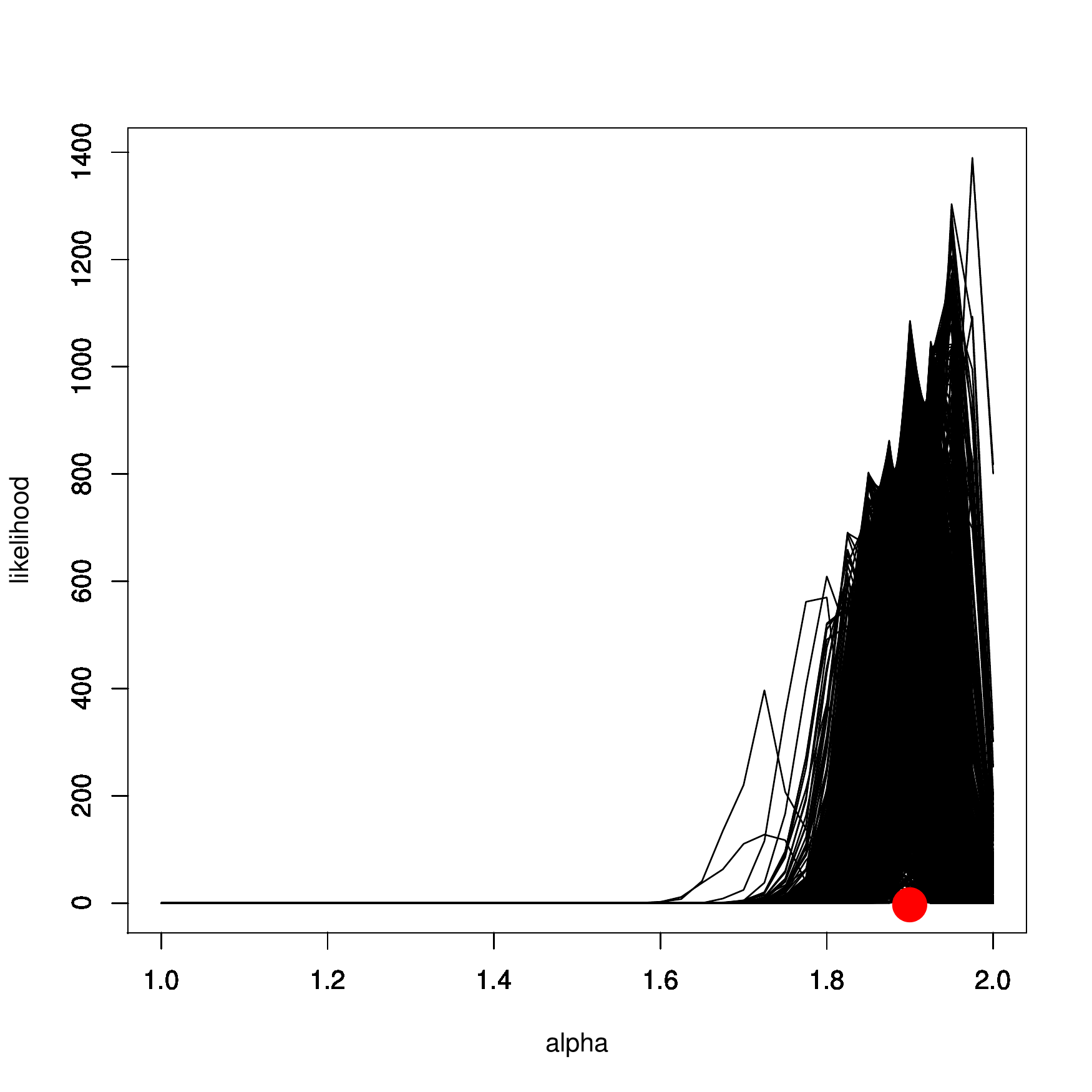}
\includegraphics[width = 0.24 \linewidth]{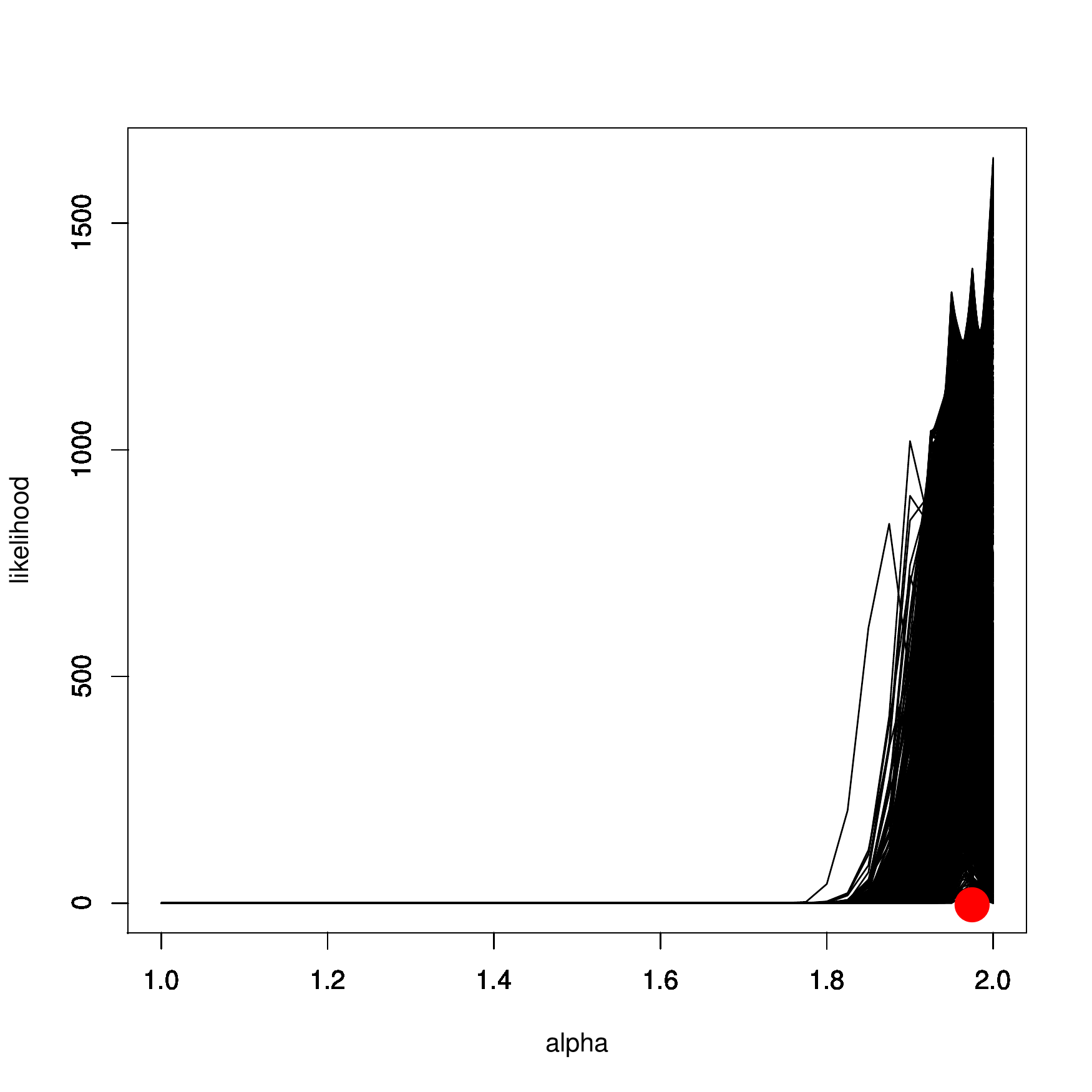}
\caption{1000 simulated likelihood curves for various values of $\alpha$, averaging over $L = 23$ loci with lumping at $k = 15$ and sample size $n = 500$. The red dots along the x-axis denote the true, data generating values.
Post-processing these graphs from the simulated data in \textsf{R} took 16 minutes on a single Intel i5-2520M 2.5 GHz processor.}
\label{params}
\end{figure}

Figure \ref{params} shows 1000 replicates of simulated likelihood functions with parameter space $\Theta_1$ from Section \ref{results}.
These have been produced by computing kernel density estimators $\hat{ P }^{ \Pi, \theta }$ for the likelihood functions as outlined in Section \ref{results}, and using these estimators in place of the true, intractable likelihood.
The curves are centered around the true values of $\alpha$ on average, showing little evidence of bias from the use of an approximate likelihood function.
However, it is clear that realisations resulting in a substantial mismatch between the truth and the maximiser of the approximate likelihood are possible, especially for lower values of $\alpha$.
Note, however, that the Kingman coalescent, $\alpha = 2$, can be excluded with high confidence as soon as $\alpha \lesssim 1.9$.
We emphasize that this comparison is different to the hypothesis tests in Section \ref{results} because the comparison is between $\alpha = 2$ and $\alpha < 2$, and not between the two model classes $\Theta_0$ and $\Theta_1$.

\begin{figure}[!ht]
\centering
\includegraphics[width = 0.5 \linewidth]{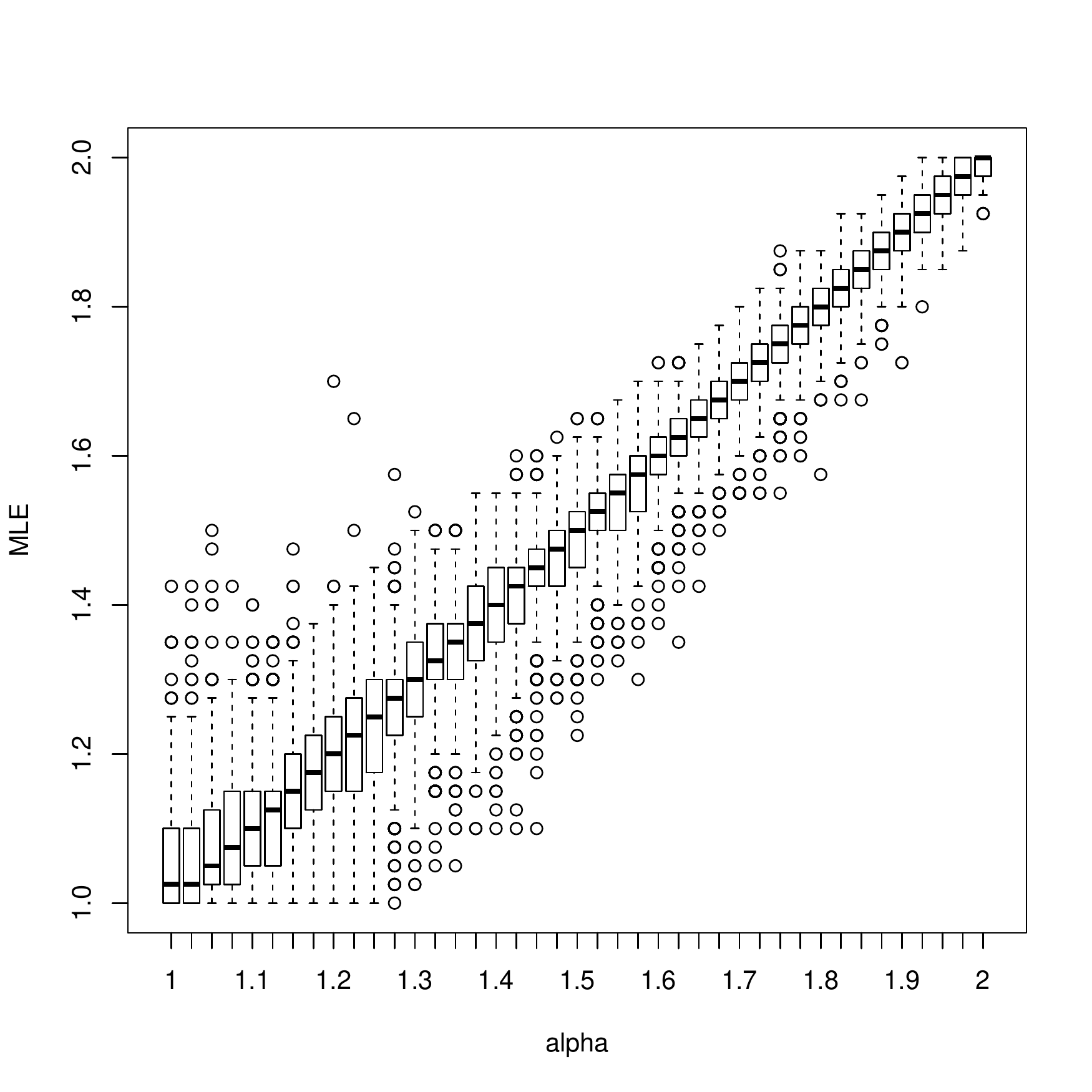}
\caption{Box plots of 1000 realisations of maximum likelihood estimators as a function of the true $\alpha$ when $L = 23$, $k = 15$ and $n = 500$ as in Figure \ref{params}.}
\label{mles}
\end{figure}

Figure \ref{mles} demonstrates the performance of the resulting approximate maxium likelihood estimator for $\alpha$.
The plot clearly centres on a straight, diagonal line confirming that the using an approximate likelihood results in no discrenable bias.
As suggested by Figure \ref{params}, the distribution of the maximum likelihood estimator has heavy tails, but Figure \ref{mles} also shows that the two central quantiles concentrate tightly around the true parameter values.

\begin{figure}[!ht]
\centering
\includegraphics[width = 0.49 \linewidth]{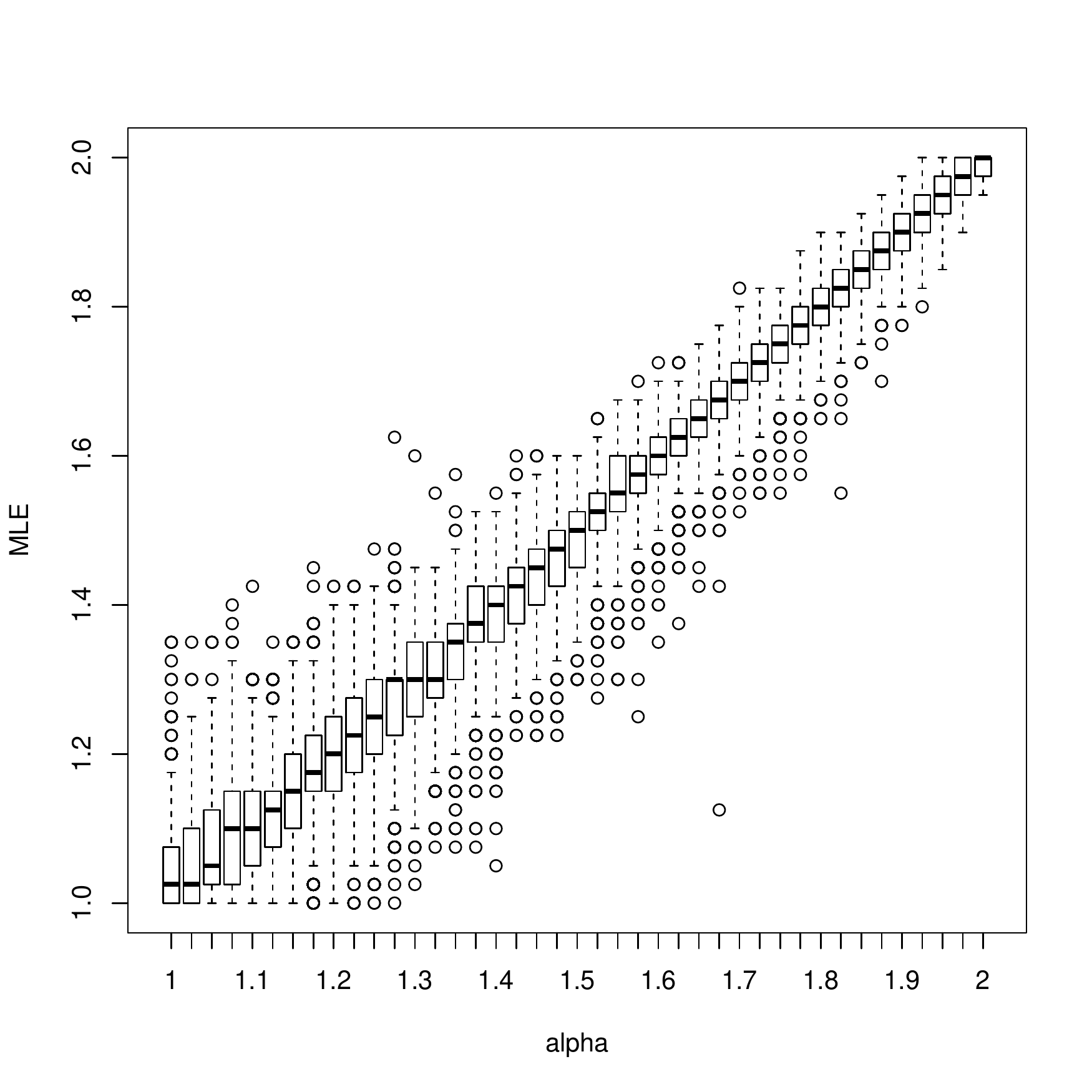}
\includegraphics[width = 0.49 \linewidth]{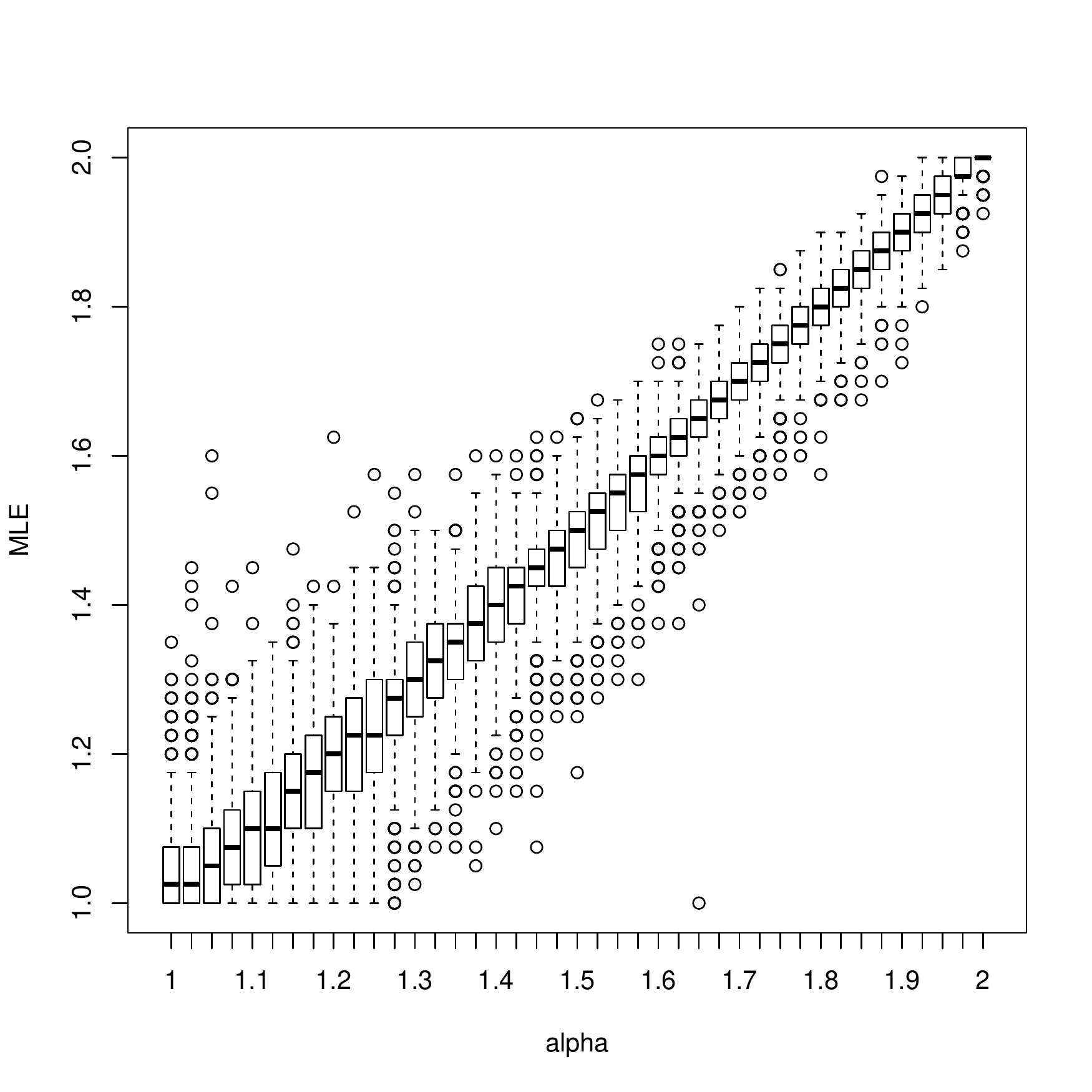}
\caption{The analysis in Figure \ref{mles} repeated using a mutation rate that is too low (left) or high (right) by a factor of ten.}
\label{mles_wrong_mut}
\end{figure}

Finally, Figure \ref{mles_wrong_mut} demonstrates that parameter inference is also robust to misspecification of the mutation rate.
The distribution of maximum likelihood estimators is essentially unchanged when the mutation rate is misspecified by a factor of 10, and in particular the use of an approximate likelihood still results in no bias.
Hence, use of the Watterson estimator is justifiable with real data sets in this context, as well as for model selection, when the true mutation is unknown.

\section{Discussion}\label{discussion}

We have shown that modelling the sampling distribution of a well chosen, low-dimensional function of the SFS based on simulated realisations yields statistical tests which can reliably distinguish multiple merger coalescents from population growth.
The same statistics can also be used for parameter inference, albeit with a somewhat lower degree of confidence.
The low dimensionality of \eqref{mean_sfs}, our function of choice, ensures that a simulation based approach is scalable.
Our tests focused on hundreds of samples and tens of loci, and the run times for both simulating the various data sets and computing all the size and power curves in Sections \ref{results} and \ref{parameters} were reasonable, even in serial and using a modest processor.
Since much of this time was spent producing independent simulation replicates, it is obvious that parallel computing can yield substantial improvement.

A drawback of our method is that it relies on the singleton class of the SFS, which is known to be prone to errors \citep{achaz08}.
In addition to improving statistical power, averaging across multiple loci helps with this problem as well.
If errors are equally likely to either miss a true singleton, or to create a false one, then the average number of singletons across multiple loci is far more robust than the singletons observed on any individual locus.
However, if errors are not symmetric, then high-quality data will be needed to make accurate inferences.
Our results demonstrate that moderate sample sizes suffice for high-power experiments, and hence resources should be invested in high accuracy sequencing of a large number of loci instead.

Robustness of both our model selection and parameter inference to misspecification of the mutation rate helps alleviate the computational cost of our method, as otherwise data sets corresponding to a grid of mutation rates would have to be simulated.
This results in either faster results, or enables the analysis of more complex coalescent families such as the $\operatorname{Beta}( \beta_1, \beta_2 )$-$\Xi$-coalescents for arbitrary $\beta_1 > 0$ and $\beta_2 > 0$.
However, it is not clear whether such coalescents can be derived from a biologically relevant individual based models, and what the corresponding real time embeddings are.
Other more complex hypotheses, such as (weak) selection under a Kingman coalescent, or spatial structure, could also be incorporated into the tests, provided that SFS data can be simulated under the corresponding models.
Off-the-shelf methods for kernel density estimation in up to six dimensions exist (c.f.~the \texttt{ks} package in \textsf{R}), which further suggests that more complex scenarios can be handled within our framework as it is not restricted to two dimensional summary statistics.

In addition to hypothesis testing and inference, our method also lends itself to design of experiments by providing estimates of likelihood functions and statistical power under various scenarios.
In particular, we have been able to show that 500 samples of 23 unlinked loci results in very high power to distinguish $\operatorname{Beta}( 2 - \alpha, \alpha)$-$\Xi$-coalescents from simple population growth models across most of the $\alpha \in [1, 2]$ parameter range.
By way of example, previous estimates have identified $\alpha = 1.5$ as a plausible value for Atlantic cod \citep[Table 2]{BBS2011}, for which our two hypotheses can be distinguished essentially with certainty.
The dramatic improvement in power yielded by increasing the number of loci being averaged in Figure \ref{loci} indicates that it is advisable to sequence as many unlinked loci as possible, in which case sample sizes in the hundreds are sufficient.
More complex questions of experimental design will require bespoke simulations run on a case by case basis, but our method serves as a clear guide for how such simulations can be conducted.

\section*{Acknowledgements}

The author is grateful to Bjarki Eldon, Jochen Blath and Matthias Birkner for their help and suggestions concerning summary statistics for multiple merger detection, and was supported by Deutsche Forschungsgemeinschaft (DFG) grant BL 1105/3-2.
Part of this work was completed while visiting Matthias Birkner at the Johannes Gutenberg Universit\"at Mainz, funded by DFG grant BI 1058/2-2.
Both grants are part of SPP Priority Programme 1590.

\section*{Appendix}

A $\Lambda$-coalescent for $L$ unlinked loci with $n_i( t )$ lineages at locus $i \in [ L ]$ can be simulated naively by setting $N := \sum_{ i = 1 }^L n_i( t )$, sampling a merger size $K$ from $\{ 2, \ldots, N \}$ proportional to weights
\begin{equation}\label{weights}
\binom{ N }{ j } \int_0^1 x^{ j - 2 } ( 1 - x )^{ N - j } \Lambda( dx ),
\end{equation}
selecting $K$ lineages uniformly at random without replacement from among the $N$ possible segments, merging all groups of lineages which occupy the same locus, and incrementing time by a step drawn from an exponential distribution with rate given by the sum of the weights \eqref{weights}.
The remaining sample size $N$ can then be updated, and the process iterated until the most recent common ancestor (MRCA) is reached, at which point the algorithm terminates.
Diploid, biparental $\Xi$-coalescent models developed in \citep{BBE13a} can also be incorporated by further randomly assigning all lineages into 4 groups, and only merging lineages which occupy the same locus and the same group.

However, normalising the weights \eqref{weights} is an $O( N )$ operation, which is expensive when the sample size and number of loci are large.
Moreover, the resulting merger is unlikely to sample two or more lineages occupying the same locus unless $\Lambda( dx )$ is concentrated near 1 (resulting in very large mergers), so that most steps of the algorithm result in no progress towards the MRCA.
In practice, this naive method was infeasibly slow for our simulation-based model selection.
Here we present a faster alternative based on rejection sampling.

The following is a modification of the Poisson construction of $\Xi$-coalescents given in \citep[Section 1.4]{BBMST09} to the present multi-locus, biparental, diploid $\Xi$-coalescents.
Consider a Poisson point process $\Upsilon$ on $[ 0, \infty ) \times [ 0, 1 ] \times [ 0, 1 ]^{ \N \times [ L ] }$ with intensity
\begin{equation*}
d t \otimes \frac{ \Lambda( d x ) }{ x^2 } \otimes U[ 0, 1 ]^{ \otimes ( [ L ] \times \N ) }.
\end{equation*}
At each point 
\begin{equation*}
( t_i, y_i, u_i ) \in \Upsilon,
\end{equation*}
that is, at the time of the $i^{ \text{th} }$ point $t_i$, each of the $n_i( t ) \le n_i$ active lineages at each locus checks whether $u_i( j, k ) \le y_i$ for each $k \in [ L ], j \in [ n_i( t ) ]$.
All lineages for which this is the case take part in a merger within the locus.
These merging lineages are the uniformly at random assigned into one of four groups, and any groups with more than one member are merged into a common ancestor.

A technical problem from the simulation point of view is that the intensity measure of our Poisson point process $\Upsilon$ is infinite for $\Lambda( dx )$ corresponding to a $\operatorname{Beta}( \beta_1, \beta_2 )$-coalescent, and hence also for a $\operatorname{Beta}( 2 - \alpha, \alpha )$-$\Xi$-coalescent for any $\alpha \geq 0$.
However, if we restrict $\Upsilon$ to the set where $u_i( j, k ) \le y_i$ for at least one $j \in [ L ]$ and two distinct $k$s from $[ n_j( t ) ]$, the restricted intensity  is finite, and given by
\begin{equation*}
\int_0^1 f( x; n_1( t ), \ldots, n_L( t ) ) \Lambda( dx ),
\end{equation*}
where
\begin{equation}\label{f_defn}
f( x; n_1( t ), \ldots, n_L( t ) ) := \frac{ 1 - \prod_{ j = 1 }^L \big( ( 1 - x )^{ n_j( t ) }+ n_j( t ) x ( 1 - x )^{ n_j( t ) - 1 } \big) }{ x^2 }.
\end{equation}
This restricted intensity corresponds to thinning $\Upsilon$ by only retaining those events in which there is at least one locus in which two lineages successfully participate in the event.

An alternative simulation mechanism is thus as follows: Given $n_1( t ), \dots, n_L( t )$, consider a new {\em thinned} Poisson point process $\tilde \Upsilon$ with intensity
\begin{equation}\label{finite_lambda}
d t \otimes \tilde \Lambda( dx ),  
\end{equation}
where $\tilde{ \Lambda }( dx ) := f( x; n_1( t ), \ldots, n_L( t ) ) \Lambda( dx )$ determines the  merging times and success probabilities $( t_i, y_i )$. 
Given $( t_i, y_i )$, which now appear with finite rate, independently and uniformly choose a {\em pair} of lineages within a locus to merge out of the
\begin{equation}\label{eq:c_def}
\binom{ n_1( t ) }{ 2 } + \dots + \binom{ n_L( t ) }{ 2 } =: c( t )
\end{equation}
within-locus pairs. 
Further mergers or merging lineages can then be added according to independent uniform coin flips with success probability $y_i$.
As before, all merging lineages within a locus have to be grouped into four uniform classes in the diploid case, and only lineages within the same locus and class can merge.
If no class contains at least two lineages, then nothing happens.

It remains to specify a way of sampling the pairs $( t_i, y_i )$ with intensity \eqref{finite_lambda}.
Note that the function $f( x; n_1( t ), \ldots, n_L( t ) )$ is maximised at $x = 0+$ for fixed $( n_1( t ), \ldots, n_L( t ) )$, where it takes the value $c( t )$ from \eqref{eq:c_def}.
Thus, pairs with the correct intensity can be generated by sampling a time $T^{ ( 1 ) } \sim \operatorname{Exp}( c( t ) )$, an independent $U^{ (1) } \sim U[ 0, 1 ]$ and a further independent $Y^{ (1) } \sim \Lambda$, and by checking whether 
\begin{equation*}
U^{ ( 1 ) } \leq \frac{ f( Y^{ ( 1 ) }; n_1( t ), \ldots, n_L( t ) ) }{ c( t ) }.
\end{equation*}
If yes, accept $( T^{ ( 1 ) }, Y^{ ( 1 ) } )$ for $( t_1, y_1 )$. 
If not, repeat this procedure, generating independent tuples $( T^{ ( 2 ) }, U^{ ( 2 ) }, Y^{ ( 2 ) } )$ etc. 
If the first success requires $k$ steps, accept $( T^{ ( 1 ) } + \dots + T^{ ( k ) }, Y^{ ( k ) } )$ for $( t_1, y_1 )$, at which point the merger event can be simulated and $n_1( t_1 ), \dots, n_L( t_1 )$, $f( \cdot; n_1( t ), \ldots n_L( t ) )$ and $c( t )$ updated accordingly. 
This procedure is then repeated to obtain $( t_2, y_2 )$, and so on, until the most recent common ancestor is reached.
\vskip 11pt
\begin{rmk}
This method is straightforward to adapt to the case of variable, deterministic population size by first generating the next coalescence time under a constant population size, and then adjusting it to the variable population size scenario using the method of \citet{DT1995}.
\end{rmk}
\vskip 11pt
\begin{rmk}
This procedure also covers the Kingman coalescent case $\Lambda = \delta_0$, in which $Y^{ ( 1 ) } \equiv 0$ and $f( 0, n_1( t ), \ldots, n_L( t ) ) \equiv c( t )$.
This means that only one pair of lineages ever merges at any given time, the coin flips for all other lineages need not be simulated, and all proposed mergers are accepted.
More general atoms at 0 can also be handled similarly by first sampling $Y^{ ( 1 ) } \sim \Lambda$, and then setting $f( 0; n_1( t ), \ldots, n_L( t ) ) = c( t )$ if $Y^{ ( 1 ) } = 0$.
\end{rmk}

For $x$ close to $0$, the function $f( x; n_1( t ), \ldots, n_L( t ) )$ is numerically unstable because of the very small number in the denominator in \eqref{f_defn}, rendering the procedure outlined above unreliable. 
Note also that it is in fact a polynomial in $x$ and for small arguments, readily represented via
\begin{equation}\label{eq:f_expand2}
f( x; n_1( t ), \ldots, n_L( t ) ) = \sum_{ j = 1 }^L\binom{ n_j( t ) }{ 2 } - 2 \sum_{ j = 1 }^m \binom{ n_j( t ) }{ 3 } x + O( x^2 ),
\end{equation} 
as can be verified as follows.
For $n \in \N$ we have 
\begin{align} 
( 1 - x )^n + n x ( 1 - x )^{ n - 1 } &= 1 + \sum_{ k = 1 }^n \binom{ n }{ k } ( -x )^k + \sum_{ j = 0 }^{ n - 1} n \binom{ n - 1 }{ j } ( -1 )^j x^{ j + 1 } \nonumber \\
&= 1 + \sum_{ k = 2 }^n ( -x )^k \left[ \binom{ n }{ k } - n \binom{ n - 1 }{ k - 1 } \right] = 1 - \sum_{ k = 2 }^n ( k - 1 ) \binom{ n }{ k } ( -x )^k \label{single_index_exp}
\end{align}
where we have used the fact that the terms for $k = 1$ and for $j = 0$ in the first line cancel, then
replaced $j = k - 1$ in the second sum in the first line and used 
\begin{equation*}
\binom{ n }{ k } - n \binom{ n - 1 }{ k - 1 } = \frac{ n! }{ k! ( n - k )! } - \frac{ n ( n - 1 )! }{ ( k - 1 )! ( n - k )! } 
= \binom{ n }{ k } ( 1 - k )
\end{equation*}
in the second line. 
Thus 
\begin{align*} 
\prod_{ j = 1 }^L \left( ( 1 - x )^{ n_j( t ) } + n_j( t ) x ( 1 - x )^{ n_j( t )-1 }\right) &= \prod_{ j = 1 }^L \left( 1 - \binom{ n_j( t ) }{ 2 } x^2 + 2 \binom{ n_j( t ) }{ 3 } x^3 + O( x^4 ) \right) \\
& = 1 - \sum_{ j = 1 }^L \binom{ n_j( t ) }{ 2 } x^2 + 2 \sum_{ j = 1 }^L \binom{ n_j( t ) }{ 3 } x^3 + O( x^4 )
\end{align*}
and subtracting from 1 before dividing by $x^2$ to obtain $f(x; n_1( t ), \ldots, n_L( t ) )$ yields \eqref{eq:f_expand2}. 

Numerical experiments were performed to determine that the polynomial approximation \eqref{eq:f_expand2} needed to be used whenever $x \leq 10^{ -4 }$.
The form of higher order terms in \eqref{single_index_exp} shows that the constant of proportionality of the $O( x^k )$ term in \eqref{eq:f_expand2} contains factors that are $O( n^k )$.
Hence, the algorithm cannot remain accurate for samples of size $n = 10^4$, as then approximation errors are $O( 1 )$ for a polynomial approximation of \emph{any} order, short of the trivial case of using the full, finite expansion which recovers the function $f( x; n_1( t ), \ldots, n_L( t ) )$ exactly.
We remark that using the full polynomial expansion of $f( x; n_1( t ), \ldots, n_L( t ) )$ to compute it exactly is also prone to numerical instability due to alternating signs.
However, the first order truncation shown in \eqref{eq:f_expand2} was found to be fast, accurate and stable for the sample sizes considered in this paper, which we have also shown to be sufficient for our aim of distinguishing multiple merger coalescents from exponential or algebraic population growth models.

We conclude this appendix by specifying how the summary statistic \eqref{mean_sfs} can be computed from data sets, as well as simulated realisations of the coalescent tree.
The form of \eqref{mean_sfs} as an average across loci implies that it is sufficient to consider a single locus in both cases.

First, consider a SNP data set of $n$ individuals sequenced at a single locus.
We assume the data is consistent with the infinite alleles model of mutation, and associate to individual $i$ the list $x_i = ( l_i( 1 ), \ldots, l_i( k_i ) )$, where $l_i( q )$ is the position of the $q^{\text{th}}$ derived allele from the left on individual $i$, and $k_i$ is the number of mutations carried by that individual.
For an location $l$ on the locus, we say $l \in x_i$ if $l = l_i( q )$ for some $q \in \{ 1, \ldots, k_i \}$, and $l \neq x_i$ otherwise.
Then the pair $( \zeta_1^{ ( n ) }, \bar{ \zeta }_k^{ ( n ) } )$ can be computed as
\begin{equation*}
( \zeta_1^{ ( n ) }, \bar{ \zeta }_k^{ ( n ) } ) = \sum_{ i = 1 }^n \sum_{ l \in x_i } \left( \prod_{ j \neq i }^n \mathds{ 1 }_{ \{ l \notin x_j \} }, \mathds{ 1 }_{ \{ \exists j_1 \neq \ldots \neq j_k : l \in x_{ j_1 }, \ldots, l \in x_{ j_k } \} } \right).
\end{equation*}

Now consider a simulated ancestral tree for $n$ individuals.
The tree consist of $q \in \{ n, \ldots, 2 n \}$ branches, which we assume are labelled with mutations.
We impose an arbitrary ordering on the branches, and denote the number of mutations on branch $i$ by $m_i$, the set of indices of the immediate child branches of branch $i$ by $c_i$, and the number of leaves subtended by branch $i$ by $s_i$.
The $s_i$'s can be computed recursively for a given tree by setting
\begin{equation*}
s_i = 
\begin{cases}
1 &\text{ if } c_i = \emptyset, \\
\sum_{ j \in c_i } s_j &\text{ otherwise}.
\end{cases}
\end{equation*}
Once the $s_i$'s are computed, the summary statistic can be expressed as
\begin{equation*}
( \zeta_1^{ ( n ) }, \bar{ \zeta }_k^{ ( n ) } ) = \sum_{ i = 1 }^q m_i ( \mathds{ 1 }_{ \{ s_i = 1 \} }, \mathds{ 1 }_{ \{ s_i \geq k \} } ).
\end{equation*}

\bibliographystyle{plainnat}
\bibliography{old_power}

\end{document}